\documentclass[acmsmall,natbib=true,screen,anonymous=False]{acmart}
\usepackage[utf8]{inputenc}

\usepackage{booktabs} 
\usepackage{algorithm, algpseudocode}
\usepackage{amsmath}
\usepackage{graphics}
\usepackage{epsfig}
\usepackage{graphicx}
\usepackage{xcolor}
\usepackage{balance}
\usepackage{multirow}
\usepackage{mathrsfs}
\usepackage{acronym}
\usepackage{placeins}
\usepackage{xcolor}
\usepackage[inline]{enumitem}
\usepackage{fancyhdr}
\usepackage{amsfonts}
\usepackage{svg}
\usepackage{threeparttable}
\usepackage{subfig}
\usepackage{float}
\usepackage{array}
\usepackage{fancyhdr}
\usepackage{bbm}

\AtBeginDocument{%
  \providecommand\BibTeX{{%
    \normalfont B\kern-0.5em{\scshape i\kern-0.25em b}\kern-0.8em\TeX}}}
    
\newcommand{\header}[1]{\vspace*{1mm}\noindent{\textbf{#1}}}

\setcopyright{acmcopyright}
\copyrightyear{2022}
\acmYear{2022} \acmVolume{1} \acmNumber{1} \acmArticle{1} \acmMonth{1}
\acmDOI{10.1145/3568954}
\acmPrice{15.00}
\acmJournal{TOIS}
\def\@journalName{ACM Transactions on Information Systems}%
\def\@journalNameShort{ACM Trans. Inf. Syst.}%
\def\@permissionCodeOne{1046-8188}%

\begin{document}
\title[On the User Behavior Leakage from Recommender Exposure]{On the User Behavior Leakage from Recommender System Exposure}
\author{Xin Xin}
\authornote{Both authors contributed equally to this research.}
\email{xinxin@sdu.edu.cn}
\affiliation{%
  \institution{Shandong University}
  \city{Qingdao}
  \country{China}
}
\author{Jiyuan Yang}
\authornotemark[1]
\email{jiyuan.yang@mail.sdu.edu.cn}
\affiliation{%
  \institution{Shandong University}
  \city{Qingdao}
  \country{China}
}

\author{Hanbing Wang}
\email{hanbing.wang@mail.sdu.edu.cn}
\affiliation{%
  \institution{Shandong University}
  \city{Qingdao}
  \country{China}
}

\author{Jun Ma}
\email{majun@sdu.edu.cn}
\affiliation{%
  \institution{Shandong University}
  \city{Qingdao}
  \country{China}
}
\author{Pengjie Ren}
\email{renpengjie@sdu.edu.cn}
\affiliation{%
  \institution{Shandong University}
  \city{Qingdao}
  \country{China}
}
\author{Hengliang Luo}
\email{luohengliang@meituan.com}
\affiliation{%
  \institution{Meituan Inc.}
  \city{Beijing}
  \country{China}
}
\author{Xinlei Shi}
\email{shixinlei@meituan.com}
\affiliation{%
  \institution{Meituan Inc.}
  \city{Beijing}
  \country{China}
}
\author{Zhumin Chen}
\email{chenzhumin@sdu.edu.cn}
\affiliation{%
  \institution{Shandong University}
  \city{Qingdao}
  \country{China}
}
\author{Zhaochun Ren}
\authornote{Corresponding author.}
\email{zhaochun.ren@sdu.edu.cn}
\affiliation{%
  \institution{Shandong University}
  \city{Qingdao}
  \country{China}
}
\renewcommand{\shortauthors}{Xin Xin and Jiyuan Yang, et al.}

\begin{abstract}
Modern recommender systems are trained to predict users potential future interactions from users historical behavior data. During the interaction process, despite the data coming from the user side recommender systems also generate exposure data to provide users with personalized recommendation slates. Compared with the sparse user behavior data, the system exposure data is much larger in volume since only very few exposed items would be clicked by the user. 
Besides, the users historical behavior data is privacy sensitive and is commonly protected with careful access authorization. 
However, the large volume of recommender exposure data which is generated by the service provider itself usually receives less attention and could be accessed within a relatively larger scope of various information seekers or even potential adversaries.

In this paper, we investigate the problem of user behavior leakage in the field of recommender systems. We show that the privacy sensitive user past behavior data can be inferred through the modeling of system exposure. In other words, \emph{one can infer which items the \textbf{user} have clicked just from the observation of current \textbf{system} exposure for this user}. Given the fact that system exposure data could be widely accessed from a relatively larger scope, we believe that the user past behavior privacy has a high risk of leakage in recommender systems. 
More precisely, we conduct an attack model whose input is the current recommended item slate (i.e., system exposure) for the user while the output is the user's historical behavior.  Specifically, we exploit an encoder-decoder structure to construct the attack model and apply different encoding and decoding strategies to verify the attack performance. 
Experimental results on two real-world datasets indicate a great danger of user behavior leakage. 
To address the risk, we propose a two-stage privacy-protection mechanism which firstly selects a subset of items from the exposure slate and then replaces the selected items with uniform or popularity-based exposure. 
Experimental evaluation reveals a trade-off effect between the recommendation accuracy and the privacy disclosure risk, which is an interesting and important topic for privacy concerns in recommender systems.


\end{abstract}


\maketitle
\acresetall

\section{Introduction}
Recommender systems have been widely applied in various online web services, e.g., online shopping~\cite{reinforce-e-commerce}, video or music platforms~\cite{nextitnet}, news recommendation~\cite{qi2020fednews}, etc., to provide users with the most interesting items. 
Generally speaking, a recommender system targets on predicting the \textbf{\emph{user's}} future behavior based on the \textbf{\emph{user's}} historical interactions. Then the recommendation list is generated by selecting the most relevant items according to the predicted user behavior. The user historical behavior data is the keystone to conduct effective recommendation. Meanwhile, it is also highly privacy sensitive since various user profile information such as gender, age, and even political orientation could be inferred from the user behavior data~\cite{yuan2020peterec}. 
As a result, the user behavior data is protected with strict department-specific access authorization or regulations such as the General Data Protection Regulation (GDPR) \cite{otto2018GDPR}.

\begin{figure}
    \captionsetup[subfloat]
    {}
    \subfloat[]{
        \label{intorduction-a}
        \includegraphics[width=0.48\textwidth]{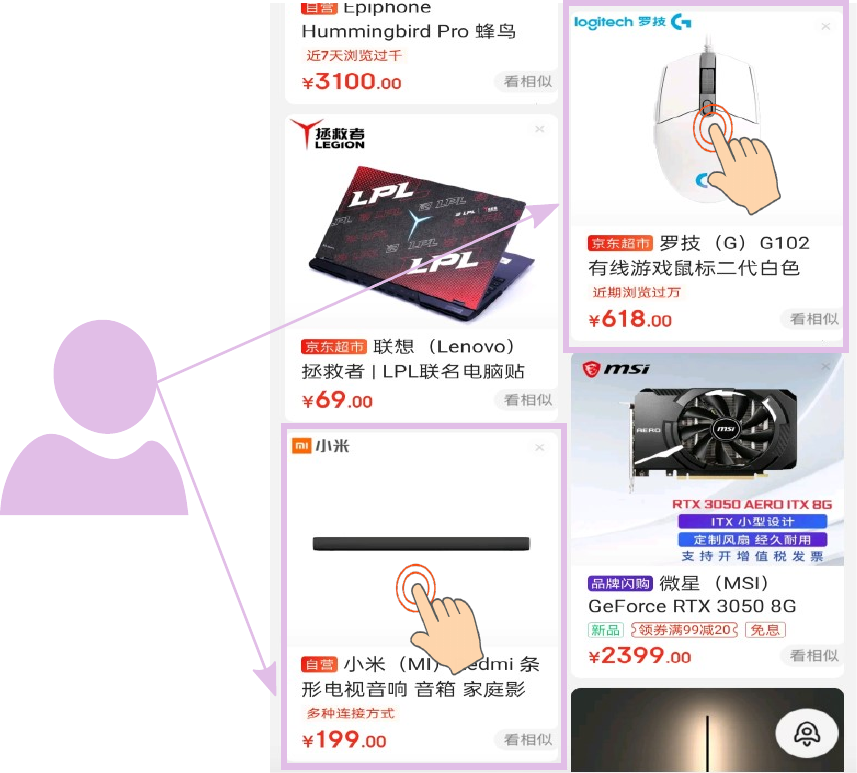}
    }
    \hspace{0.5cm}
    \subfloat[]{
        \label{intorduction-b}
        \includegraphics[width=0.33\textwidth]{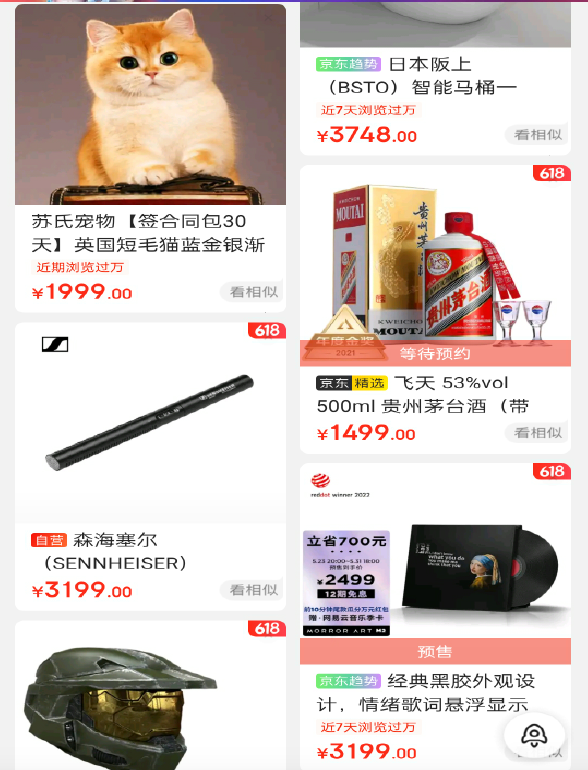}
    }
    \caption{Two illustrative examples of recommender exposure data. (a) shows that the user interacts with two items in the exposed item slate while (b) shows that there is no observed interactions between the exposed items and the user.}
    \label{fig:intorduction}
\end{figure}

During the interaction process between users and recommenders, despite the data coming from the user feedback, the system also generates personalized slates of items which are pushed to the user as the recommendation service. Such a kind of \textbf{\emph{system}} behavior data is also known as recommender system exposure. 
Fig. \ref{fig:intorduction} gives illustrative examples of the exposure data.
Compared with the sparse user behavior data, the system exposure data is much larger in volume since only very few exposed items would be clicked by the user. Besides, although service provides could adopt different strategies to infuse items or advertisements into the exposed list, the system exposure data still reflects users historical behavior patterns in a large degree. 
However, the large volume of system exposure receives much less attention compared with user behavior data, and could usually be accessed in a relatively larger scope which contains various information seekers or even adversaries \cite{sotto2010privacy}. For example, different departments may share the system behavior logs to perform cross-domain collaboration. \cite{jeckmans2013privacyinrec} shows that there exists security concerns  regarding the protection of recommender exposure data.


In this paper, we investigate the problem of user behavior leakage in the field of recommender systems. We aim to answer the following questions:
\begin{enumerate}[label={(\arabic*)}]
\item Is there a risk that the \textbf{\emph{user}} historical behavior privacy can be inferred from the \textbf{\emph{system}} behavior (i.e., recommender exposure) data?
    \item If there is the risk, how to conduct the attack model?
    \item How to design a protection mechanism to downgrade the user behavior leakage risk?
\end{enumerate}
%
There are plenty of previous works focusing on the privacy concern in recommender systems. For example, \cite{DBLP:journals/tissec/KimKKYSK18,doi:10.1177/15501477211061250} use cryptography to mask the user profile and behavior data. \cite{DBLP:journals/tissec/KimKKYSK18,doi:10.1177/15501477211061250} introduce differential privacy \cite{DBLP:journals/tkde/ShinKSX18,DBLP:conf/sigir/GaoHLJ020,DBLP:conf/www/ZhangY0HC021} to prevent user profile leakage. 
\cite{qi2020fednews,wu2021fedgnn,lin2020fedrecexplicit,yang2019federated,DBLP:journals/corr/abs-2205-11857,wang2021fast} utilize federated learning to perform local computing on edge devices.
\cite{DBLP:journals/corr/abs-2005-10322,beigi2020privacy-aware} utilize adversarial learning to promote recommendation models security. 
\cite{DBLP:conf/ccs/ZhangRWRCHZ21} investigates the membership inference attack against recommender systems.
However, such existing works only focus on the protection of \textbf{\emph{user}} data. While in this paper, we target on a new privacy preserving scenario, which is how the system behavior data exposes user privacy.
Such a scenario poses a new attack and protection view from the \textbf{\emph{recommender}} perspective.


More precisely, we propose a framework containing a black-box attack model \cite{shokri2017membershipblackbox} to infer user past behavior privacy from system exposure; and a protection mechanism to downgrade the privacy leakage risk. Here the black-box model means that we don't need to know or explicitly model the recommendation algorithm.
For the attack model, we adopt an encoder-decoder architecture. The input of the attack model is the exposed items from the system over a period. Note that the input data just comes from the recommender perspective which means that we don't need to know which items in the exposed list are interacted by the user.
Then the encoder is utilized to map the input system behavior data to a latent representation. Here we utilize three different encoding strategies including mean pooling, max pooling and self-attention \cite{Transformer} based encoding\footnote{In this paper, we utilize three simple encoding strategies to further demonstrate the risk of privacy leakage since the attack can be performed without trivial design and complex computation. We leave more advanced encoding methods for the attack model as one of our future works.}.
Based on the latent representation of system behavior, we propose two decoding methods to infer the privacy of user past behavior, namely point-wise decoding and sequence-wise decoding. Point-wise decoding treats the inference as a multi-label classification task with each past interaction as a label. Sequence-wise decoding further considers the sequential order of user past behavior. We utilized three different sequential models to conduct sequence-wise decoding, including Gated Recurrent Units (GRU), Long-Short Term Memory Network (LSTM), and attention-based transformer decoder.
We conduct experiments on two real-world datasets and empirical results indicate a great danger of user behavior leakage. In other words, \emph{potential adversaries can infer which items the user have clicked before just from the observation of current system exposure for this user.} To alleviate the leakage risk, we propose a two-stage protection mechanism which first selects a subset of items from the system exposure and then replaces the selected items with uniform or popularity-based exposure.
Experimental evaluation reveals a trade-off effect between the recommendation accuracy and the privacy disclosure risk. We hope this work can raise more community concern regarding the protection of system data other than just focusing on the protection from the user perspective. To summarize, the main contributions of this work are:
\begin{itemize}[leftmargin=*,nosep]
    \item We point out a new privacy disclosure risk in recommender systems which is that the user historical behavior privacy could be inferred from the system exposure.
    \item We propose an attack model to perform user privacy inference. Experimental results on two real-world datasets indicate a great danger of privacy leakage.
    \item We propose a  protection mechanism to alleviate the privacy risk. Empirical evaluation reveals a trade-off effect between the  recommendation accuracy and the privacy leakage risk.
\end{itemize}

\section{Related work}
\label{sec:related work}
In this section, we provide a literature review of the research about recommendation, the research about exposure data, and the privacy concern in recommender systems.
\subsection{Recommendation}
A recommender system is an information filtering system, targeting on predicting user future preference based on historical user interactions with the system \cite{DBLP:conf/ccs/ZhangRWRCHZ21}. Collaborative filtering is one of the most successful recommendation methods, which is based on the similarity of users or items \cite{koren2009mf}. The key idea of collaborative filtering is that the user preference of an item can be predicted from similar user preference over similar items \cite{koren2009mf}. Typical collaborative filtering methods includes matrix factorization-based methods \cite{koren2009mf,ning2011slim,kabbur2013fism} or neighbourhood-based methods \cite{sarwar2001itemcf}. Besides, content-based recommendation \cite{DBLP:conf/adaptive/PazzaniB07} which utilizes the metadata of users and items (e.g., descriptions of item attributes and user profiles) to generate recommendation is also applied widely. 

Recently, deep learning-based recommendation methods have become a hot research topic. \cite{he2017ncf} proposed neural collaborative filtering, which utilizes a multi-layer preceptron (MLP) to model high-order user-item interaction signals. Plenty of research has been proposed in this research line, such as NFM \cite{he2017nfm}, CDAE\cite{wu2016collaborative} and Wide\&Deep \cite{cheng2016wide}. The key idea is to use deep learning to increase model expressiveness. Besides, graph neural networks also demonstrated their capability in recommendation since the user-item interactions can be naturally represented by interaction graphs.
Plenty of graph recommendation models have emerged, such as HOP-Rec \cite{yang2018hop}, NGCF \cite{wang2019neural} and LightGCN \cite{he2020lightgcn}. There are also other recommendation methods such as casual inference-based methods and reinforcement learning-based methods \cite{googlewsdmoffpolicycorrection,xin2020self,DBLP:conf/sigir/ZhangF0WSL021}. We don't elaborate them all since this work focuses on the attack and protection of user past behavior privacy, other than specific recommendation models.

Another closely related sub-field is sequential recommendation, which aims to predict the user's next interesting item from previous interacted items (in an interaction session). Early works for sequential recommendation are based on Markov Chain~\citep{DBLP:conf/www/RendleFS10} and factorization methods \cite{gmf}. Recently, plenty of deep learning-based recommendation models also emerged, including recurrent neural network (RNN)-based methods \cite{DBLP:journals/corr/HidasiKBT15}, convolutional neural network (CNN)-based methods \cite{nextitnet}, and the renowned attention-based methods \cite{SASRec,sun2019bert4rec}. 
The main difference between our attack task and sequential recommendation lies in the following folds:
\begin{itemize}[leftmargin=*,nosep]
    \item The input of our attack model is the \emph{current system} exposure data while the input for sequential recommendation is the \emph{previous user} interactions.
    \item The output of our attack model is the \emph{past user} behavior while the output of sequential recommendation is the predicted \emph{user future} behaviours.
    \item The system usually exposes a slate of items simultaneously which means that the input for our attack model could have no strong sequential signals. 
\end{itemize}

\subsection{Recommender Exposure}
Compared with the sparse user feedback data, the large volume of exposure data generated by the recommender receives relatively less research attention.
The exposed items are selected by the recommender and thus can be regarded as a kind of system behavior data.  
Existing research regarding the exposure data mainly focuses on negative sampling from exposed items or exposure debiasing \cite{DBLP:journals/corr/abs-2010-03240}.

To perform effective item ranking, negative training instances are necessary to provide comparison signals. 
To this end, the recommender exposure data contains rich information about the negative preference of users since only very few exposed items would be interacted by the user.
\cite{ding2018improved} proposed to use the exposed but not interacted items as negative examples.
\cite{DBLP:conf/ijcai/DingQ00J19,wang2020reinforced} proposed  reinforced negative samplers which can generate exposure-alike negative instances instead of directly choosing from exposure data. 
Generally speaking, the exposed but not interacted items can be seen as a kind of hard negative examples which can help to boost the ranking performance. 
However, \cite{10.1145/3184558.3186905,10.1145/3124791.3124794} verified that 
both easy negative examples and hard negative examples are important for the model training. 
Negative sampling based on only exposure data would also downgrade the model performance. 
The mixture of exposed items and uniformly sampled items could be a good negative sampling strategy.


Besides, exposure bias also known as ``previous model bias'' \cite{DBLP:conf/sigir/LiuCDHP020} since the exposure data generation is affected by previous recommendation policies, is one of the biggest sources of bias in recommendation \cite{DBLP:conf/sigir/LiuCDHP020}. Exposure bias happens since only parts of specific items are exposed by the system so non-exposed and non-interacted items do not always represent user negative preference. More specifically, an unobserved interaction could only be attributed to the user unawareness of the item because the item is not exposed to the user. 
Therefore, regarding non-interacted items as negative samples could lead to misunderstanding of user true preference and sub-optimal performance. 
There are works focusing on alleviate the effect of exposure bias.
\cite{DBLP:conf/wsdm/SaitoYNSN20} proposed to use the missing-not-at-random assumption to perform debiasing.
\cite{10.1145/3306618.3314288} proposed an exposure-based propensity matrix factorization framework to counteract the exposure bias. 
\cite{DBLP:conf/sigir/ChenDQ0XCLY21} constructed a general debiasing framework and proposed an automatic debiasing method for recommendation system based on meta learning. How to model the exposure data distribution and conduct exposure debiasing has become an emerging hot topic.

To conclude, existing research regarding recommender exposure mainly focuses on negative sampling and debiasing. However, this work focuses on a new perspective which is how the system exposure data leaks user behavior privacy.  


\subsection{Privacy Security in Recommender Systems}
Recent research shows that various user sensitive information such as age, gender, occupation, and even political orientation can be inferred from user-item interactions \cite{yuan2020peterec}. As a result, plenty of laws and regulations regarding the privacy protection has been established for online web services, especially for recommender systems \cite{otto2018GDPR}. Privacy security has obtained rapidly growing research interests in both academia and industry.

Existing privacy preserving recommendation methods can be categorized into cryptography-based methods, differential privacy-based methods, and federated learning-based methods. The cryptography methods attempt to mask the user data, for example through fully homomorphic encryption ~\cite{DBLP:journals/tissec/KimKKYSK18}.
~\cite{doi:10.1177/15501477211061250} presented a privacy recommendation algorithm in which the data on the server are encrypted and embedded to
reduce the readability of the data. 
Differential privacy-based methods aim to introduce random perturbations as the noisy signal to the user data \cite{DBLP:journals/tkde/ShinKSX18,DBLP:conf/sigir/GaoHLJ020}. 
For example, \cite{DBLP:journals/tkde/ShinKSX18} developed matrix factorization algorithms under local differential privacy. \cite{DBLP:conf/www/ZhangY0HC021} proposed differential private graph convolutional networks to protect users’ sensitive data against the attribute inference attack and provide high-quality recommendation at the same time. \cite{DBLP:journals/corr/abs-2005-10322} utilized gender obfuscation for user profiles to protect user privacy.  
\cite{beigi2020privacy-aware} conducted privacy-aware recommendation with adversarial training  which aims to both protect users against attribute inference attacks and preserve the quality of recommendation.
In recent years, federated learning-based methods have become a hot research topic with the prevalence of more and more edge devices with computation capacity. The key idea of federated learning-based methods is the distributed model training \cite{yang2019federated}. The original data is not transferred between the server and the device to avoid leakage. The transferred communication message is based on model gradients \cite{yang2019federated,ammad-ud-din2019federated}. However, the model gradients can also be utilized to infer the original data. As a result, \cite{lin2020fedrecexplicit,DBLP:journals/expert/SecureFMF,DBLP:conf/recsys/stronger} proposed to combine federated learning with cryptography methods or differential privacy. There is also research focusing on improving the efficiency of federated learning-based recommender systems \cite{muhammad2020fedfast,khan2021payload}. Besides, federated learning has also been applied to domain-specific recommendation tasks, such as news recommendation \cite{qi2020fednews}, graph-based recommendation \cite{wu2021fedgnn} and sequential recommendation \cite{han2021deeprec}. 

Furthermore, \cite{DBLP:conf/ccs/ZhangRWRCHZ21} conducted the research focusing on the membership inference attack against recommender systems. Their attack targets on determining whether a user's behavior data is used by the target recommender and then they proposed a simple yet effective protection mechanism to address the privacy concern. 

To conclude, the above existing privacy preserving methods in recommender systems almost focus on the protection of user behavior data. However, the system behavior data (i.e., the system exposure data) is less explored for the privacy concern. Due to the fact that the interactions between users and recommenders naturally form a closed loop, we argue that the system behavior data should also be investigated to protect the user privacy, especially when there is data sharing between different recommender systems.
To the best of our knowledge, our work is the first attempt which raises the privacy concern from the system perspective other than the commonly investigated user perspective.

\section{The Attack Model }

\label{sec:method}
In this section, we first present notations and the task formulation of the user behavior privacy attack in this work. Then we describe the encoding and decoding detail of the attack model. The task of user behavior privacy attack is to infer user historical behavior given system exposure data. In this paper, we exploit a simple encoder-decoder architecture to further demonstrate the risk of privacy leakage since the attack can be performed without trivial design and complex computation.

\subsection{Notations and the Attack Task Formulation}
\label{subsec:problem formulation}
\begin{figure*}[t]
  \centering
  \includegraphics[width=1.0\linewidth]{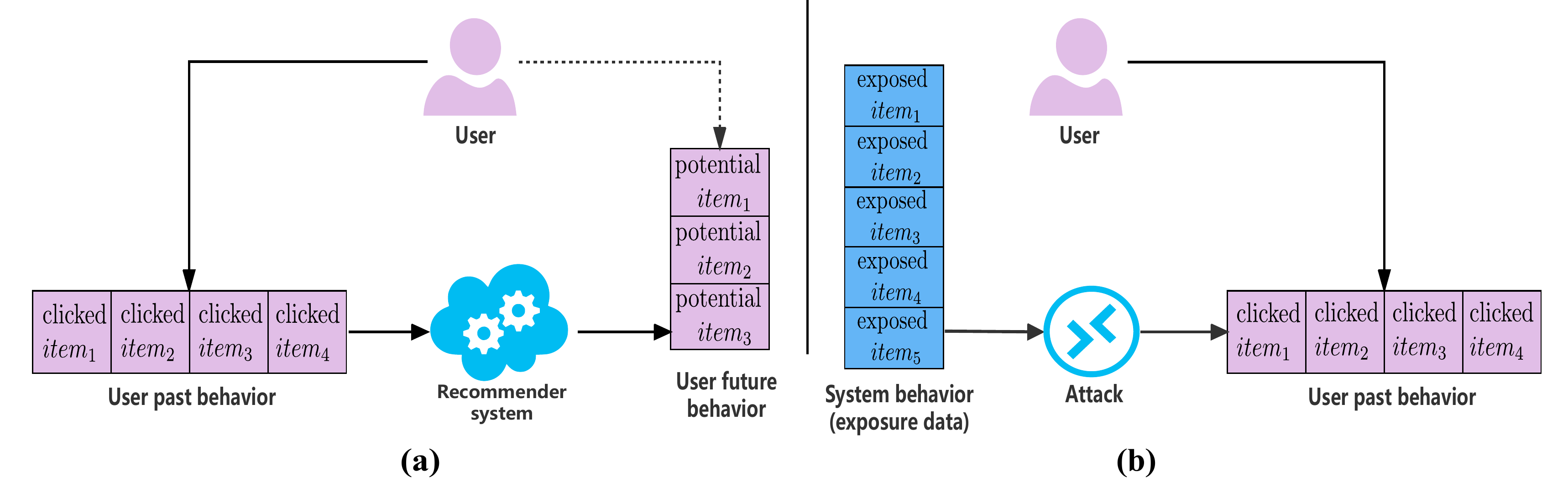}
  \caption{A recommender system (a) aims to predict the user's future behavior from the historical user interactions. In this paper, the attack scenario (b) focuses on inferring the privacy of user's past behavior from the system behavior data.}
  \label{fig:task}
\end{figure*}
Let $\mathcal{U}$ and $\mathcal{I}$ denote the user set and the item set, respectively.
During the service of a recommender system, the user $u$ produces a series of user behaviors (e.g., view, clicks, or purchases) based on the items exposed by the system. 
We use $B^u$=\{$B^u_1$, $B^u_2$, ..., $B^u_{|B^u|}$\} denotes the set of user behavior sequences of user $u$, where $B^u_i$=($b_1$, $b_2$, ..., $b_{M}$) with $b_j \in \mathcal{I}$ is a specific user behavior sequence. This is regarded as the user privacy in this work.
The set of recommended item slates for user $u$ is represented by $E^u$=\{$E^u_1$, $E^u_2$, ..., $E^u_{|E^u|}$\}, where $E^u_i$=($e_1$, $e_2$, ..., $e_{N}$) with $e_j \in \mathcal{I}$ denotes a specific exposed item slate. This is the input data for the attack model of this work. As shown in Fig. \ref{fig:task}(b),
the task of the attack model is to infer $B_i^u$ from the observation of $E_i^u$, which can be formulated as the estimation of 
\begin{equation}
\label{eq:attacktask}
p(b_1, b_2, ..., b_{M}|e_1, e_2, ...,e_{N}).
\end{equation}
The common sequential recommendation task which aims to predict the next interesting item from previous user interactions can be formulated as the estimation of $p(b_{t+1}|b_1, b_2, ..., b_t)$ as shown in Fig. \ref{fig:task}(a).
We can see that there are substantial differences between our attack task and the sequential recommendation task. 
The input and output of our attack task are from different information sources, which are systems and users, respectively. While for sequential recommendation, the input and output only focus on the user perspective. Besides, as the input of our attack task, the system exposure data $E_i^u$ just comes from the system side, which means that the attack model doesn't know which items in $E_i^u$ were clicked by the user. Generally speaking, only very few of items in $E_i^u$ could be interacted by the user. 
Finally, the recommender usually exposes a slate of items simultaneously which means that the items in $E_i^u$ could have
no strong sequential orders.

Table \ref{notation} summarizes the important notations used in this paper.
\begin{table}
\caption{The glossary table.}
\label{notation}
\scalebox{1.0}{
\begin{tabular}{l|l}
\toprule 
Notations & Description                                                                 \\ \midrule
$\mathcal{U},\mathcal{I}$       & the user and item set                                                            \\ \midrule
\multirow{2}{*}{$B^u$} & the user behavior sequence set of user $u$,\\
&$B^u$=\{$B^u_1$, $B^u_2$, ..., $B^u_{|B^u|}$\}  \\ \midrule
\multirow{2}{*}{$B^u_i$}
 & a specific user behavior sequence of user $u$,\\
&$B^u_i$=($b_1$, $b_2$, ..., $b_{M}$) with $b_j \in \mathcal{I}$ \\ \midrule
\multirow{2}{*}{$E^u$} & the set of exposed item slates for user $u$,\\
&$E_u$=\{$E^u_1$, $E^u_2$, ..., $E^u_{|E^u|}$\}    \\ \midrule
\multirow{2}{*}{$E^u_i$}       & a specific exposed item slate for user $u$,\\
&$E^u_i$=($e_1$, $e_2$, ..., $e_{N}$) with $e_j \in \mathcal{I}$  \\ \midrule
$M$       & the length of user behavior sequence  \\ \midrule
$N$       & the size of exposed item slate \\ \midrule
$d$&the item embedding size \\ \bottomrule
\end{tabular}}
\end{table}
\subsection{Privacy Attack}
\begin{figure*}
  \centering
  \includegraphics[width=1\linewidth]{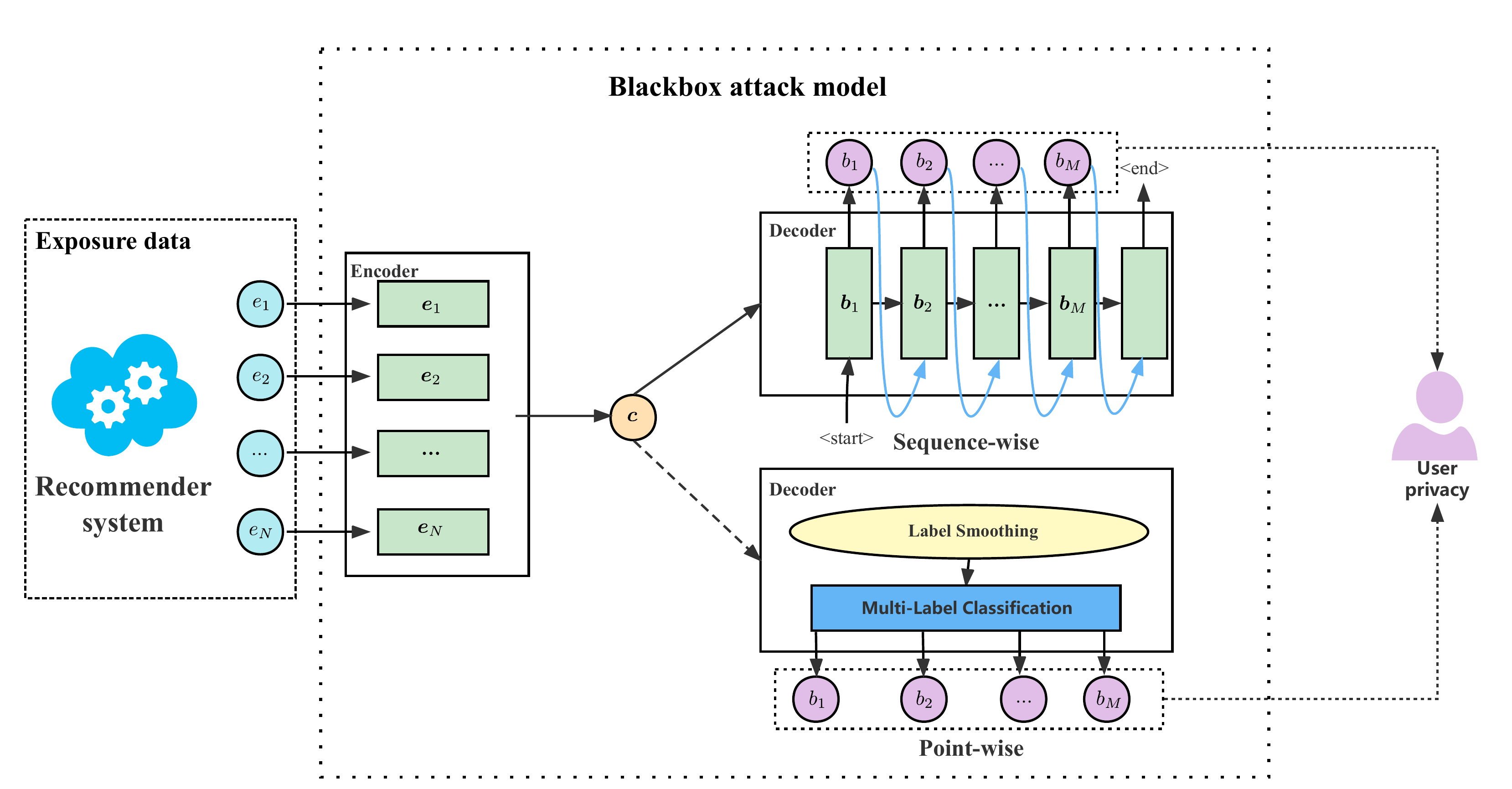}
  \caption{The overall attack model structure. The encoder aims to map the input of system exposure $E_i^u$ to a latent representation $\mathbf{c}$. Then the user privacy $B_i^u$ could be inferred through point-wise decoding or sequence-wise decoding.}
  \label{fig:attackarchitecture}
\end{figure*}

To perform the attack task, we exploit an encoder-decoder architecture. 
The encoder aims to convert the input of current system exposure $E_i^u$ into a latent representation.
Then the decoder attempt to infer the past user behavior privacy $B_i^u$ from the encoded representation.
Fig. \ref{fig:attackarchitecture} illustrates the overall architecture of the attack model.


\subsubsection{\textbf{Encoder}}
\label{encoder}
Due to the fact that the recommender system usually exposes a slate of items simultaneously, so we don't consider the sequential order of items in $E_i^u$. In this work, we utilize three simple encoding strategies including mean pooling, max pooling and self-attention based encoding to further demonstrate the risk of privacy leakage since the attack can be conducted without trivial design and complex computation. We leave more advanced encoding methods for the attack model as one of our future works.


\header{Mean pooling-based encoding.}
The pooling technique has been widely used in the field of computer vision to produce a summary statistic of the input and reduce the spatial dimension.
Given the input system exposure $E_i^u=(e_1,e_2,...e_N)$, we
first embed each item $e_j$ into a dense representation $\mathbf{e}_j \in \mathbb{R}^d$ where $d$ denotes the embedding size. This can be done through a simple embedding table lookup operation. Then we utilize mean pooling to seize the mean effect of the recommender exposed item feature vectors as the slate-level representation:
\begin{equation}
    \mathbf{c}_{mean}= \frac{1}{N}\sum_{j=1}^{N} {\mathbf{e}_j} \in \mathbb{R}^d.
\end{equation}

\header{Max pooling-based encoding.}
While the mean pooling-based encoding attempts to encode the mean effect of system behavior, the max pooling-based encoding forces the encoder to retain only the most useful exposed item features. We select the highest neuron activation value across the whole embedding space:

\begin{equation}
    c_{max}(i) = \max_{j=1,...,N} {e_j(i)}, \qquad i = 1,...,d,
\end{equation}
where $c_{max}(i)$ denotes the $i$-th element of the max pooling encoding representation $\mathbf{c}_{max} \in \mathbb{R}^d$. $e_j(i)$ is the $i$-th element of the item embedding $\mathbf{e}_j$.

\header{Self-attention based encoding.}
\begin{figure}
  \centering
  \includegraphics[width=0.65\textwidth]{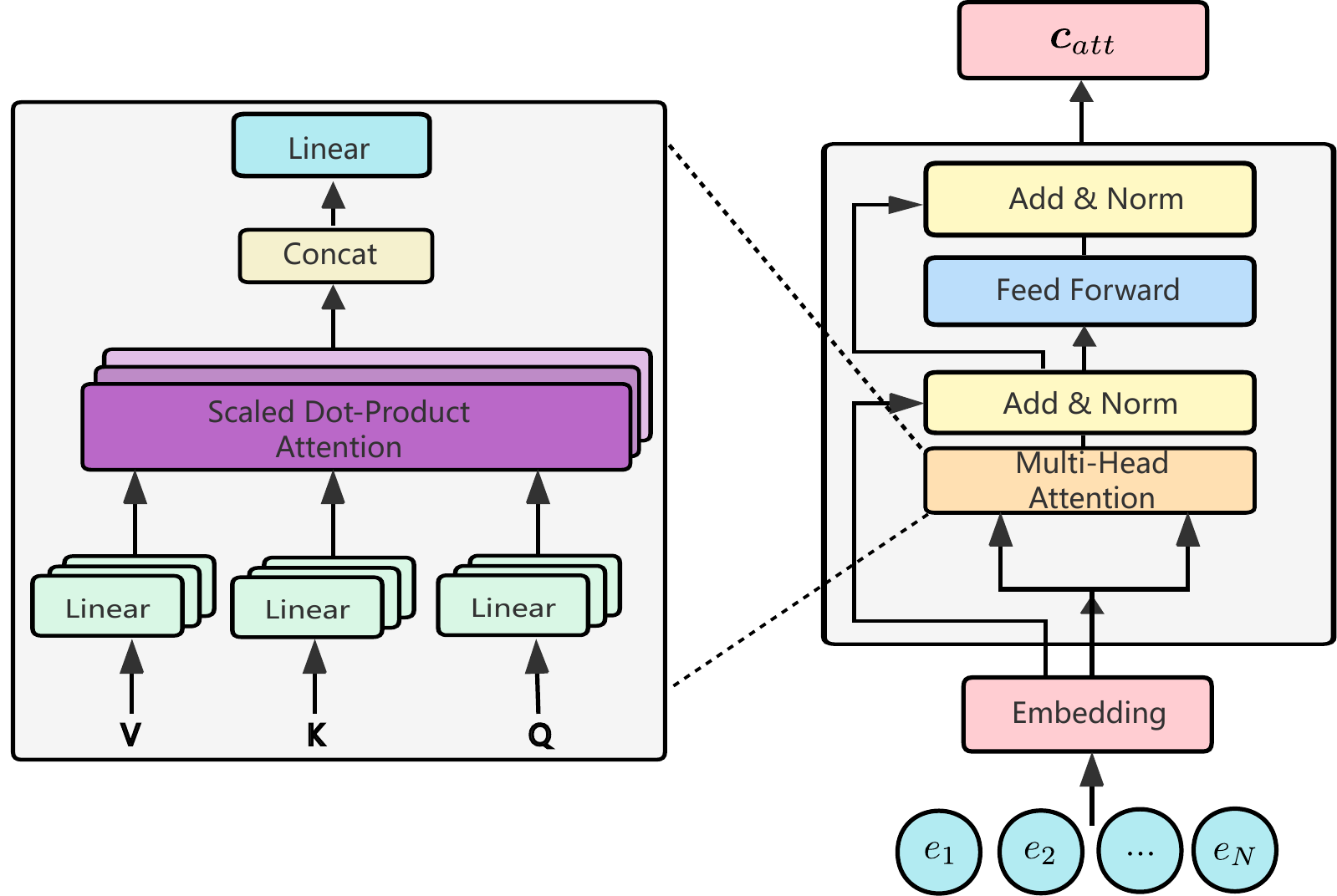}
  \caption{Self-attention based encoding.}
  \label{fig:TrmEncoder}
\end{figure}
While the mean pooling and max pooling are rather simple encoding methods, we also attempt to exploit the successful transformer encoder \cite{Transformer} to learn the latent representation of the system exposure data. 
The transformer encoder is based on the self-attention mechanism, which is highly efficient and capable of uncovering semantic patterns of the input data.
We illustrate the structure of the self-attention based encoding in Fig. \ref{fig:TrmEncoder}.
As discussed before, the recommender often exposes a slate of items simultaneously, so we don't introduce the position embeddings in our attack encoder. 

The self-attention based encoder contains \emph{multi-head attention} and \emph{feed-forward network}. 
The multi-head attention adopts scaled dot-product attention at each head to learn the importance of input items. 
The dot-product based attention is formulated as 
\begin{equation}
\label{eq:attention}
    Attention(\mathbf{Q},\mathbf{K},\mathbf{V})=softmax(\frac{\mathbf{Q}\mathbf{K}^T}{\sqrt{d}})\mathbf{V},
\end{equation}
where $\mathbf{Q,K,V}$ denote the queries, keys and values, respectively. 
The attention operation computes a weighted sum of values according to the weights calculated from the correlations between the query and the key. The scale factor $\sqrt{d}$ is used to normalize the computed correlations to avoid overly large inner product. Self-attention uses the same objects as queries, keys, and values. Then the multi-head attention on the input item embeddings is formulated as:
\begin{equation}
 \begin{aligned}
     \label{headi}
     MHA(\mathbf{E})&=concat\{head_1, head_2, . . . , head_{h} \}\mathbf{W}^O \text{, where}\\
     head_i &= Attention(\mathbf{W}^Q_i\mathbf{E}, \mathbf{W}^K_i\mathbf{E}, \mathbf{W}^V_i\mathbf{E}).
 \end{aligned}
 \end{equation}
$\mathbf{E}=[\mathbf{e}_1,\mathbf{e}_2,...\mathbf{e}_N]\in \mathbb{R}^{N\times d}$ is the stacked embedding matrix of the exposed items.
$\mathbf{W}^Q_i,\mathbf{W}^K_i,\mathbf{W}^Q_i$, and $\mathbf{W}^O$ are trainable parameters. $h$ is the number of heads.

To avoid overfitting and enable a more stable learning without vanishing or exploding gradient issues, we also introduce dropout layers, residual connection, and layer normalization.
The representation after multi-head attention is formulated as 
\begin{equation}
\label{eq:attention}
    \tilde{\mathbf{E}}=\mathbf{E}+dropout(MHA(LayerNorm(\mathbf{E})))\in \mathbb{R}^{N\times d}.
\end{equation}
After that, a two layer of feed-forward network (FFN) is utilized to increase the encoder capacity. 
Then, the latent representation after self-attention based encoding is formulated as 
\begin{equation}
     \mathbf{C}_{att} = \mathbf{\tilde{E}}+dropout(FFN(LayerNorm(\mathbf{\tilde{E}}))) \in \mathbb{R}^{N\times d}.
 \end{equation}
Besides, we insert a CLS token into the original input of $(e_1,e_2,..e_N)$ and then the corresponding vector of the CLS token in $\mathbf{C}_{att}$ can also be regarded as the final encoded vector representation of system exposure. 
 
\subsubsection{\textbf{Decoder}}
\label{decoder}
Based on the encoded latent representation of system behavior, we propose two decoding strategies to infer the privacy of user past behavior, namely point-wise decoding and sequence-wise decoding.
More specifically, for sequence-wise decoding, we utilize three notable sequential models including GRU, LSTM, and the attention-based transformer decoder.

\header{Point-wise decoding.}
Point-wise decoding treats the inference as a multi-label classification task with each past interaction as a label.
The decoder is composed of a fully-connected layer and the classification probability can be defined as 
\begin{equation}
    \mathbf{q} =[q_1,q_2,..q_{|\mathcal{I}|}]= softmax(\sigma(\mathbf{W}_{d} \mathbf{c} + \mathbf{b}))\in \mathbb{R}^{|\mathcal{I}|},
\end{equation}
where $\mathbf{q}$ denotes the classification probability, $\sigma$ is the activation function, $\mathbf{c}$ is $\mathbf{c}_{mean}$, $\mathbf{c}_{max}$ or the CLS token vector in $\mathbf{C}_{att}$ according to different encoding methods. $\mathbf{W}_d\in \mathbb{R}^{d\times |\mathcal{I}|}$ and $\mathbf{b}\in \mathbb{R}^{|\mathcal{I}|}$ are trainable parameters.

Then we use \emph{label smoothing regularization}\cite{DBLP:conf/cvpr/SzegedyVISW16,DBLP:conf/cvpr/HeZ0ZXL19} to generate the ground-truth label for the multi-label classification task.
Label smoothing prevents the network from becoming over-confident. It encourages the fully-connected layer to make a finite output and  generalize better. 
Assume the ground-truth user behavior sequence is $B_i^u = (b_1,b_2,...,b_{M})$, the ground-truth probability after label smoothing is defined as 
\begin{equation}
    y_j=
\begin{cases}
(1-\varepsilon)/M & \text{if\,} j \in  B_i^u\\
\varepsilon/(|\mathcal{I}|-M) & \text{otherwise,}
\end{cases}
\end{equation}
where $\varepsilon$ is a small constant.

Finally, the training loss function for point-wise decoding is formulated as 
\begin{equation}
    \mathcal{L}(\mathbf{y},\mathbf{q})= - \sum_{j=1}^{|\mathcal{I}|}y_j\text{log}q_j.
\end{equation}
Empirically, we found that label smoothing dramatically downgrade the risk of model overfitting.

\header{Sequence-wise decoding.}
Point-wise decoding doesn't consider the sequential order of user past behaviors. In this subsection, we describe the detail to use sequence-wise decoding to infer the user behavior privacy in a reversed order (i.e., from the most recently $b_M$ to the earliest $b_1$).
According to subsection \ref{encoder}, we have obtained a latent representation for the system exposure data. 
We then use three different sequential models to perform the inference, including GRU \cite{DBLP:conf/emnlp/ChoMGBBSB14},LSTM \cite{DBLP:conf/interspeech/SakSB14}, and the transformer decoder \cite{Transformer}. 

More precisely, during the training process, we process the user behavior sequence $B_i^u=(b_1,b_2,...,b_{M})$ reversely and complement $B_i^u$ with two special tokens as  $({<start>},b_{M},...,b_2,b_1,{<end>})$. 
Then we put $B^{'u}_i=({<start>},b_{M},...,b_2,b_1)$ as the input of the decoder and the output is shifted to $(b_{M},...,b_2,b_1,{<end>})$. 
We use $\mathbf{Q}$ $\in$ $\mathbb{R}^{(M+1)\times |\mathcal{I}|}$
to denote the output classification probability of the decoder as 
\begin{equation}
    \mathbf{Q} = SeqDecoder(\mathbf{C}, B^{'u}_i),
\end{equation}
where $SeqDecoder$ represents the three different sequential models.
$\mathbf{C} \in \mathbb{R}^{N\times d}$ is the stacked $\mathbf{c}_{mean}$, $\mathbf{c}_{max}$ or $\mathbf{C}_{att}$ according to different encoding methods. In the following, we briefly introduce the  three sequential decoding models.

\emph{(1) LSTM-based decoder.} LSTM is a famous sequence modeling neural network. We utilize the most common LSTM unit which is composed of a cell, an input gate, an output gate and a forget gate. The cell can remember values over arbitrary time intervals and the above three gates regulate the flow of information into and out of the cell\cite{DBLP:conf/interspeech/SakSB14}. More precisely, for each time step $t (1\leq t \leq M+1)$, the output classification probability  $\mathbf{q}_t \in \mathbb{R}^{|\mathcal{I}|}$ from LSTM-based decoder is formulated as
	\begin{equation}
	\begin{aligned}
	    \mathbf{i}_t = &\sigma(\mathbf{W}_i[\mathbf{h}_{t-1};\mathbf{b}_{t-1}] +\mathbf{p}_i)\\
	    \mathbf{f}_t = &\sigma(\mathbf{W}_f[\mathbf{h}_{t-1};\mathbf{b}_{t-1}] +\mathbf{p}_f)\\
	    \tilde{\mathbf{g}}_t =&tanh(\mathbf{W}_g[\mathbf{h}_{t-1};\mathbf{b}_{t-1}] + \mathbf{p}_g)\\
	    \mathbf{g}_t =& \mathbf{i}_t \odot \tilde{\mathbf{g}}_t + \mathbf{f}_t \odot \mathbf{g}_{t-1}\\
	    \mathbf{o}_t =& \sigma(\mathbf{W}_o[\mathbf{h}_{t-1};\mathbf{b}_{t-1}] +\mathbf{p}_o)\\
	    \mathbf{h}_t =& \mathbf{o}_t \odot tanh(\mathbf{g}_t)\\
	    \mathbf{q}_t =& softmax(FFN(\mathbf{h}_t)),\\
	\end{aligned}
	\end{equation}
where $\mathbf{b}_{t-1}$ is the embedding of ($t$-1)-th item (i.e., $b_{t-1}$) in the decoder. $\mathbf{h}_{t-1}$ denote the output hidden state of ($t$-1)-th step. The $\mathbf{i_t}, \mathbf{f_t}, \mathbf{o_t}$ represent the three gates' activation vector, respectively. $\mathbf{g}_t$ and $\tilde{\mathbf{g}}_t$ are cell state vectors. $\mathbf{W}_i, \mathbf{W}_f, \mathbf{W}_g, \mathbf{W}_o, \mathbf{p}_i, \mathbf{p}_f,  \mathbf{p}_g$, and $\mathbf{p}_o $ are trainable parameters. Specially, at the first training time step, the input $\mathbf{b}_0$ is the embedding of the special token ${<start>}$.

\emph{(2) GRU-based decoder.} GRU is also a notable gating-based recurrent neural network. Compared with LSTM, it's relatively simpler with fewer parameters.
For each time step $t (1\leq t \leq M+1)$, the output classification probability $\mathbf{q}_t \in \mathbb{R}^{|\mathcal{I}|}$ from GRU-based decoder is formulated as
	\begin{equation}
	\begin{aligned}
	    \mathbf{z}_t = &\sigma(\mathbf{W}_z[\mathbf{h}_{t-1};\mathbf{b}_{t-1}] +\mathbf{p}_z)\\
	    \mathbf{r}_t = &\sigma(\mathbf{W}_r[\mathbf{h}_{t-1};\mathbf{b}_{t-1}] +\mathbf{p}_r)\\
	    \tilde{\mathbf{h}}_t=&tanh(\mathbf{W}_h[\mathbf{r}_t \odot \mathbf{h}_{t-1};\mathbf{b}_{t-1}] + \mathbf{p}_h)\\
	    \mathbf{h}_t =& (1 - \mathbf{z}_t) \odot  \mathbf{h}_{t-1} + \mathbf{z}_t \odot \tilde{\mathbf{h}}_t\\
	    \mathbf{q}_t=& softmax(FFN(\mathbf{h}_t)),
	\end{aligned}
	\end{equation}
where $\mathbf{b}_{t-1}$ and $\mathbf{h}_{t-1}$ are similar to LSTM-based decoder. $\mathbf{z}_t$ and $\mathbf{r}_t$ represent the ``update gate'' vector and the ``reset gate'' vector, respectively.  $\mathbf{W}_z, \mathbf{W}_r, \mathbf{W}_h, \mathbf{p}_z, \mathbf{p}_r$, and $\mathbf{p}_h$ are learnable parameters. Similar with the LSTM-based decoder, the input $\mathbf{b}_0$ is the embedding of the special token ${<start>}$ at the first training time step.

\emph{(3) Attention-based transformer decoder.} We don't elaborate the mathematical detail again since the attention-based transformer decoder shares similar calculation with the attention-based encoding method. However, we still need to claim the following differences
\begin{itemize}
    \item Unlike the encoder, position embeddings are introduced in the decoder to distinguish the order of user past behaviors and introduce the sequential signals.
    \item Due to the nature of attacking user past behavior sequences, the model should consider only the last $j$ items when inferring the ($j$-1)-th item. As a result, a casual mask is introduced in the decoder to  modify the attention computation as masked multi-head attention.
\end{itemize}

Different with the point-wise decoding, in sequence-wise decoding we define a cross-entropy loss for each output position to perform the classification. 
The final loss function for sequence-wise decoding can be formulated as
\begin{equation}
    \mathcal{L}(\mathbf{Q},\mathbf{Y})=- \sum_{m=1}^M\sum_{j=1}^{|\mathcal{I}|}y_{mj}\text{log}q_{mj},
\end{equation}
where $y_{mj}$ and $q_{mj}$ are ground-truth probability and the $(m,j)$-th entry of the computed probability matrix $\mathbf{Q}$ (i.e., the stack of $\mathbf{q}_t$). Note that in the inference stage, we cannot obtain the ground-truth user behaviors like the training process. As a result, we use the inference results of previous steps as the input for the next inference step.

\section{Privacy Protection}
In the previous section, we describe the attack model to infer user past behavior privacy from system exposure data. Here we describe a two-stage protection mechanism which targets on alleviating the privacy risk. 
The reason that user behavior can be inferred from system exposure data lies in the fact that the items exposed by the system have inherent relationships with the previous interacted items of the user (e.g., similarity or sequential connections). So a simple solution is to infuse random items as the noisy signal into the system exposure to obfuscate the relationships. The key idea is similar with differential privacy \cite{sotto2010privacy}. However the difference is that existing differential privacy-based methods add noise to the user data while our protection targeting on introducing noise into the system exposure. 

The proposed protection mechanism consists of two stages namely position selection and item replacement. At the first stage, we decide which exposure positions would be replaced using random-based or similarity-based position selection method. After that, we utilize uniform-based or popularity-based replacement strategy to sample items to replace the exposed items on corresponding selected positions. Fig.\ref{fig:protection} shows the illustration of the protection mechanism.


\subsection{\label{Stage1}\textbf{Position Selection}} 
\begin{figure*}
  \centering
  \includegraphics[width=1.0\linewidth]{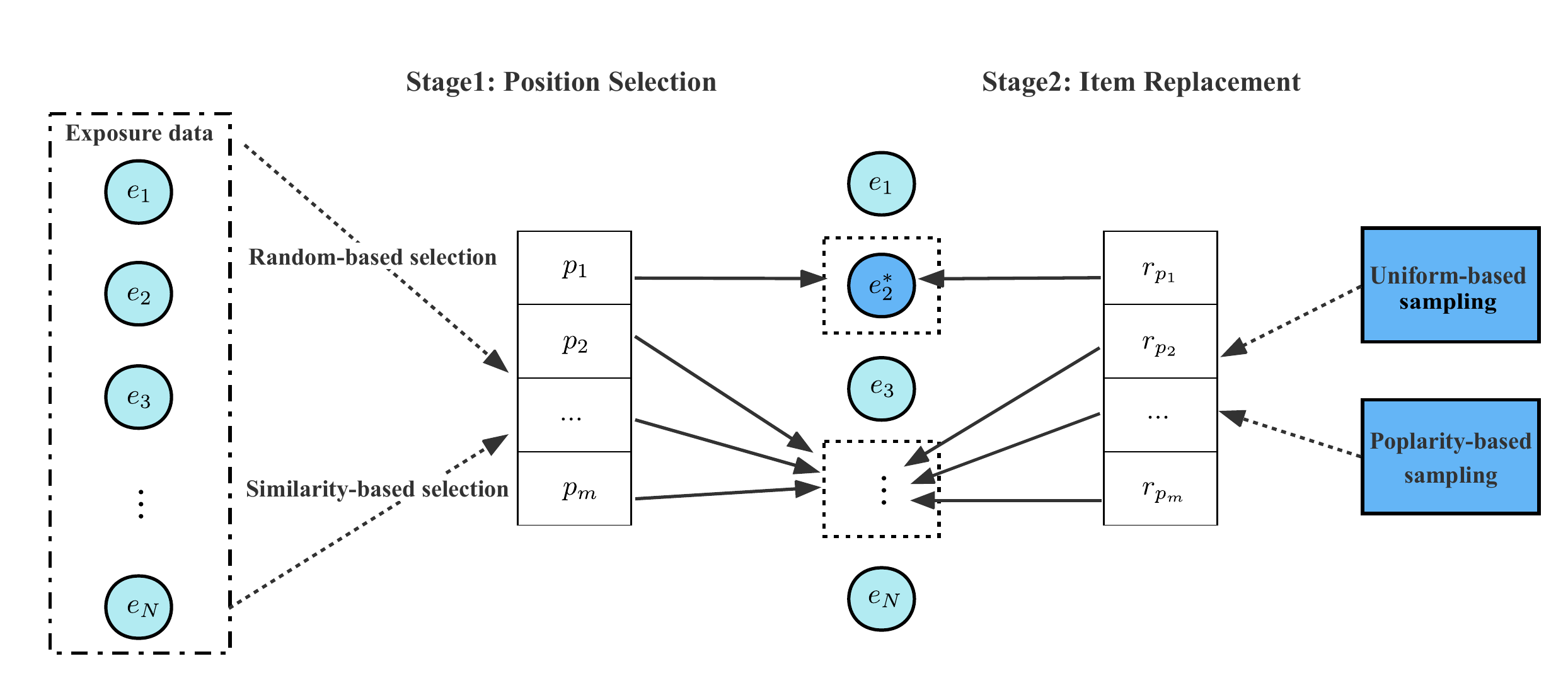}
  \caption{Privacy protection includes two stages: position selection and item replacement. Position selection decides which exposure positions would be replaced using random-based or similarity-based strategies. Item replacement replaces the exposed items on previously selected positions with uniform or popularity sampled items.}
  \label{fig:protection}
\end{figure*}

At this stage, we design two methods: the random-based method and the similarity-based method to choose which positions in the exposed item slates would be replaced, as shown in the left part of Fig. \ref{fig:protection}. 

\header{Random-based position selection.} In this method, we randomly select positions of the exposed item slate according to a uniform distribution. Given the exposed item slate $E_i^u$=$(e_1,e_2,...,e_N) $ and a replacement proportion $L$. We randomly select $m=\lceil N*L \rceil$ positions as $p$=\{$p_1,p_2,...,p_m$\},  where $p_i \leq N$. 

\header{Similarity-based position selection.} 
The above random-based position selection could lead to a situation that potential future interacted items are switched out and thus downgrades the recommendation accuracy.  
In similarity-based position selection, we select positions according to the similarity between a specific exposed item and the overall slate representation (i.e., the representation center of all exposed items within a slate). 
Given the exposed item slate $E_i^u$, we utilize the item embeddings obtained from the recommendation model(i.e., SASRec\cite{SASRec}) pre-trained using user historical behavior data and perform mean pooling to seize the mean effect of the user historical behavior.
The user behavior representation of user $u$ is defined as 
\begin{equation}
    \label{equation-slate-mean}
    \mathbf{b}_{u}= \frac{1}{|B^u|}\sum_{j=1}^{|B^u|} {\mathbf{b}^u_j} \in \mathbb{R}^d.
\end{equation}
Then we calculate the softmax cosine similarity between the above user behavior representation $\mathbf{b}_{u}$ and each exposed item $\mathbf{e}_i$ as
\begin{equation}
    s(u, i) = softmax(\frac{\mathbf{e}_i \cdot \mathbf{b}_{u}}{max(\left \| \mathbf{e}_i \right \|_2 \cdot \left \| \mathbf{b}_{u} \right \| _2,\omega )}),
\end{equation}
where $s(u, i)$ represents the probability similarity between $\mathbf{e}_i$ and the user behavior, and $\omega$ is a small constant. 

Since $\mathbf{b}_{u}$ denotes the overall user historical behavior representation, which can be further regarded as the user preference, we thus believe that replacing dissimilar positions could help us to maintain a good recommendation performance.
As a result, we assign the probability of selecting a position to be replaced according to the value of $s(u, i)$ and positions with smaller $s(u, i)$ are more likely to be replaced. Finally, we select $m=\lceil N*L \rceil$ positions as $p$=\{$p_1,p_2,...,p_m$\} according to the calculated similarity.

\subsection{\textbf{Item Replacement}}
After position selection, we investigate two item replacement strategies based on uniform exposure and popularity exposure, as shown in the right part of Fig.\ref{fig:protection}.

\header{Uniform replacement.}
In this strategy, we randomly sample items from the whole item set according to a uniform distribution to replace the exposed items of the system.
For the selected positions $p$=\{$p_1,p_2,...,p_m$\} discussed in section \ref{Stage1}, we uniformly sample a replacement item set \{$ r_{p_1},r_{p_2},...,r_{p_m}$\}.
After that,  the system exposure data can be updated as
\begin{equation}
\label{protection:random}
    E_i^{*u}=(e_1^*,..,e_j^*,..,e_N^*), e_j^*=
\begin{cases}
r_{p_f}\sim Unif(\mathcal{I}) & \text{if\,} j=p_f \text{,}\\
e_j & \text{otherwise,}
\end{cases}
\end{equation}
where $ f \leq m $.
We take $E_i^{*u}$ as the input data for the trained attack model to perform a new attack inference to verify how the replacement affect the attack model performance. Besides, the new recommendation accuracy can be defined as 
\begin{equation}
\label{accnew}
    acc_u^*=\frac{|E^{*u}\cap B^u|}{|E^{*u}|}.
\end{equation}
Obviously, the uniform replacement would also affect the recommendation accuracy.

\header{Popularity replacement.}
In this strategy, we sample the replacement item set according to the 
item popularity. 
Same as discussed before, given the exposed item slate $E_i^u$ and the replacement proportion $L$,
we select the positions as $p$=\{$p_1,p_2,...,p_m$\} as described in  section \ref{Stage1}.
Then we sample a replacement item set \{$ r_{p_1},r_{p_2},...,r_{p_m}$\} according to the item popularity.
Here we adopt two kinds of popularity, the overall popularity and the in-batch popularity. The overall popularity means that the popularity distribution is computed over all items while the in-batch popularity means that the popularity is computed over the items in the current inference batch.
Then, the system exposure data can be updated as
\begin{equation}
\label{protection:popular}
    e_j^*=
\begin{cases}
r_{p_f}\sim Pop(\mathcal{I}) \text{ or } Pop(\mathcal{I}_b)  & \text{if\,} j=p_f \text{,}\\
e_j & \text{otherwise,}
\end{cases}
\end{equation}
where $ f \leq m $, and $\mathcal{I}_b$ denotes the item set in the batch.
Similarly, we take $E_i^{*u}$ as the input data for the trained attack model to perform a new attack inference to verify how the popularity-based replacement affect the attack model performance and the recommendation accuracy.

\section{Experiments}
\label{sec:experiments}
In this section, we conduct experiments to verify the user behavior privacy leakage risk in recommender systems\footnote{We release our code at \url{https://github.com/nancheng58/On-the-User-Behavior-Leakage-from-Recommender-System-Exposure}}. We aim to answer the following experimental questions:

\begin{enumerate}[leftmargin=*, label=RQ\arabic*]
\item How does the attack model perform? Whether there is a substantial user behavior leakage risk in recommender systems?

\item How does the number of exposed items affect the attack performance? What's the proper setting to conduct the attack?

\item How does the protection mechanism affect the attack performance and recommendation accuracy?
\end{enumerate}

\subsection{Dataset Description}
The experiments were conducted on two real-world datasets Zhihu\footnote{\url{https://github.com/THUIR/Zhihu-Dataset}}~\citep{hao2021largescale} and MIND\footnote{\url{https://msnews.github.io/}}~\citep{DBLP:conf/acl/WuQCWQLLXGWZ20}. Both of the two datasets contain user behavior data (e.g., clicks) and system behavior data (e.g., exposed impressions). 
Table \ref{Datasets} summarizes the statistics of the two datasets.

\begin{table}
    \centering
    \begin{threeparttable}
    \caption{Dataset statistics. The impressions can be seen as system behavior data while the clicks can be regarded as the user behavior privacy.}
    \label{Datasets}
    \begin{tabular}{p{3cm}<{\centering}p{2.2cm}<{\centering}p{2.2cm}<{\centering}}
    \toprule
    Dataset & Zhihu & MIND\cr
    \midrule
    \#users&7,963 & 94,057\cr
    \#items&64,573& 34,376\cr
    \#impressions&1,000,026&8,584,442\cr
    \#clicks&271,725& 347,727\cr
    \bottomrule
  \end{tabular}
    \end{threeparttable}
\end{table}

\header{Zhihu}. This dataset is collected from a large-scale knowledge-acquisition platform. The original data contains question information, answer information and the user profile. Here we focus on the answer recommendation scenario. In the serving period of the recommender, a slate of answers (i.e., the items in our setting) are exposed to the user.
This can be seen as the system behavior data. The user may click some of the answers for more detail information, which can be seen as the user behavior data. The dataset contains the show time and click time of all answers (0 for non-click answers).
More precisely, given a timestamp $t$ and a user $u$, the $M$ answer clicks just before $t$ are seen as the past user behavior privacy. The $N$ system exposure impressions just following the timestamp $t$ are regarded as the input for the attack model.
Finally, the dataset contains 1,000,026 system impressions and 271,725 user clicks over 64,573 items of 7,963 users. With the default setting of $M=5$ and $N=10$, we can get a total of 213,172 pairs of [$B_i^u$|$E_i^u$].

\header{MIND}. This dataset focuses on the news recommendation scenario, which is collected from the system logs of a news website. The data we use contains 8,584,442 recommended impressions for 94,057 users over 34,376 news with 347,727 total user clicks.
Each impression log contains a slate of recommended news and historical user click behaviors of this user before the impression.
We sort the historical user clicks according to the timestamp. Then the historical clicked news are regarded as the user behavior privacy and the news in the recommended impressions are seen as the system behavior data.
Finally, we can get a total of 291,595 pairs of [$B_i^u$|$E_i^u$] with the default setting of $M=5$ and $N=10$.

\subsection{Evaluation Protocols}
\subsubsection{Attack evaluation.}
We adopt cross-validation to evaluate the performance of the attack model.
The ratio of training, validation, and test set is 8:1:1.
We randomly sample the data of 80\% users as the training set. Then the data of 10\% users is used for validation and the rest 10\% users are regarded as the test users. Such user-based data splits can effectively avoid the potential information shortcuts between training and test.
For validation and test, the evaluation is done by providing the attack model with item slates exposed by the system and then checking the rank of the $M$ ground-truth items in the inference results.
We adopt Recall to evaluate the attack performance. Let $\tilde{B}_i^u@k$ denote the top-$k$ inference output of the attack model\footnote{$\tilde{B}_i^u@k$ contains $k\times M$ items. For point-wise decoding, $\tilde{B}_i^u@k$ is the top-($k\times M$) items of the inference results since point-wise decoding doesn't consider the sequential order of user behavior. For sequence-wise decoding, $\tilde{B}_i^u@k$ is composed of the top-$k$ items of all $M$ inference positions.}.
Recall@$k$ measures how many ground-truth user behaviors are included in $\tilde{B}_i^u@k$, which is formulated as  
\begin{equation}
	\label{Recall}
	\text{Recall}@k=\frac{|\tilde{B}_i^u@k \cap B_i^u|}{M}.
\end{equation}
We then report the average Recall@$k$ across the whole test user set as the final results.
Besides, we also report normalized discounted cumulative gain (NDCG) and mean reciprocal rank (MRR), which are weighted versions of Recall assigning higher weights to the top-ranked positions of the inference result lists. 


\subsubsection{Protection evaluation.}
To verify the effect of the proposed protection mechanism, we first compute $E_i^{*u}$ for user $u$ in test users according to Eq.(\ref{protection:random}) or Eq.(\ref{protection:popular}).
Then we feed $E_i^{*u}$ to the pre-trained attack model to calculate the new attack metrics (i.e., Recall, NDCG, and MRR). 
Besides, we also compute the new recommendation accuracy for user $u$ according to Eq.(\ref{accnew}). Then we report the overall recommendation accuracy as the average of all users in the test set.

\subsection{Hyperparameter Settings}
We conduct our experiments with a batch size of $400$ pairs of system exposure data and user behavior data (i.e.,[$B^u_i$|$E^u_i$]). The sizes of $B^u_i$ and $E^u_i$ are set as $M=5$ and $N=10$, respectively without special mention. 
The item embedding size is set as $d=128$. 
We train all models with the Adam optimizer\cite{DBLP:journals/corr/KingmaB14}. The learning rate is set as 0.001. The dropout rate is tuned to 0.1.
For attention-based encoding and transformer-based decoder, the hyperparameters of the multi-head self-attention are set as 2 heads with a total of 128 hidden neurons. The hidden size of the FFN is also set as 128. For LSTM-based decoder and GRU-based decoder, the depth of recurrent layer is set to 1. We utilize the weight sharing technique \cite{DBLP:conf/eacl/PressW17, DBLP:conf/iclr/InanKS17} to tie the weights of the encoder item embedding and softmax layer item embedding in the decoder. For label smoothing, the $\varepsilon$ is set to the  $1/{|\mathcal{I}|}$. Each experiment is conducted 3 times and the average result is reported.

\subsection{Attack Performance (RQ1)}

\begin{table*}
    \centering
    \begin{threeparttable}
    \caption{{Attack performance with Point-wise decoding. Boldface denotes the highest score. Rec is short for Recall. ``Mean'', ``Max'', and ``Att'' denote three encoding strategies of mean-pooling, max-pooing and self-attention based encoding, respectively.}}
    \label{pointwise}
    \begin{tabular}{p{1.1cm}<{\centering}p{1cm}<{\centering}p{0.8cm}<{\centering}p{1.1cm}<{\centering}p{0.9cm}<{\centering}p{0.85cm}<{\centering}p{1.3cm}<{\centering}p{1.1cm}<{\centering}p{0.9cm}<{\centering}p{1.3cm}<{\centering}p{1cm}<{\centering}}
    \toprule
    Datasets&Encoder&Rec@5&NDCG@5&MRR@5&Rec@10&NDCG@10&MRR@10&Rec@20&NDCG@20&MRR@20\cr
    \midrule
    \multirow{3}{*}{Zhihu}
    &Mean   &0.0673& 0.0403&0.0315&0.1282&0.0597&0.0394&0.2372&0.0870&0.0467\cr
    &Max  &0.0652& 0.0393&0.0309&0.1236&0.0580&0.0385&0.2318&0.0850&0.0458\cr
    &Att &\textbf{0.0737}&\textbf{0.0439}&\textbf{0.0342}&\textbf{0.1364}&\textbf{0.0640}&\textbf{0.0424}&\textbf{0.2493}&\textbf{0.0922}&\textbf{0.0500}\cr
    \hline
    \multirow{3}{*}{MIND}
    &Mean   &\textbf{0.3998}&\textbf{0.2416}&\textbf{0.1898}&\textbf{0.6684}&\textbf{0.3282}&\textbf{0.2255}&\textbf{0.8828}&\textbf{0.3829}&\textbf{0.2408}\cr
    &Max  &0.3984&0.2406&0.1889&0.6631&0.3259&0.2241&0.8762&0.3804&0.2394\cr
    &Att  &0.3977& 0.2399&0.1883&0.6683& 0.3271&0.2242&0.8818&0.3817&0.2395\cr
    \bottomrule
    \end{tabular}
    \end{threeparttable}
\end{table*}

\begin{table*}
    \centering
    \begin{threeparttable}
    \caption{{Attack performance with LSTM-based decoders. Boldface denotes the highest score. Rec is short for Recall. ``Mean'', ``Max'', and ``Att'' denote three encoding strategies of mean-pooling, max-pooing and self-attention based encoding, respectively.}}
    \label{LSTM}
    \begin{tabular}{p{1.1cm}<{\centering}p{1cm}<{\centering}p{0.8cm}<{\centering}p{1.1cm}<{\centering}p{0.9cm}<{\centering}p{0.85cm}<{\centering}p{1.3cm}<{\centering}p{1.1cm}<{\centering}p{0.9cm}<{\centering}p{1.3cm}<{\centering}p{1cm}<{\centering}}
    \toprule
    Datasets&Encoder&Rec@5&NDCG@5&MRR@5&Rec@10&NDCG@10&MRR@10&Rec@20&NDCG@20&MRR@20\cr
    \midrule
    \multirow{3}{*}{Zhihu}
    &Mean &0.0916& 0.0579&0.0470&0.1634&0.0809&0.0563&0.2868&0.1118&0.0647\cr
    &Max  &0.1102& 0.0690&0.0556&0.1930&0.0955&0.0664&0.3301&0.1299&0.0757\cr
    &Att  &\textbf{0.1198}&\textbf{0.0744}&\textbf{0.0596}&\textbf{0.2063}&\textbf{0.1021}&\textbf{0.0709}&\textbf{0.3495}&\textbf{0.1380}&\textbf{0.0806}\cr
    \hline
    \multirow{3}{*}{MIND}
    &Mean &\textbf{0.5861}&\textbf{0.3591}&\textbf{0.2852}&\textbf{0.8772}&\textbf{0.4541}&\textbf{0.3249}&0.9809&\textbf{0.4811}&\textbf{0.3327}\cr
    &Max  &0.5657&0.3435&0.2711&0.8719&0.4432&0.3127
&\textbf{0.9837}&0.4724&0.3212\cr
    &Att  &0.5335&0.3353&0.2707&0.8158&0.4269&0.3087&0.9606&0.4644&0.3194\cr
    \bottomrule
    \end{tabular}
    \end{threeparttable}
\end{table*}

\begin{table*}
    \centering
    \begin{threeparttable}
    \caption{{Attack performance with GRU-based decoders. Boldface denotes the highest score. Rec is short for Recall. ``Mean'', ``Max'', and ``Att'' denote three encoding strategies of mean-pooling, max-pooing and self-attention based encoding, respectively.}}
    \label{GRU}
    \begin{tabular}{p{1.1cm}<{\centering}p{1cm}<{\centering}p{0.8cm}<{\centering}p{1.1cm}<{\centering}p{0.9cm}<{\centering}p{0.85cm}<{\centering}p{1.3cm}<{\centering}p{1.1cm}<{\centering}p{0.9cm}<{\centering}p{1.3cm}<{\centering}p{1cm}<{\centering}}
    \toprule
    Datasets&Encoder&Rec@5&NDCG@5&MRR@5&Rec@10&NDCG@10&MRR@10&Rec@20&NDCG@20&MRR@20\cr
    \midrule
    \multirow{3}{*}{Zhihu}
    &Mean &0.0875& 0.0542&0.0434&0.1578&0.0767&0.0526&0.2841&0.1084&0.0612\cr
    &Max  &\textbf{0.1009}& \textbf{0.0628}&\textbf{0.0504}&\textbf{0.1792}&\textbf{0.0878}&\textbf{0.0606}&\textbf{0.3114}&\textbf{0.1210}&\textbf{0.0695}\cr
    &Att  &0.0833& 0.0512&0.0408&0.1523&0.0733&0.0497&0.2744&0.1038&0.0580\cr
    \hline
    \multirow{3}{*}{MIND}
    &Mean &0.5384&0.3265&0.2575&0.8429&0.4255&0.2987&0.9739&0.4596&0.3086\cr
    &Max  &\textbf{0.5586}&\textbf{0.3471}&\textbf{0.2782}&\textbf{0.8506}&\textbf{0.4422}&\textbf{0.3178}&\textbf{0.9748}&\textbf{0.4745}&\textbf{0.3271}\cr
    &Att  &0.5414&0.3415&0.2763&0.8110&0.4291&0.3126&0.9561&0.4665&0.3233\cr
    \bottomrule
    \end{tabular}
    \end{threeparttable}
\end{table*}

\begin{table*}
    \centering
    \begin{threeparttable}
    \caption{{Attack performance with transformer decoders. Boldface denotes the highest score. Rec is short for Recall. ``Mean'', ``Max'', and ``Att'' denote three encoding strategies of mean-pooling, max-pooing and self-attention based encoding, respectively.}}
    \label{Trm Decoder}
    \begin{tabular}{p{1.1cm}<{\centering}p{1cm}<{\centering}p{0.8cm}<{\centering}p{1.1cm}<{\centering}p{0.9cm}<{\centering}p{0.85cm}<{\centering}p{1.3cm}<{\centering}p{1.1cm}<{\centering}p{0.9cm}<{\centering}p{1.3cm}<{\centering}p{1cm}<{\centering}}
    \toprule
    Datasets&Encoder&Rec@5&NDCG@5&MRR@5&Rec@10&NDCG@10&MRR@10&Rec@20&NDCG@20&MRR@20\cr
    \midrule
    \multirow{3}{*}{Zhihu}
    &Mean     &\textbf{0.1453}&\textbf{0.0946}&\textbf{0780}&\textbf{0.2359}&\textbf{0.1235}&\textbf{0.0898}&\textbf{0.3748}&\textbf{0.1585}&\textbf{0.0992}\cr
    &Max  &0.1426& 0.0920&0.0755&0.2339&0.1213&0.0874&0.3725&0.1561&0.0968\cr
    &Att  &0.1344&0.0850&0.0699&0.2293&0.1161&0.0822&0.3721&0.1520&0.0919\cr
    \hline
    \multirow{3}{*}{MIND}
    &Mean &0.5742&0.3714&0.3051&\textbf{0.8232}&0.4524&0.3389&\textbf{0.9549}&0.4864&0.3485\cr
    &Max  &\textbf{0.5758}&\textbf{0.3729}&\textbf{0.3066}&0.8208&\textbf{0.4526}&\textbf{0.3398}&0.9519&\textbf{0.4864}&\textbf{0.3494}\cr
    &Att  &0.5647&0.3639&0.2983&0.8119&0.4443&0.3316&0.9498&0.4798&0.3417\cr
    \bottomrule
    \end{tabular}
    \end{threeparttable}
\end{table*}

Table \ref{pointwise} show the attack performance with point-wise decoding on the two datasets. We can see that for point-wise decoding, the self-attention based encoding achieves much better attack performance than mean-pooling and max-pooling based encoding on the Zhihu dataset.
While for the MIND dataset, the three different encoding methods achieve similar attack performance.

The attack performance of the three sequence-wise decoding methods are shown in Table \ref{LSTM}, Table \ref{GRU}, and Table \ref{Trm Decoder}.
We can see from Table \ref{LSTM} that for LSTM-based decoder, the self-attention based encoding achieves the highest attack performance on the Zhihu dataset. While on the MIND dataset, the mean-pooling based encoding achieves much better performance than max-pooling and self-attention based encoding.
For GRU-based decoder performance shown in Table \ref{GRU}, we can see that the max-pooling encoding performance is better than the other two encoding methods on both the two datasets. 
The reason could be that the max-pooling encoding can select the most important features for GRU-based decoder. 
Table \ref{Trm Decoder} shows the attack results of the transformer decoder. 
We can see that for the transformer decoder, the mean-pooling encoding achieves the best result on the Zhihu dataset. On the MIND dataset, the max-pooling encoding achieves the highest attack performance. 

Through the comparison between point-wise decoding (i.e., Table \ref{pointwise}) and sequence-wise decoding (i.e., Table \ref{LSTM}, Table \ref{GRU}, and Table \ref{Trm Decoder}), we can see the sequence-wise decoding achieves much better performance than point-wise decoding with the combination of all three encoders. Such results indicates that considering the sequential order of user behaviours could further improve the attack model performance.


Regarding the encoding methods, we can see that on the Zhihu dataset, self-attention based encoding achieves the best performance when using point-wise decoding and the LSTM-based decoder. While for the MIND dataset, the mean-pooling encoding achieves the highest attack scores with point-wise decoding and the LSTM-based decoder. It demonstrates that for different datasets, we should choose different encoders to conduct the attack. Besides, we can see the compared with the powerful self-attention based encoding, the much simpler mean-pooling and max-pooling can still achieve comparable and even better attack performance. It further demonstrate the risk of privacy leakage since the attack can be performed without trivial and complex encoding methods.

Besides, we can see that the attack model performs much better on MIND other than Zhihu. The reason could be that the Zhihu dataset comes from an answer recommendation scenario. Each answer has an underlying latent question. As a result, the attack model could encounter difficulty to infer the change of underlying questions, leading to relatively lower attack model performance than the MIND dataset.


Taking an overall look at the results of Table \ref{pointwise}, Table \ref{LSTM}, Table \ref{GRU}, and Table \ref{Trm Decoder}, we can see that on the Zhihu dataset, the attack model can achieve the highest 23.59\% recall in the top-10 list of the inference results. It indicates a great danger that more than 20\% of user privacy can be exactly inferred from top-10 attack results. 
Regarding the MIND dataset, we can see that the privacy risk leakage further increases to the highest 87.72\% recall@10 and 98.09\% recall@20. 

To validate this common risk in the recommendation, we don't focus on the trivial model design and even though utilizing the simple encoder-decoder architecture also can achieve great attack performance. The reason is that exposure data includes enough information to infer the user’s historical behavior, even the mean-pooling encoder can learn the representation of exposure data easily.

To conclude, we can see that there is substantial user privacy leakage in recommender systems. A large amount of user past behavior privacy can be inferred from system exposure data through the simple encoder-decoder based attack model.

\subsection{Hyperparameter Study (RQ2)}
\begin{figure}
    \captionsetup[subfloat]
    {}
    \centering
    \subfloat[Point-wise decoding]{
    \label{rq2-point-att-zhihu}
    \includegraphics[width=0.24\textwidth,height=0.3001\textwidth]{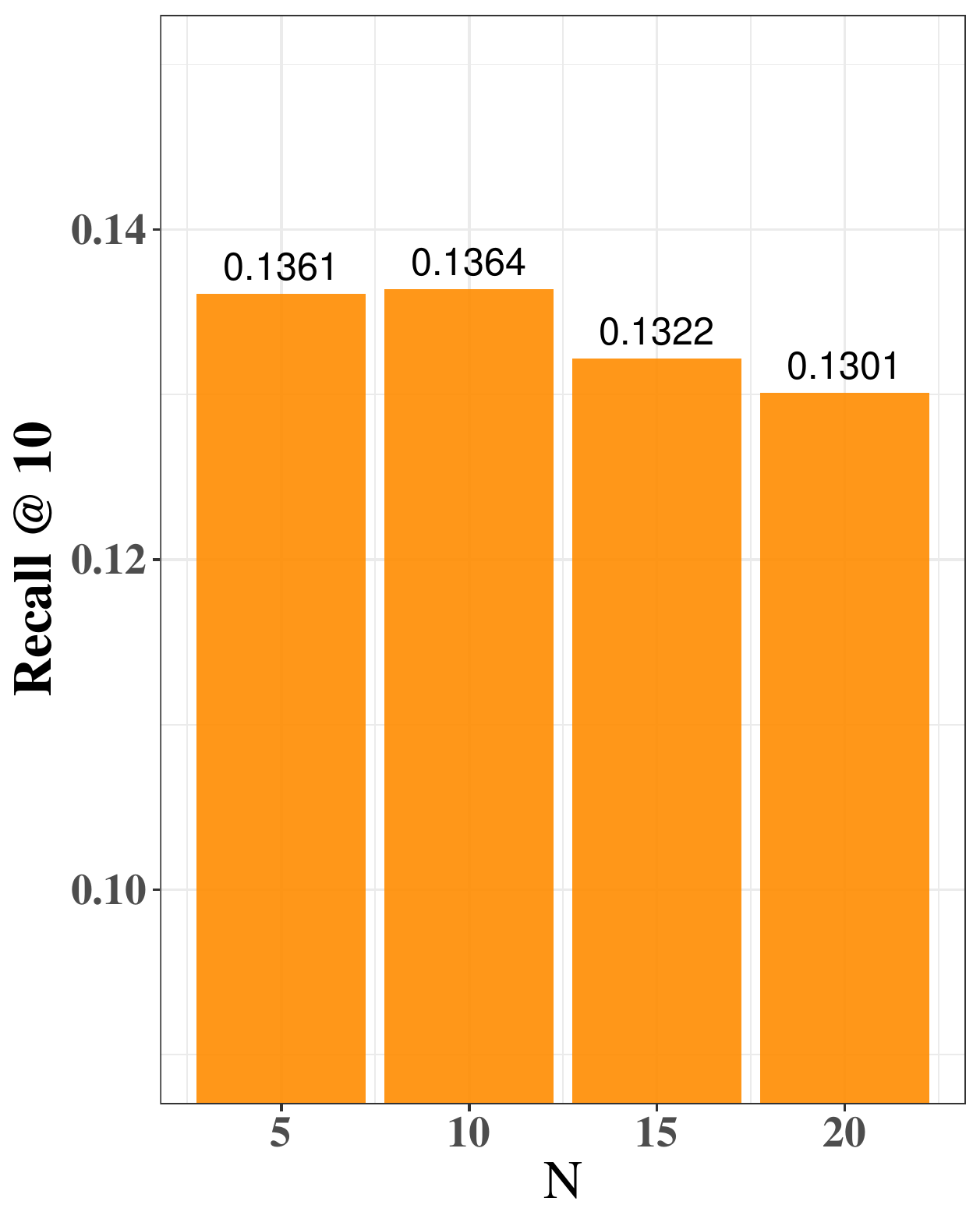}}
    \subfloat[LSTM-based decoder]{
    \label{rq2-lstm-att-zhihu}
    \includegraphics[width=0.24\textwidth,height=0.3001\textwidth]{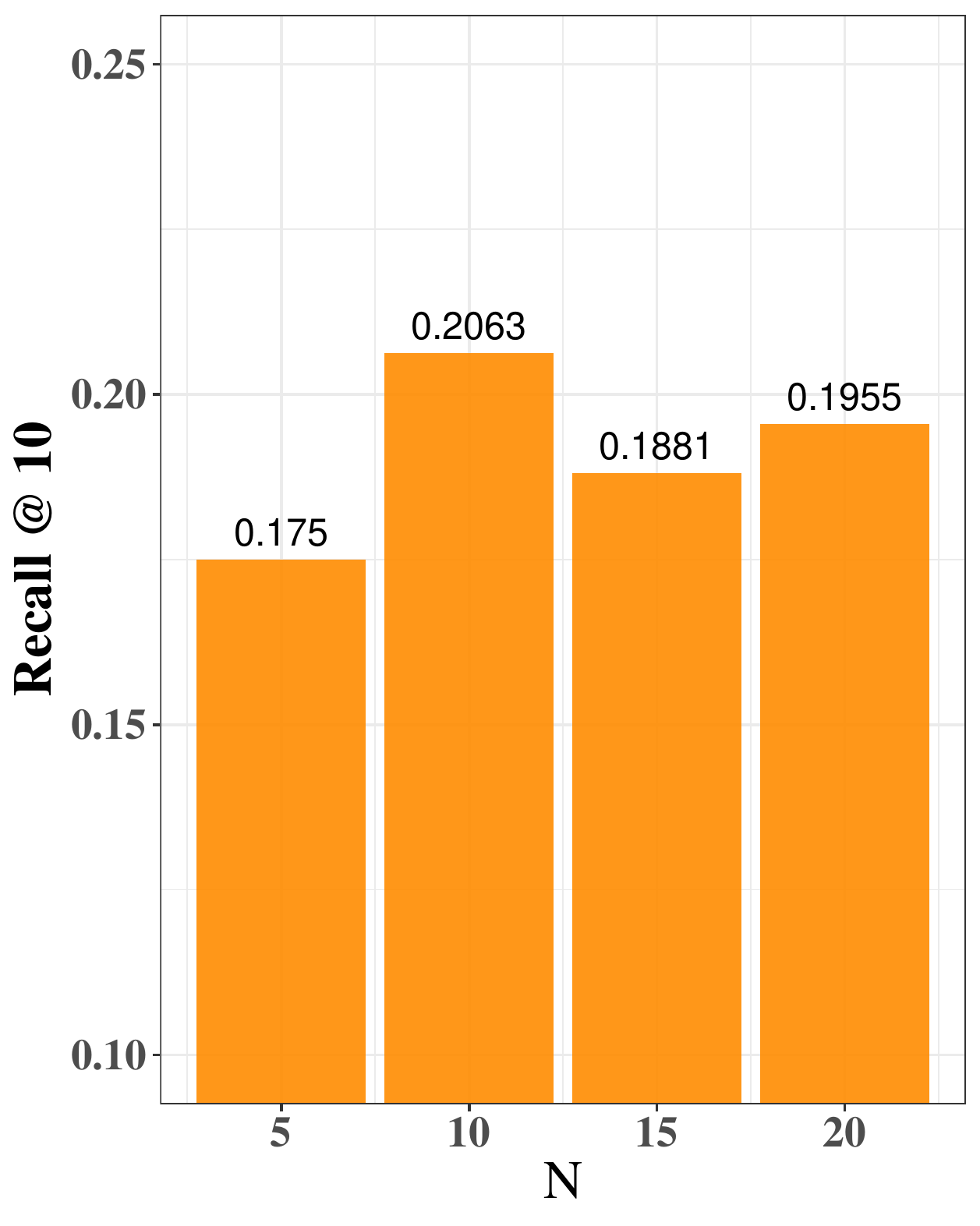}}
    \subfloat[GRU-based decoder]{
    \label{rq2-gru-att-zhihu}
    \includegraphics[width=0.24\textwidth,height=0.3001\textwidth]{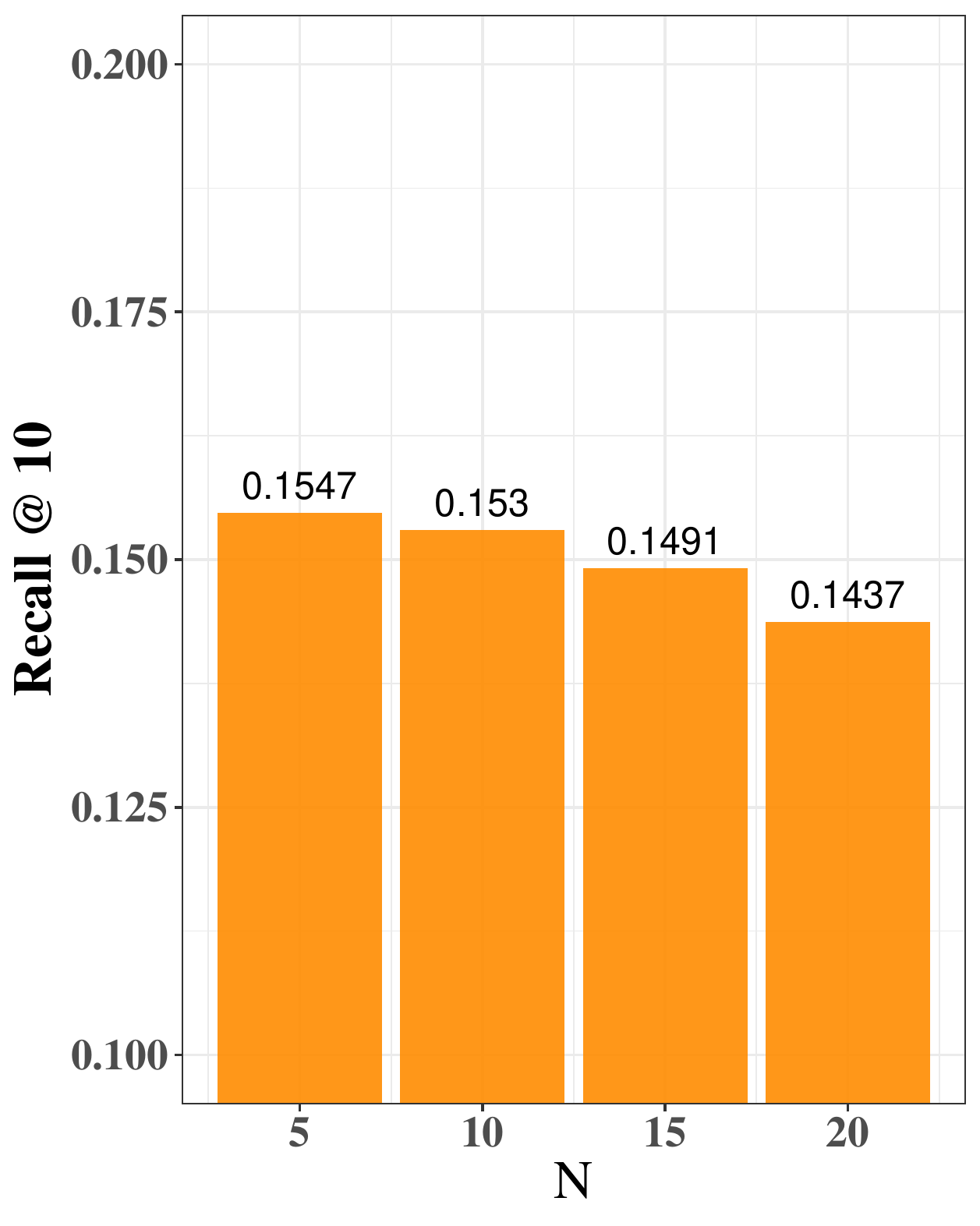}}
    \subfloat[Transformer decoder]{%
    \label{rq2-seq-att-zhihu}
    \includegraphics[width=0.24\textwidth,height=0.3001\textwidth]{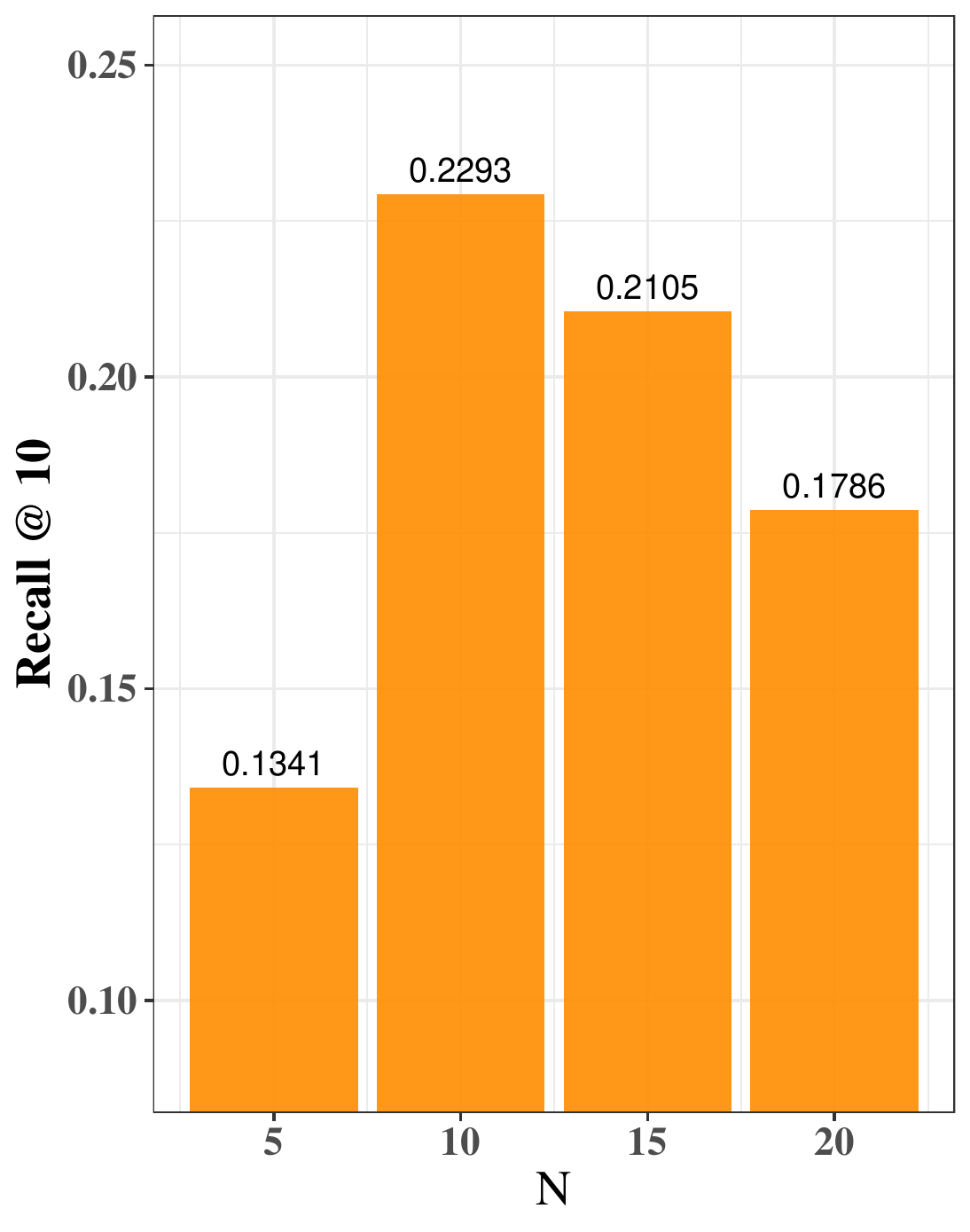}}
    \caption{Attack performance with different $N$ on Zhihu.}
    \label{rq2-zhihu}
\end{figure}
\begin{figure}
    \captionsetup[subfloat]
    {}
    \centering
    \subfloat[Point-wise decoding]{
    \label{rq2-point-att-mind}
    \includegraphics[width=0.24\textwidth,height=0.3001\textwidth]{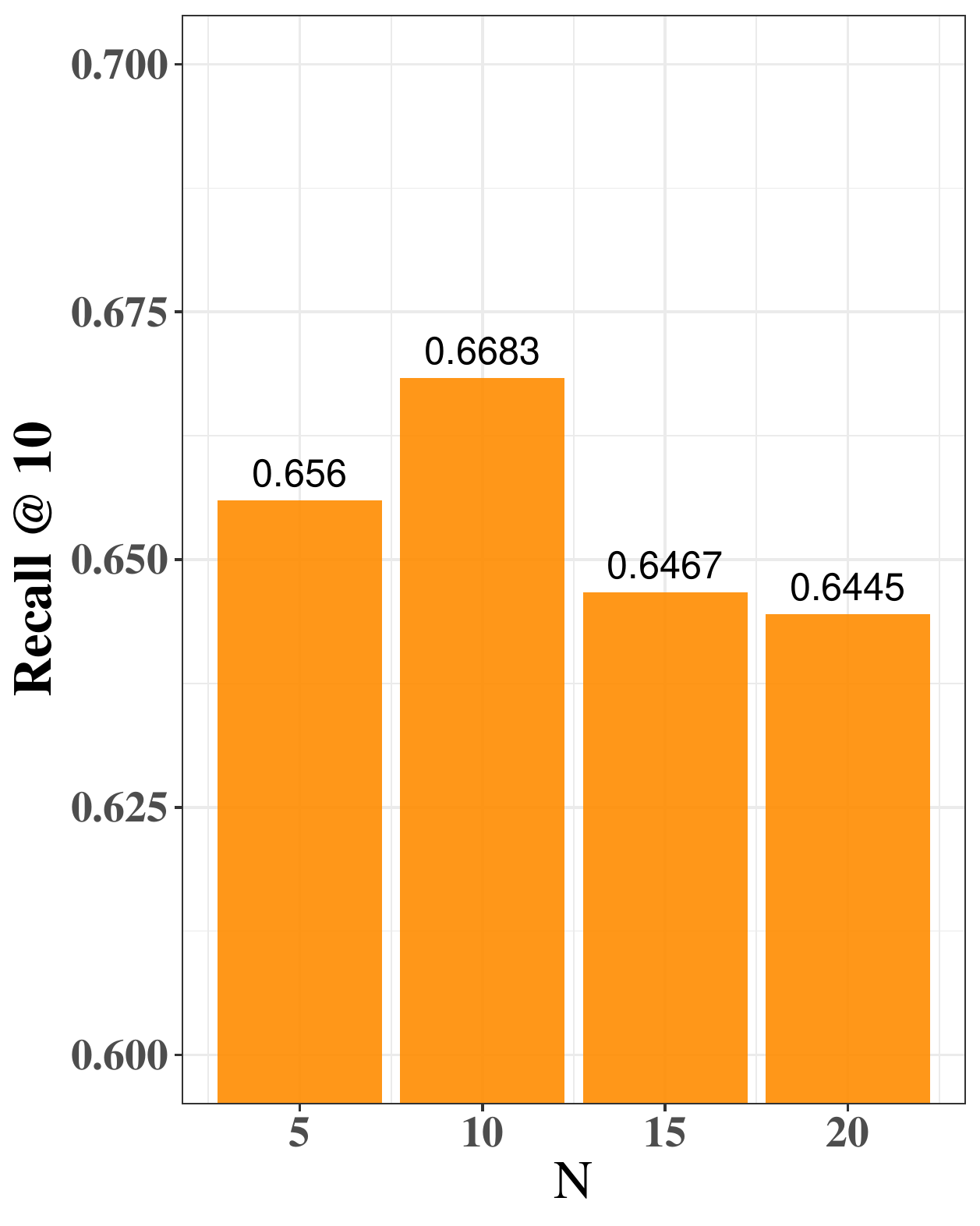}}
    \subfloat[LSTM-based decoder]{
    \label{rq2-lstm-att-mind}
    \includegraphics[width=0.24\textwidth,height=0.3001\textwidth]{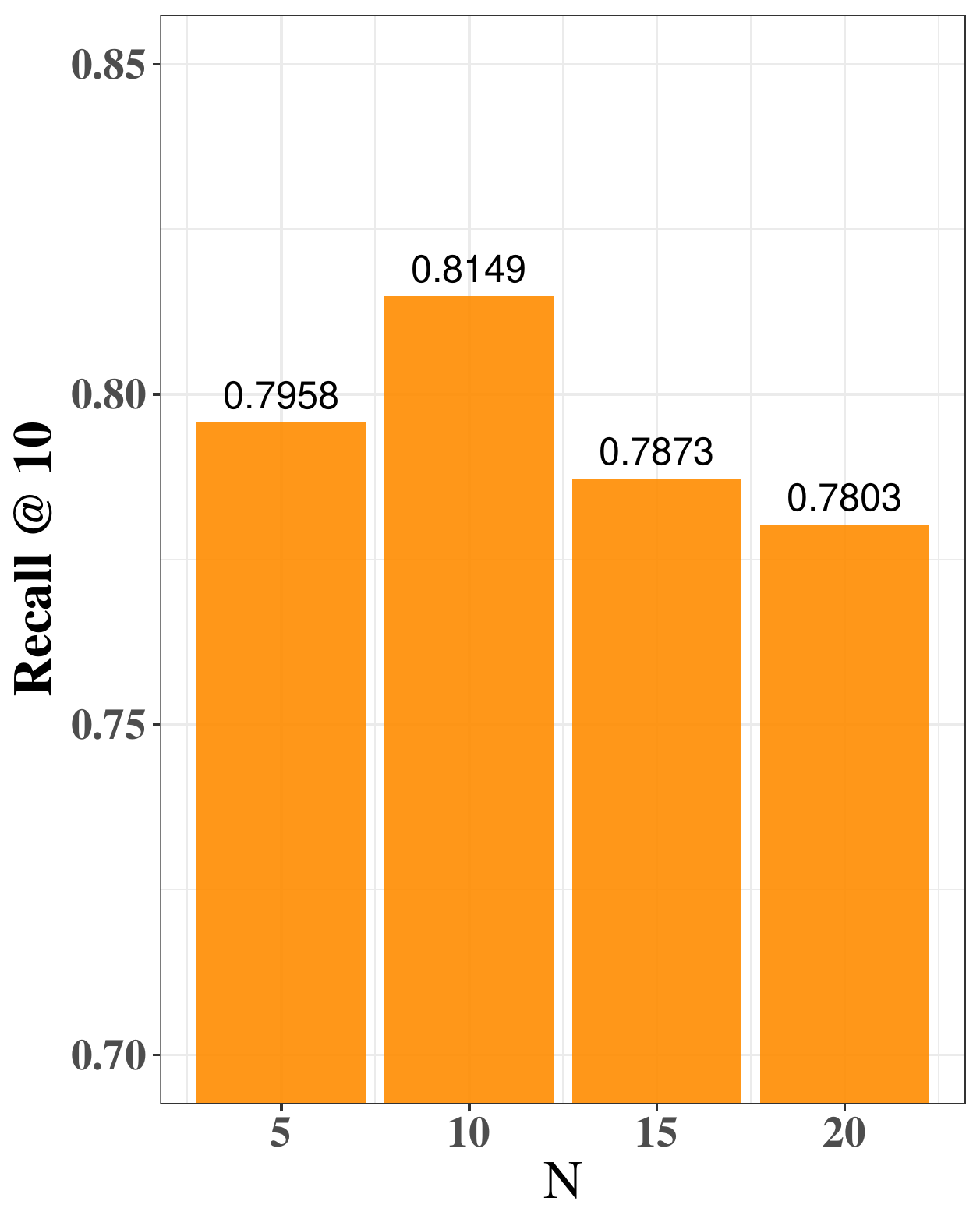}}
    \subfloat[GRU-based decoder]{
    \label{rq2-gru-att-mind}
    \includegraphics[width=0.24\textwidth,height=0.3001\textwidth]{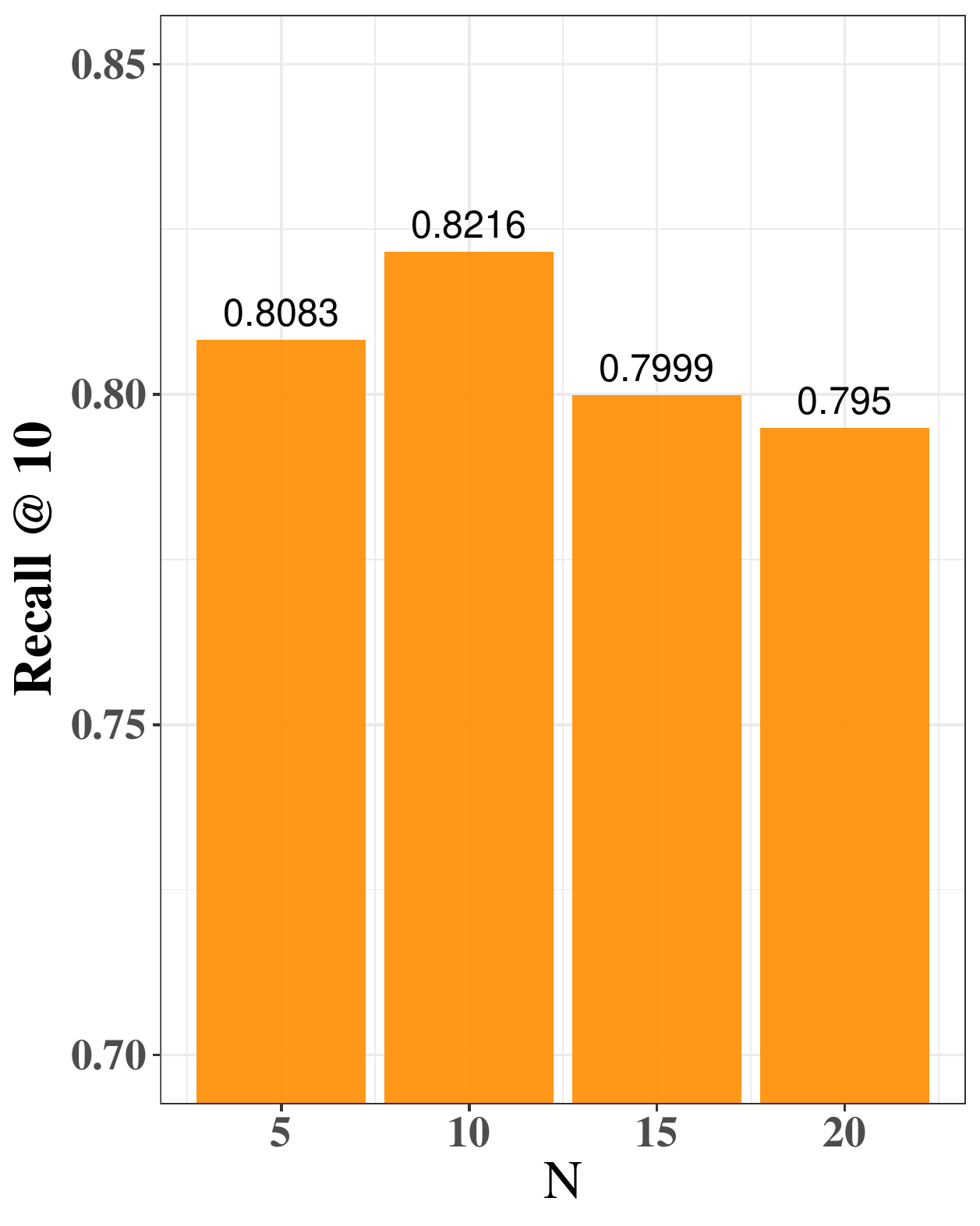}}
    \subfloat[Transfrmer decoder]{%
    \label{rq2-seq-att-mind}
    \includegraphics[width=0.24\textwidth,height=0.3001\textwidth]{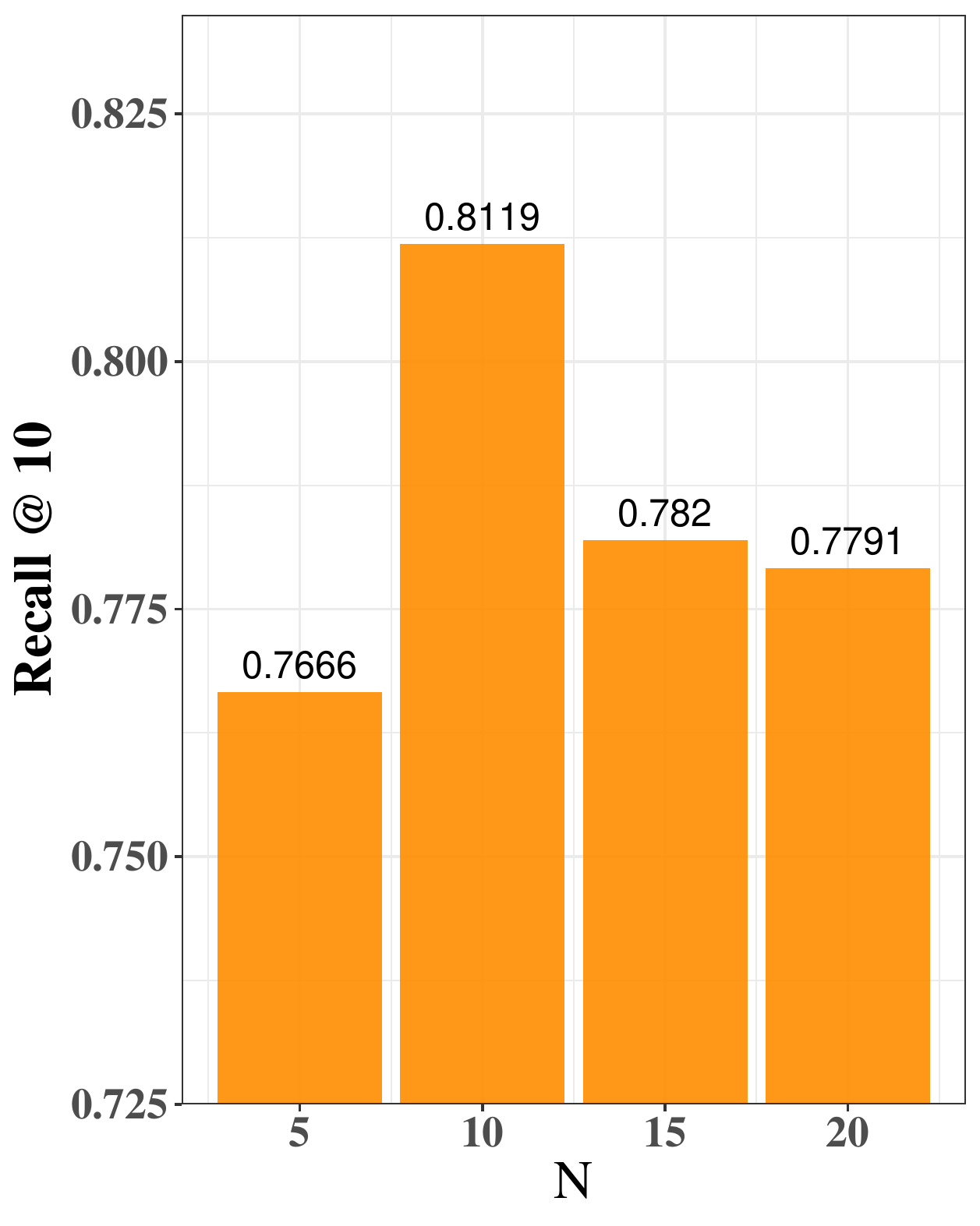}}
    \caption{Attack performance with different $N$ on MIND.}
    \label{rq2-mind}
\end{figure}

In this subsection, we conduct experiments to see how the number of exposed items affect the attack model performance. We illustrate the results when using the self-attention based encoding method. The results of the other two encoding methods show the similar trend.

Fig. \ref{rq2-point-att-zhihu}, Fig. \ref{rq2-lstm-att-zhihu}, Fig. \ref{rq2-gru-att-zhihu}, and Fig. \ref{rq2-seq-att-zhihu} show the attack performance (Recall@10) of point-wise decoding, LSTM-based decoders, GRU-based decoders, and transformer decoders on Zhihu, respectively. 
We can see that when using point-wise decoding, different exposed item numbers lead to similar attack performance. However, when using sequence-wised decoding, the attack performance is much better and varies much more with different $N$. The reason could be that the recommendation scenario of Zhihu is answer recommendation. Each answer belongs to a latent question. Due to the fact that point-wise decoding doesn't model the sequential order of user behavior, so point-wise decoding cannot capture the change of underlying questions of the answers, leading to a more flat and lower attack performance. However, the sequence-wise decoding performs the inference based on the previous inference results; such a method could have some capability to learn the change of latent underlying questions from the previous inference results. As a result, the sequence-wise decoding achieves a much higher attack performance with the proper setting of $N$.

Fig. \ref{rq2-point-att-mind}, Fig. \ref{rq2-lstm-att-mind}, Fig. \ref{rq2-gru-att-mind}, and  Fig. \ref{rq2-seq-att-mind} show the attack performance (Recall@10) of point-wise decoding and three different sequence-wise decoding methods on the MIND dataset, correspondingly. We can see that on the news recommendation dataset, 
both the point-wise decoding and sequence-wise decoding methods have  more various attack performance with the different setting of $N$. 

Besides, we can see that on both datasets, the best attack performance is achieved when $N=10$. This observation indicates that with fewer exposed items, the attack model may not learn strong signals to infer user behavior privacy. However, a too large number of exposed items could also introduce extra noise, which further confuses the model and downgrades the attack performance. $N=10$ could be a proper setting for the attack.

\subsection{Effect of Protection (RQ3)}
\begin{figure*}
    \captionsetup[subfloat]
    {}
    \centering
    \subfloat[Effect on Recall of the attack model]{
    \label{rq3-rec-zhihu-rand}
    \includegraphics[width=0.33\textwidth,height=0.33\textwidth]{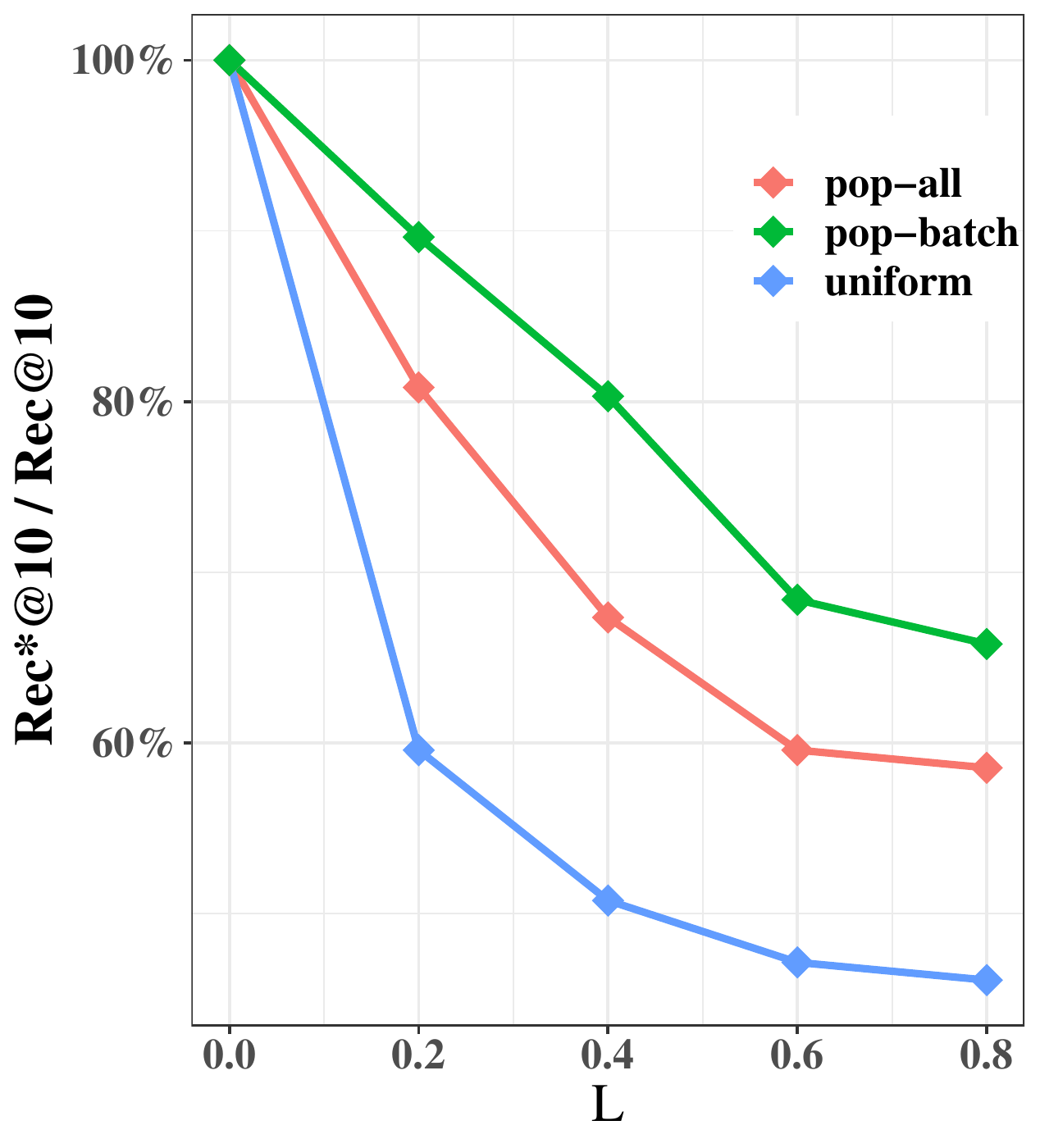}}
    \subfloat[Effect on NDCG of the attack model]{%
    \label{rq3-ndcg-zhihu-rand}
    \includegraphics[width=0.33\textwidth,height=0.33\textwidth]{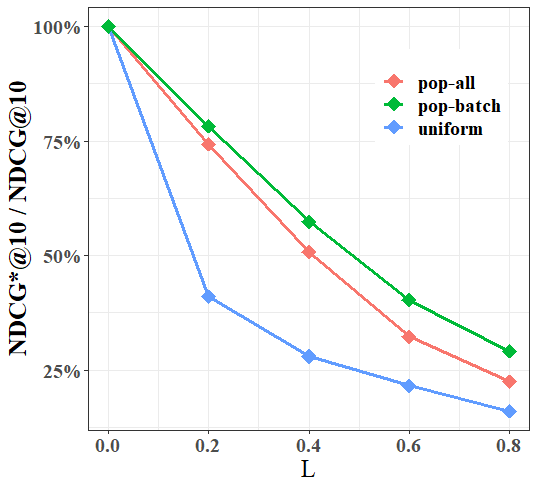}}
    \subfloat[Effect on recommendation accuracy]{%
    \label{rq3-acc-zhihu-rand}
    \includegraphics[width=0.33\textwidth,height=0.33\textwidth]{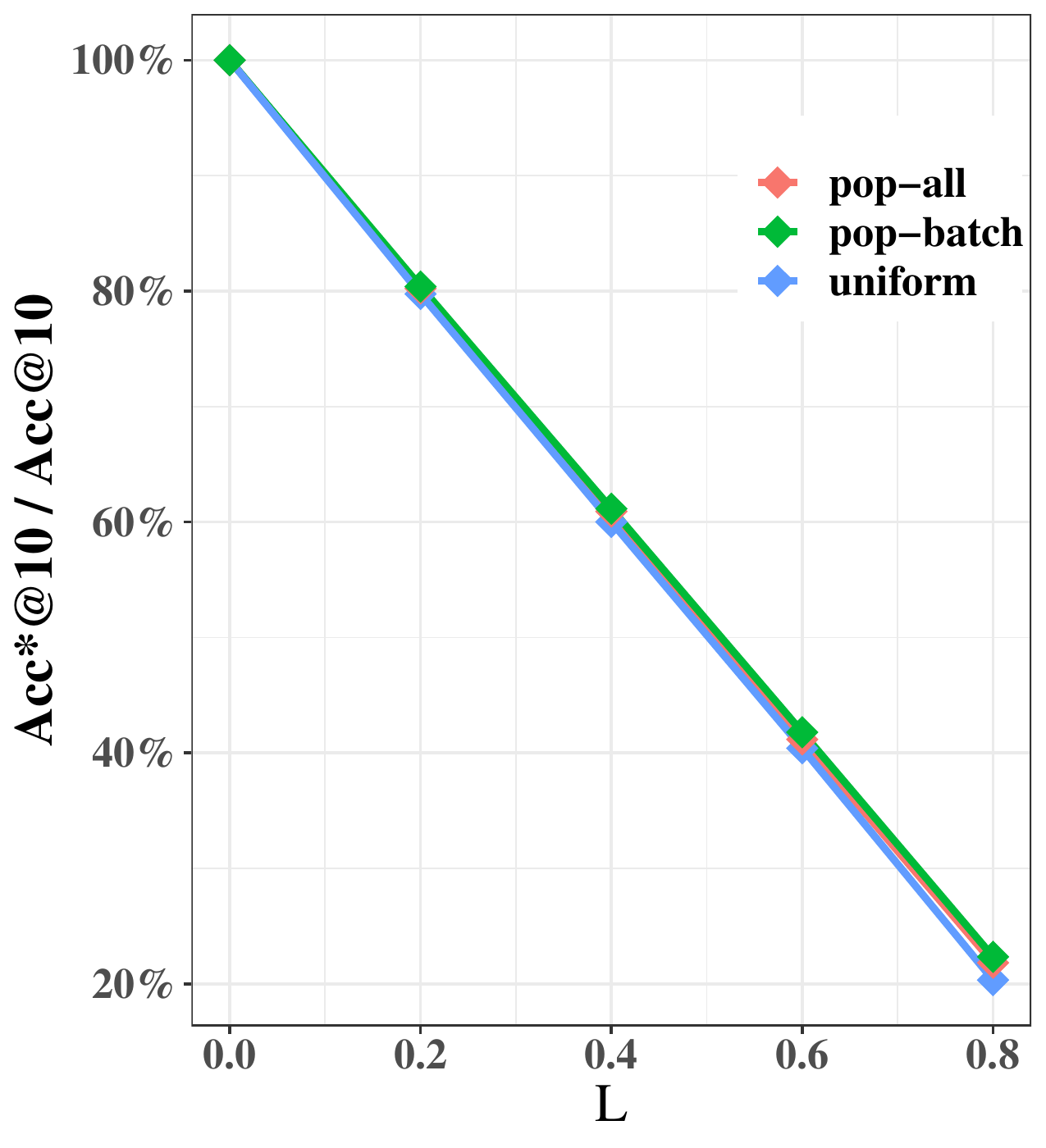}}
    \caption{Effect of the protection mechanism with random-based position selection on Zhihu. L is the proportion of replacement. Rec*, NDCG* and Acc* denote the new attack recall, new attack NDCG and new recommendation accuracy, respectively.}
    \label{rq3-zhihu-rand}
\end{figure*}
\begin{figure*}
    \captionsetup[subfloat]
    {}
    \centering
    \subfloat[Effect on Recall of the attack model]{
    \label{rq3-rec-mind-rand}
    \includegraphics[width=0.33\textwidth,height=0.33\textwidth]{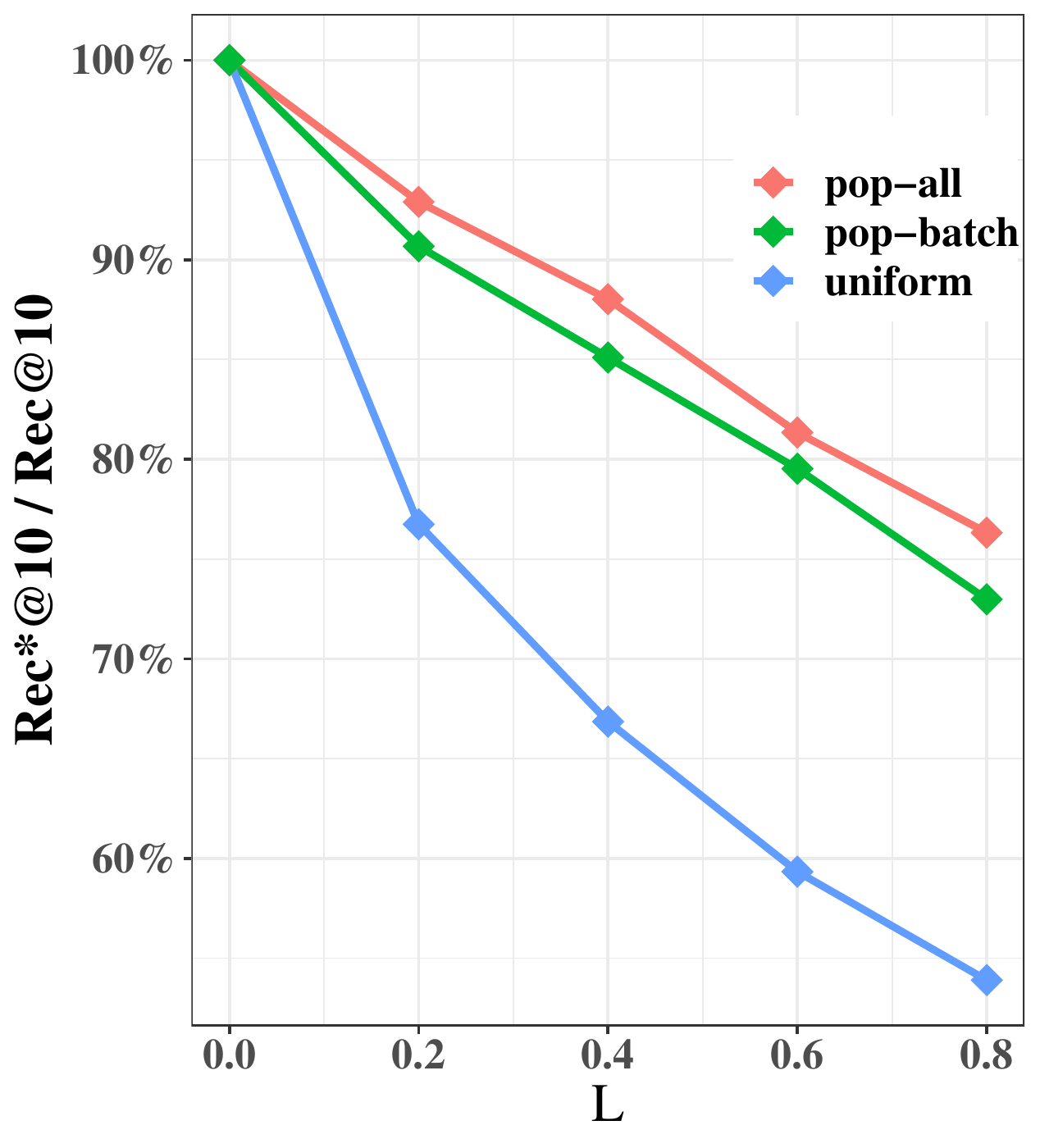}}
    \subfloat[Effect on NDCG of the attack model]{%
    \label{rq3-ndcg-mind-rand}
    \includegraphics[width=0.33\textwidth,height=0.33\textwidth]{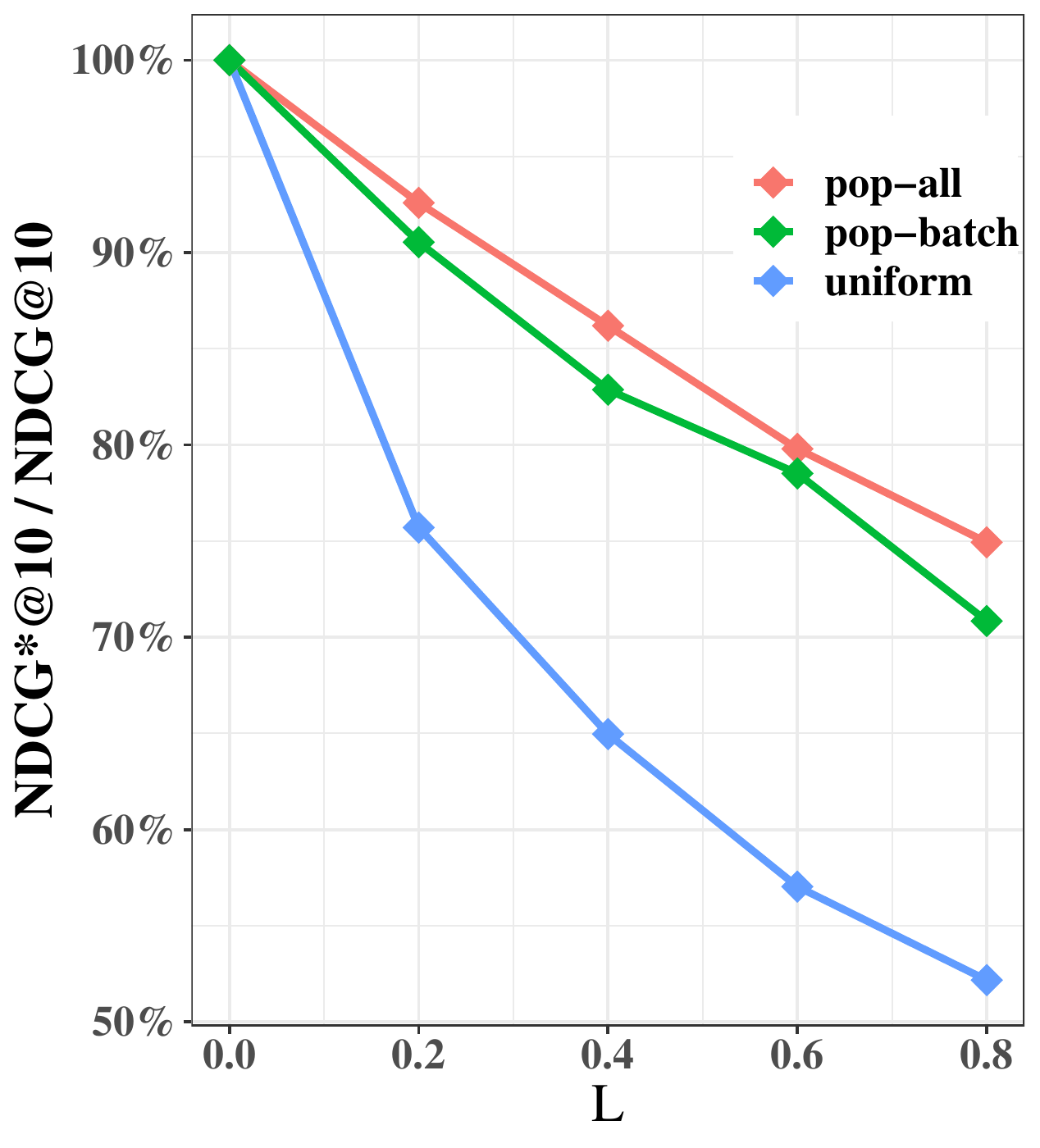}}
    \subfloat[Effect on recommendation accuracy]{%
    \label{rq3-acc-mind-rand}
    \includegraphics[width=0.33\textwidth,height=0.33\textwidth]{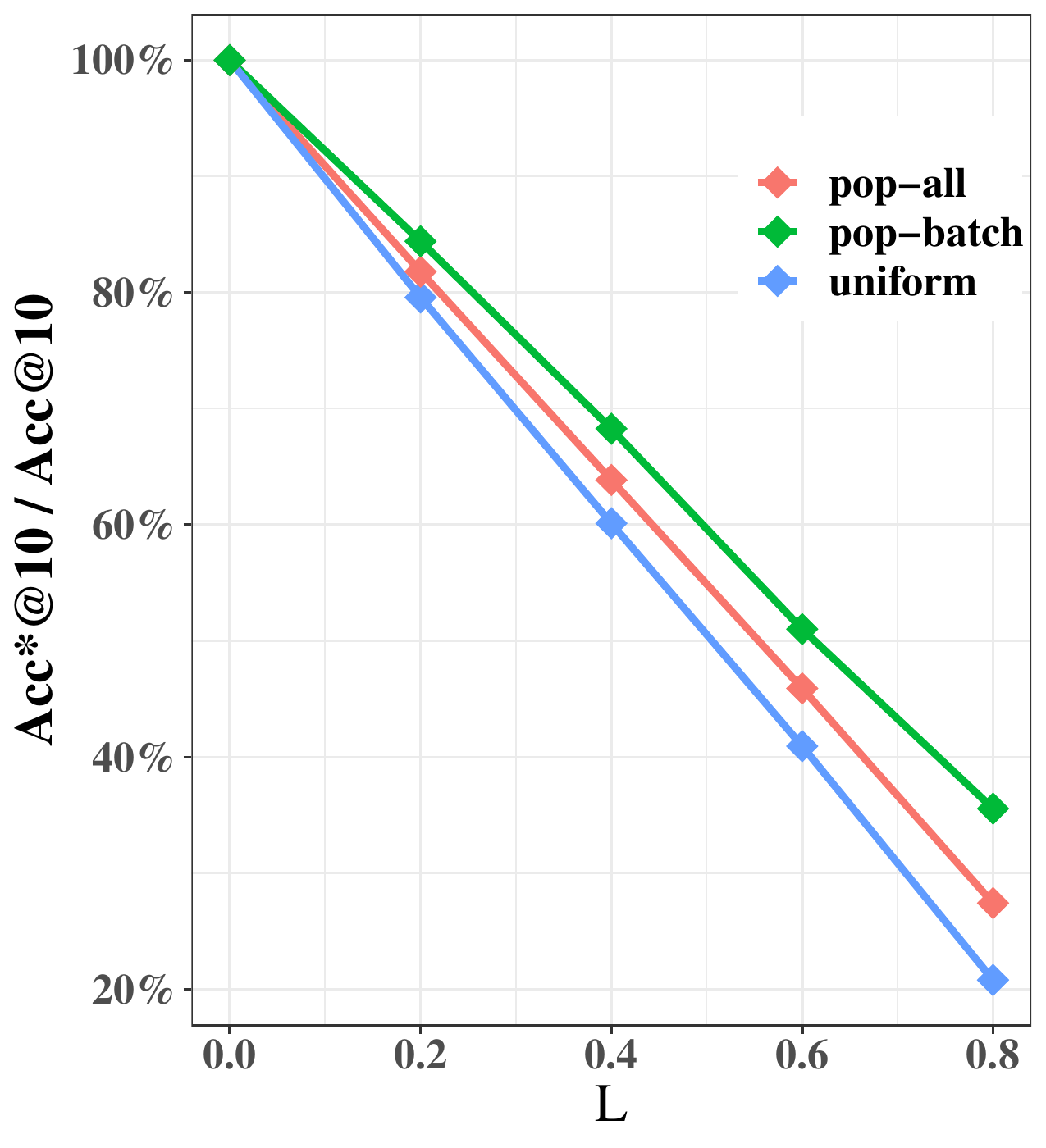}}
    \caption{Effect of the protection mechanism with random-based position selection on MIND. L is the proportion of replacement. Rec*, NDCG* and Acc* denote the new attack recall, new attack NDCG and new recommendation accuracy, respectively.}
    \label{rq3-mind-rand}
\end{figure*}

\begin{figure*}
    \captionsetup[subfloat]
    {}
    \centering
    \subfloat[Effect on Recall of the attack model]{
    \label{rq3-rec-zhihu-similarity}
    \includegraphics[width=0.33\textwidth,height=0.33\textwidth]{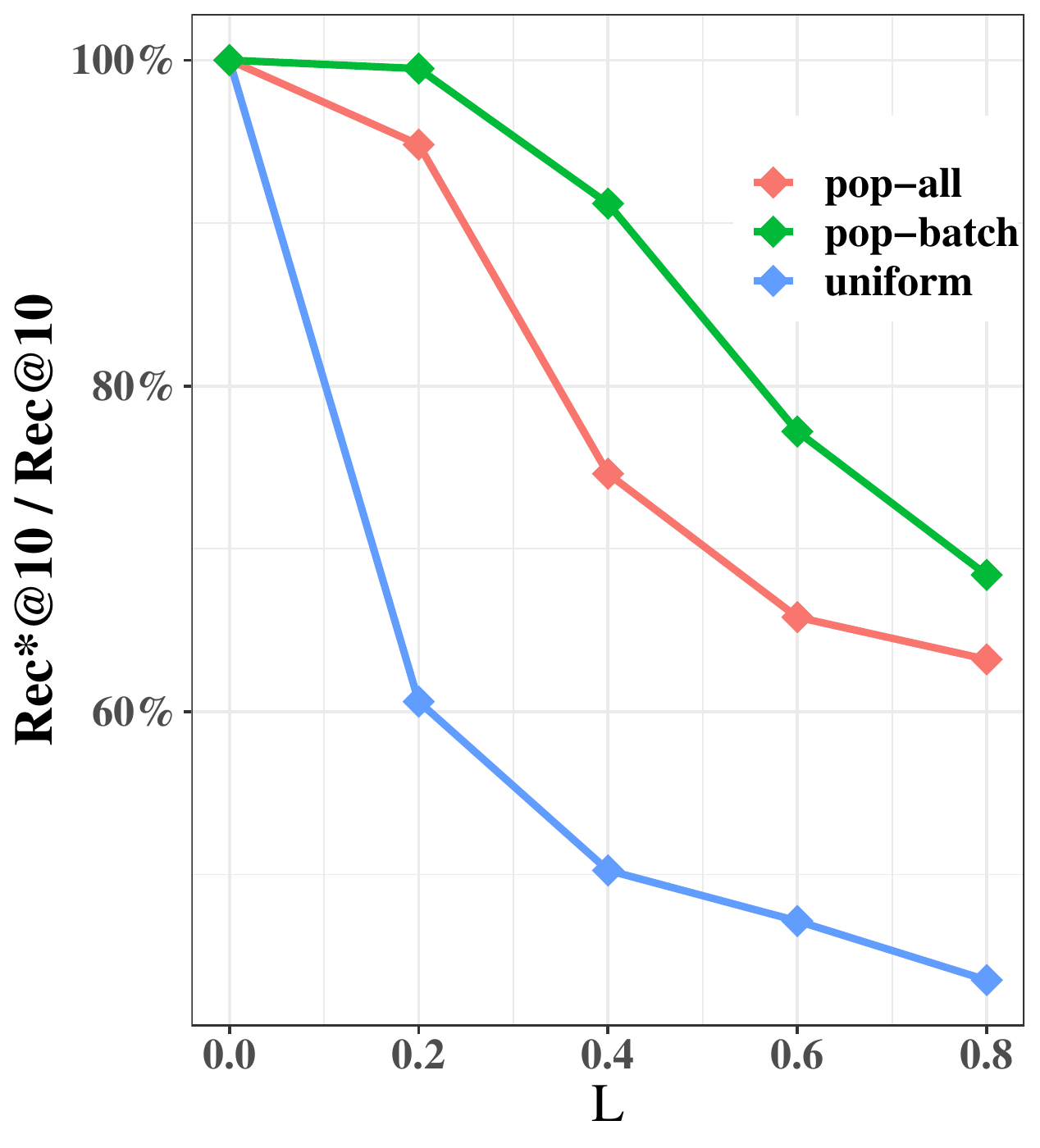}}
    \subfloat[Effect on NDCG of the attack model]{%
    \label{rq3-ndcg-zhihu-similarity}
    \includegraphics[width=0.33\textwidth,height=0.33\textwidth]{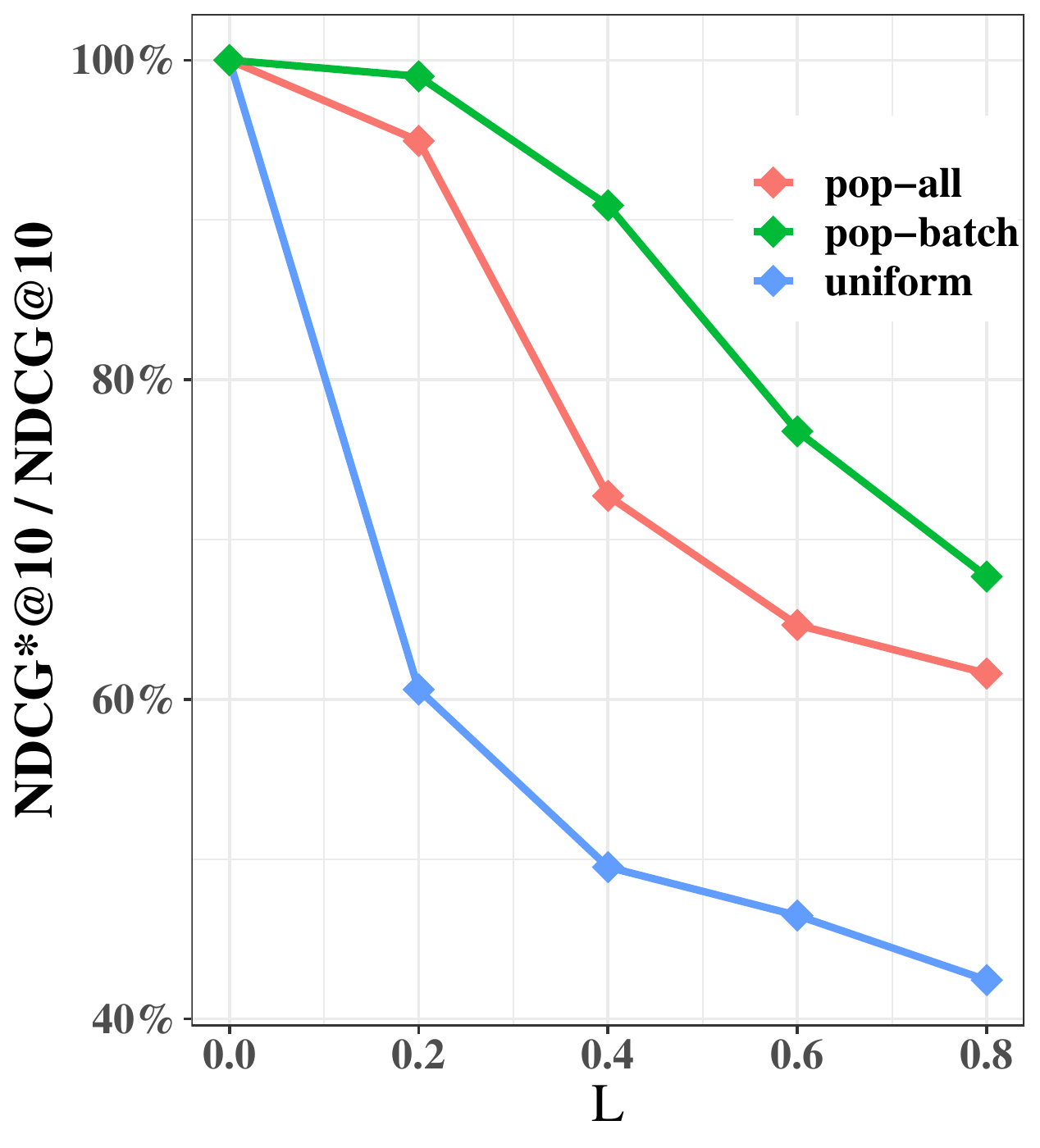}}
    \subfloat[Effect on recommendation accuracy]{%
    \label{rq3-acc-zhihu-similarity}
    \includegraphics[width=0.33\textwidth,height=0.33\textwidth]{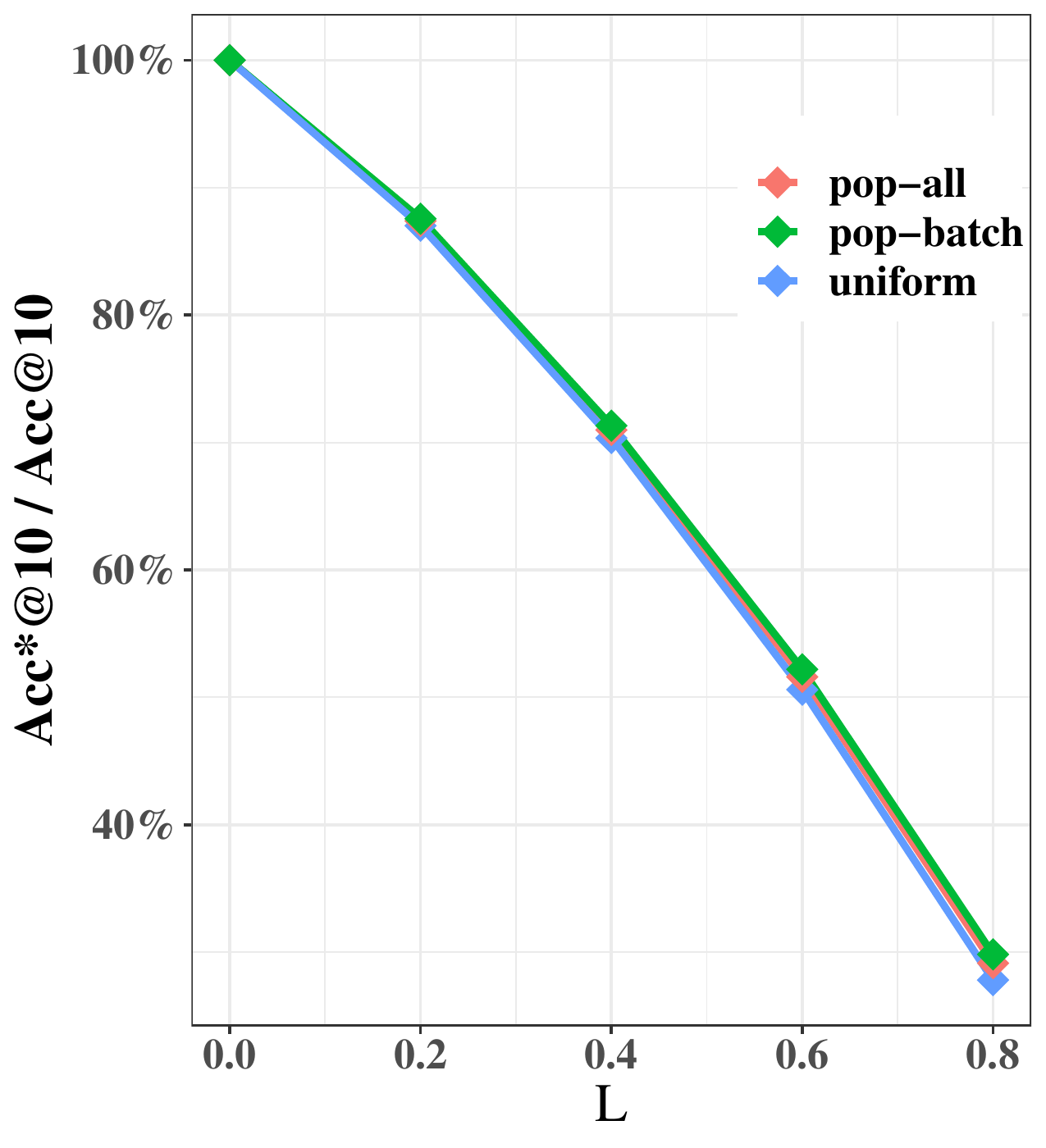}}
    \caption{Effect of the protection mechanism with similarity-based position selection on Zhihu. L is the proportion of replacement. Rec*, NDCG* and Acc* denote the new attack recall, new attack NDCG and new recommendation accuracy, respectively.}
    \label{rq3-zhihu-sim}
\end{figure*}
\begin{figure*}
    \captionsetup[subfloat]
    {}
    \centering
    \subfloat[Effect on Recall of the attack model]{
    \label{rq3-rec-mind-similarity}
    \includegraphics[width=0.33\textwidth,height=0.33\textwidth]{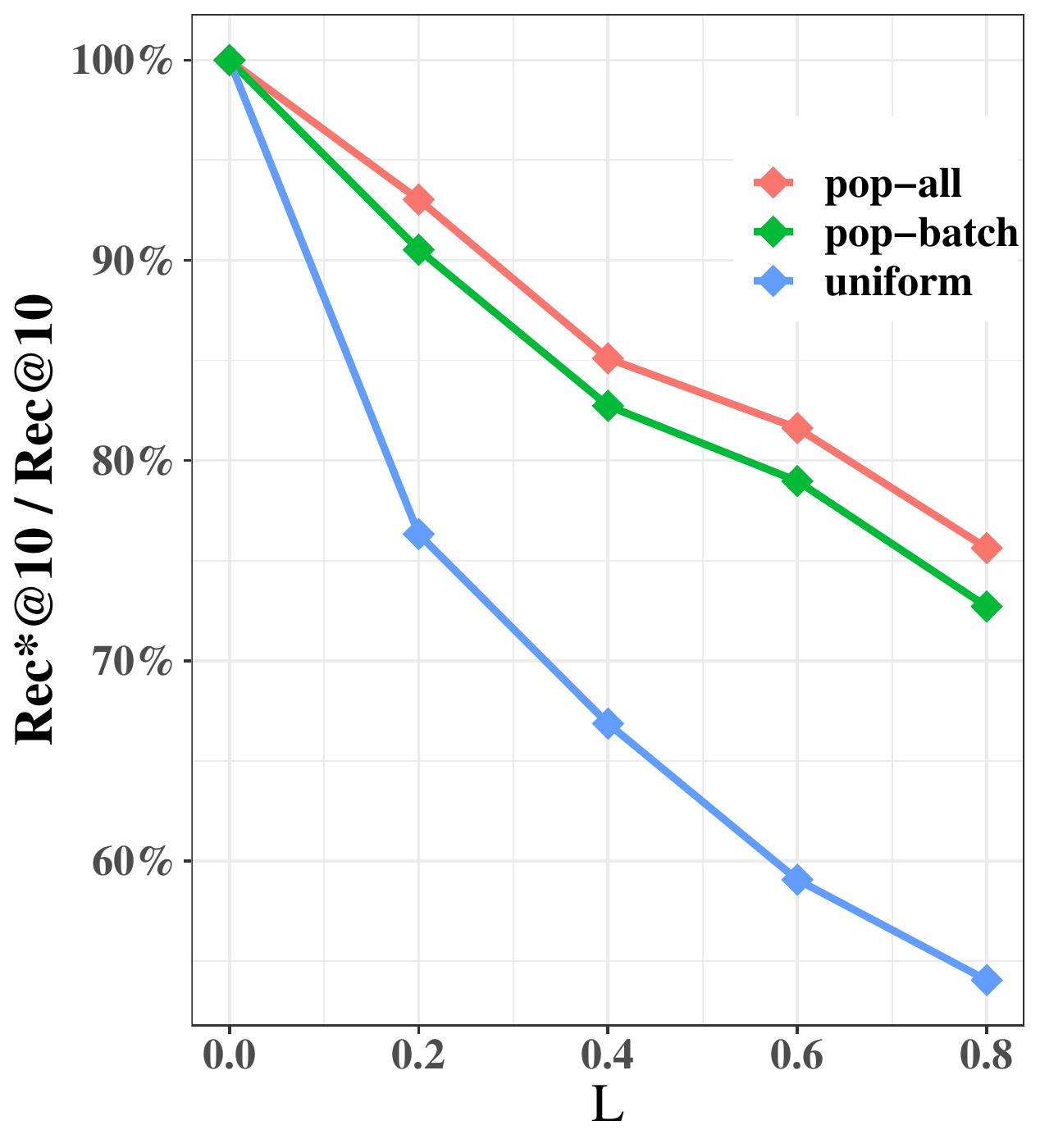}}
    \subfloat[Effect on NDCG of the attack model]{%
    \label{rq3-ndcg-mind-similarity}
    \includegraphics[width=0.33\textwidth,height=0.33\textwidth]{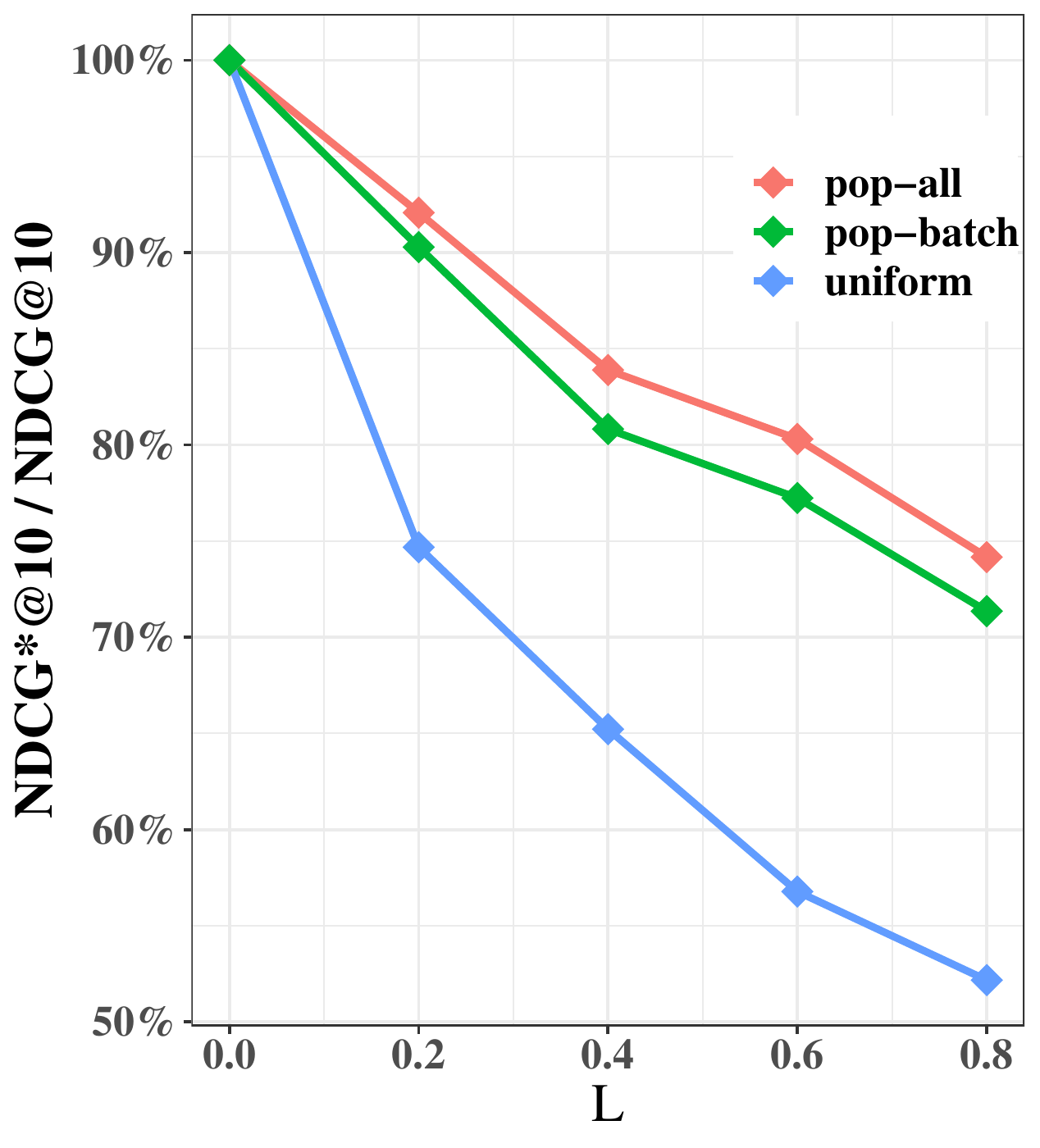}}
    \subfloat[Effect on recommendation accuracy]{%
    \label{rq3-acc-mind-similarity}
    \includegraphics[width=0.33\textwidth,height=0.33\textwidth]{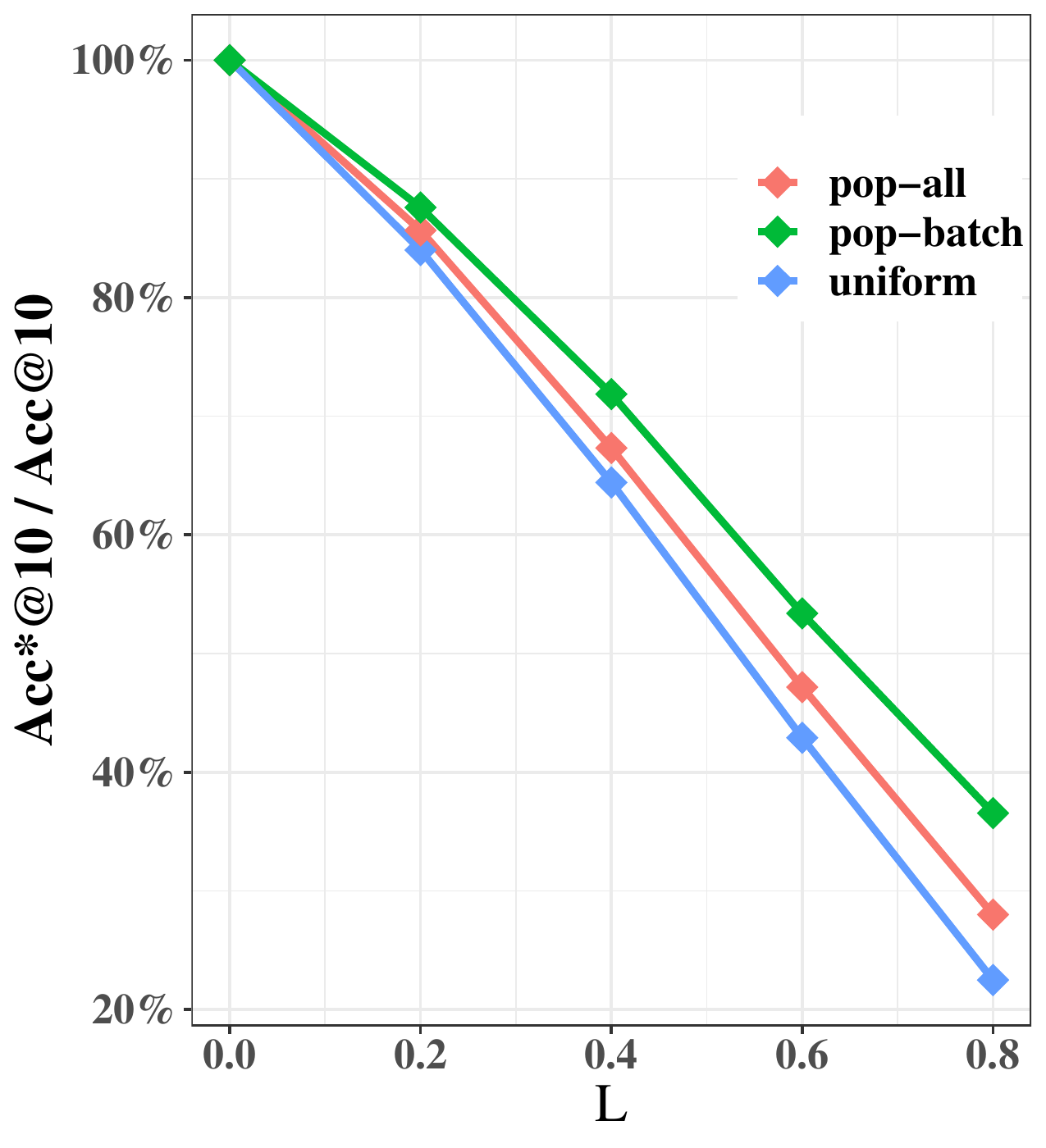}}
    \caption{Effect of the protection mechanism with similarity-based position selection on MIND. L is the proportion of replacement. Rec*, NDCG* and Acc* denote the new attack recall, new attack NDCG and new recommendation accuracy, respectively.}
    \label{rq3-mind-sim}
\end{figure*}

\begin{figure*}
    \captionsetup[subfloat]
    {}
    \centering
    \subfloat[Effect on Recall of the attack model]{
    \label{rq3-rec-zhihu-similarityornot}
    \includegraphics[width=0.33\textwidth,height=0.33\textwidth]{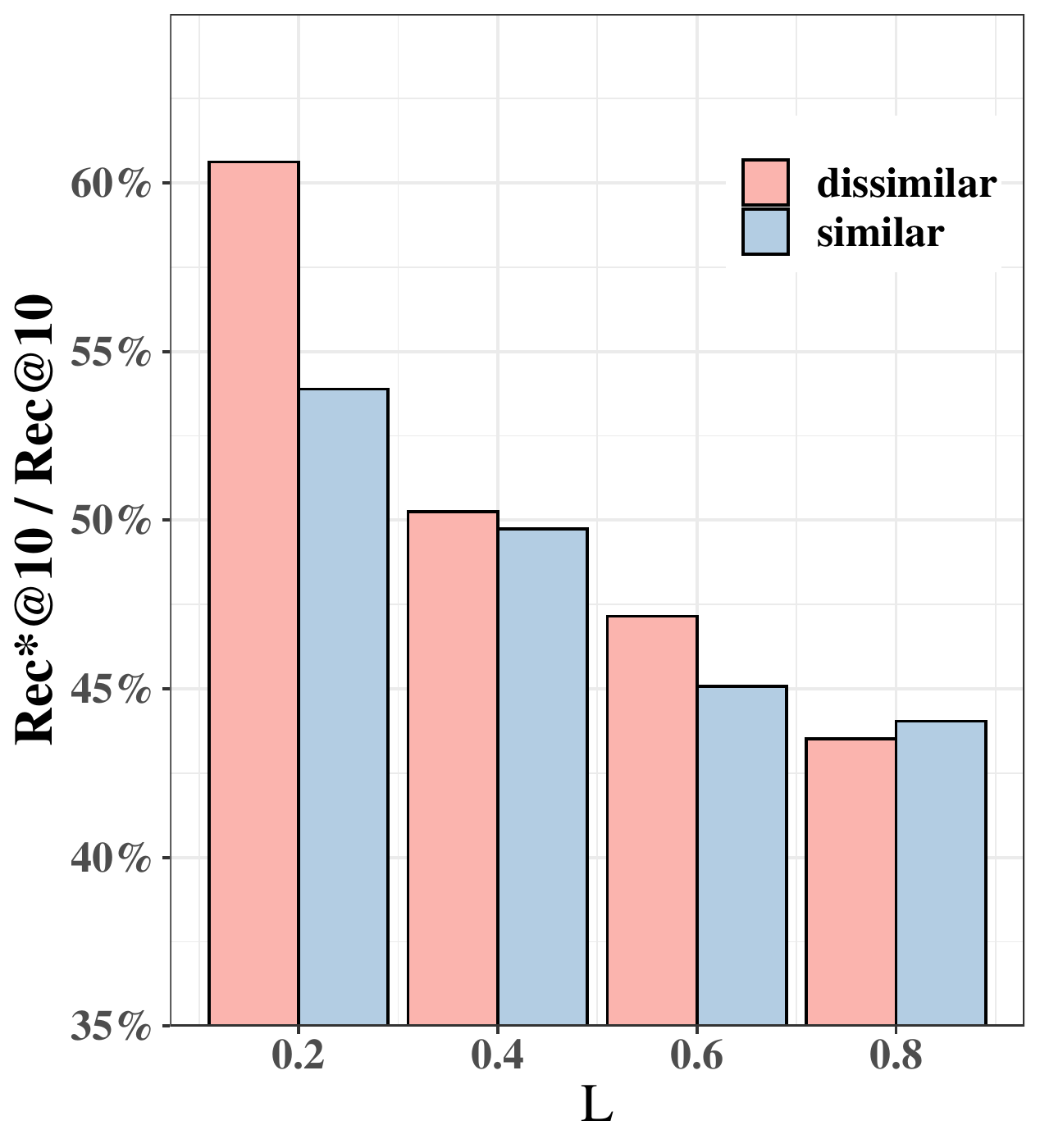}}
    \subfloat[Effect on NDCG of the attack model]{%
    \label{rq3-ndcg-zhihu-similarityornot}
    \includegraphics[width=0.33\textwidth,height=0.33\textwidth]{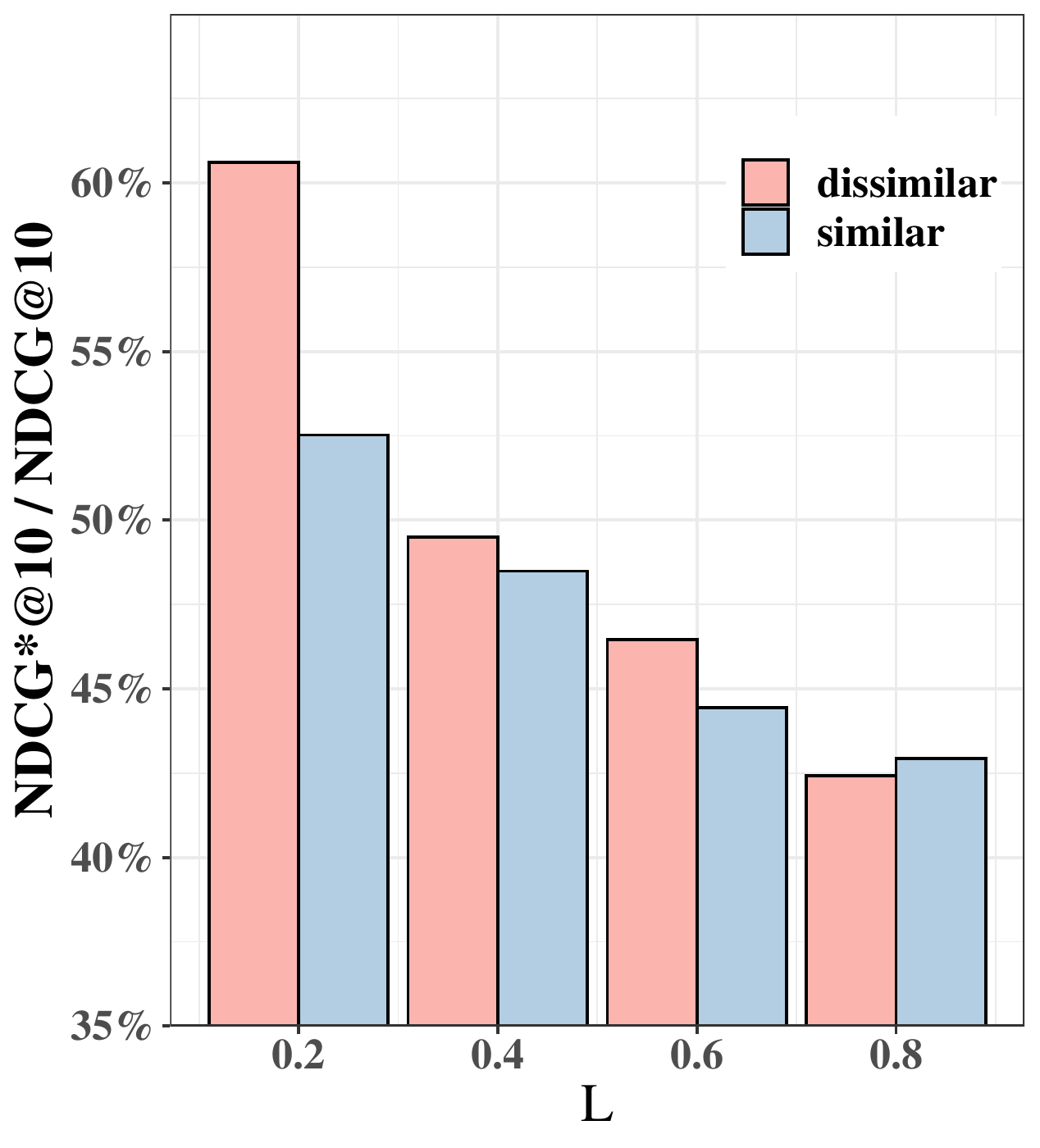}}
    \subfloat[Effect on recommendation accuracy]{%
    \label{rq3-acc-zhihu-similarityornot}
    \includegraphics[width=0.33\textwidth,height=0.33\textwidth]{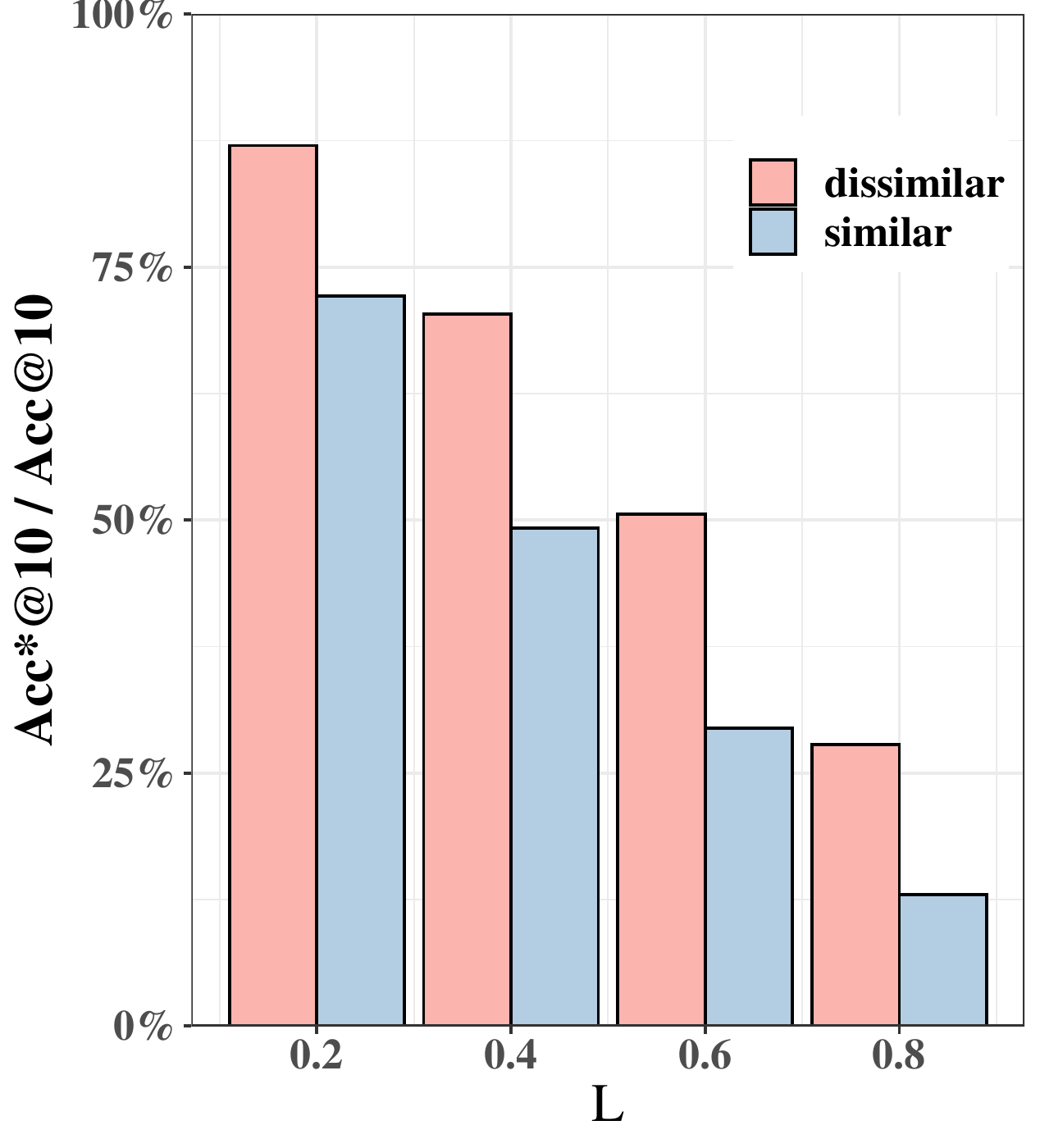}}
    \caption{Effect of the similarity-based position selection protection mechanism with uniform-based item replacement on Zhihu. L is the proportion of replacement. Rec*, NDCG* and Acc* denote the new attack recall, new attack NDCG and new recommendation accuracy, respectively.}
    \label{rq3-zhihu-similarityornot}
\end{figure*}
\begin{figure*}
    \captionsetup[subfloat]
    {}
    \centering
    \subfloat[Effect on Recall of the attack model]{
    \label{rq3-rec-mind-similarityornot}
    \includegraphics[width=0.33\textwidth,height=0.33\textwidth]{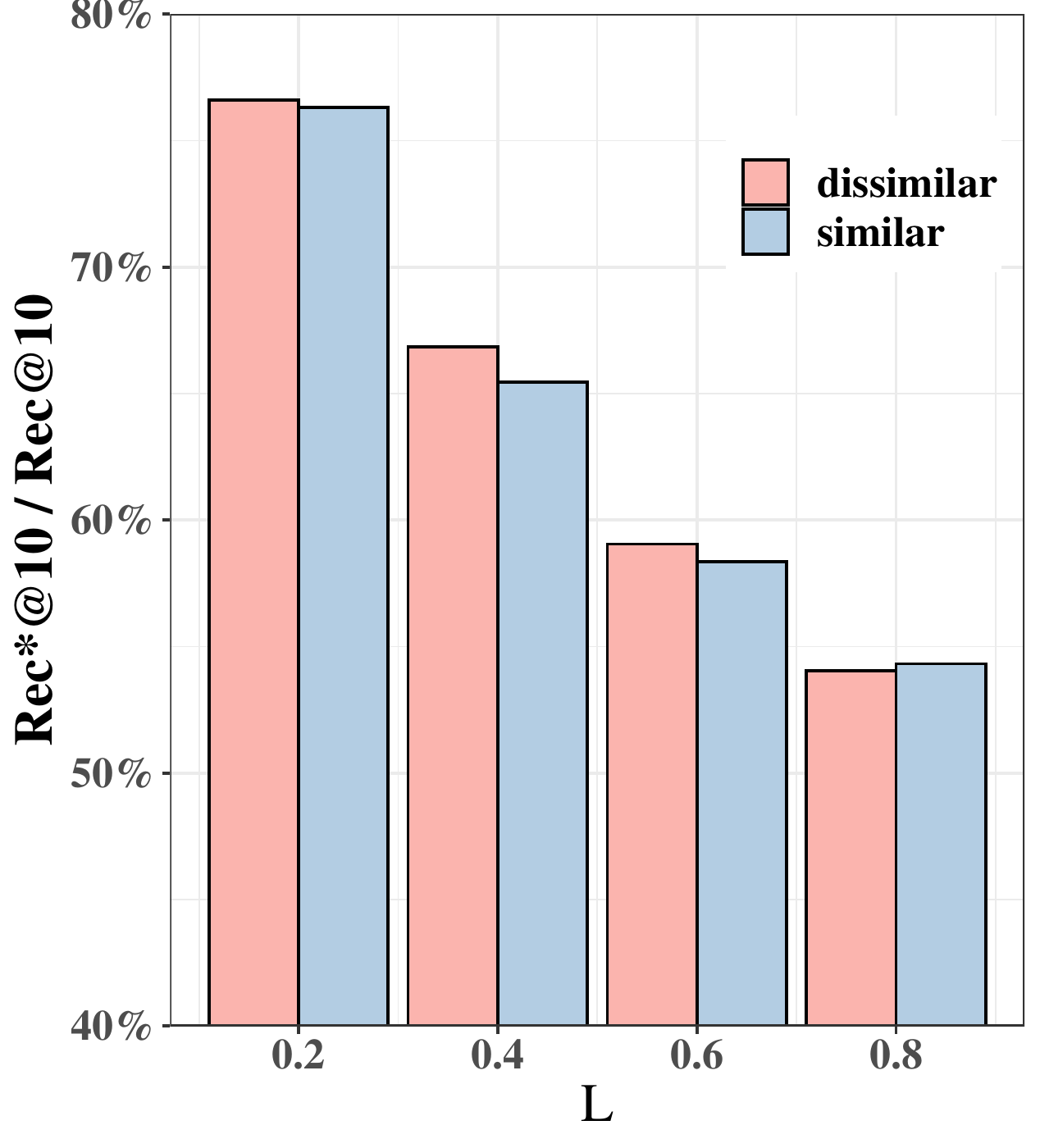}}
    \subfloat[Effect on NDCG of the attack model]{%
    \label{rq3-ndcg-mind-similarityornot}
    \includegraphics[width=0.33\textwidth,height=0.33\textwidth]{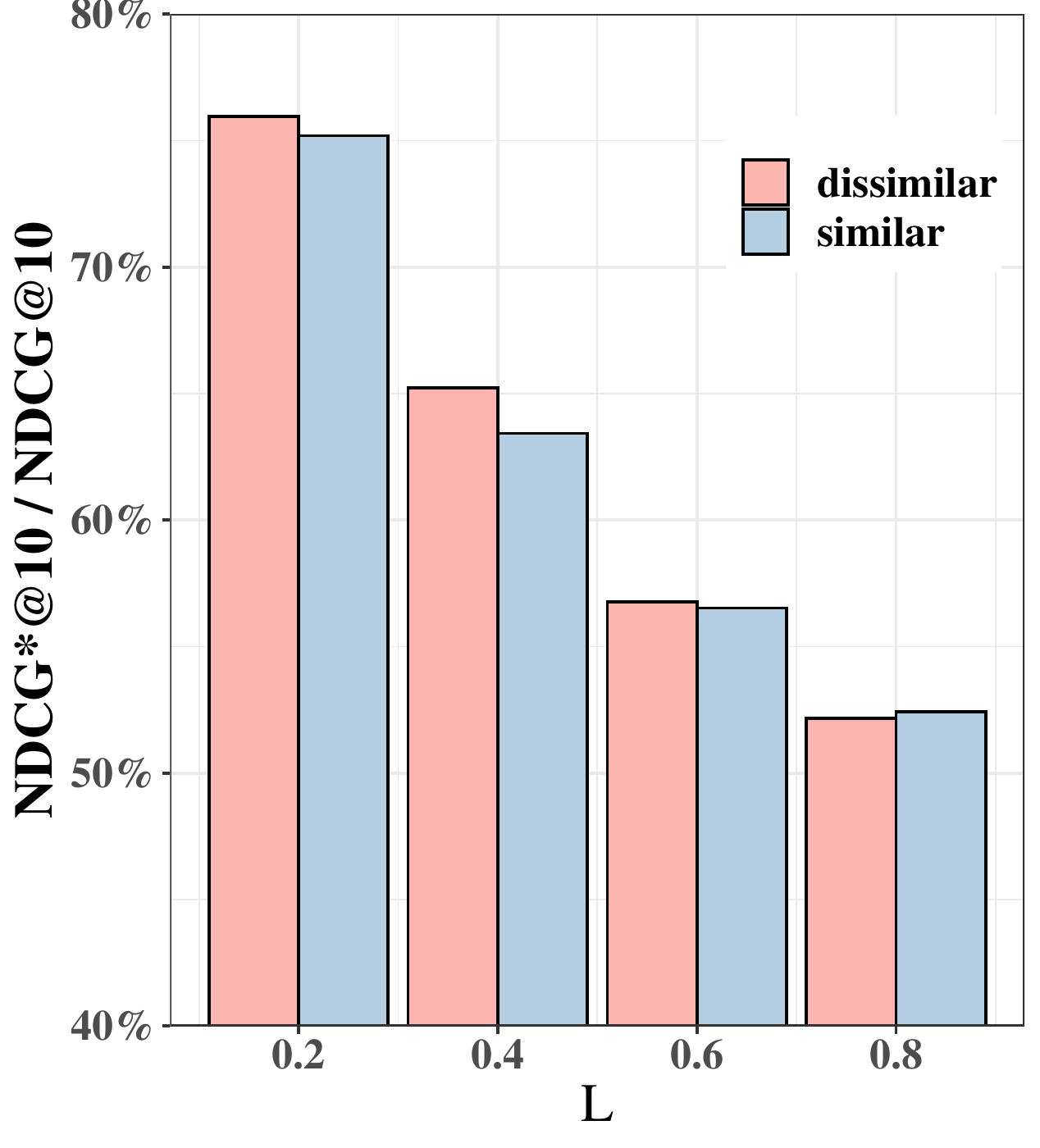}}
    \subfloat[Effect on recommendation accuracy]{%
    \label{rq3-acc-mind-similarityornot}
    \includegraphics[width=0.33\textwidth,height=0.33\textwidth]{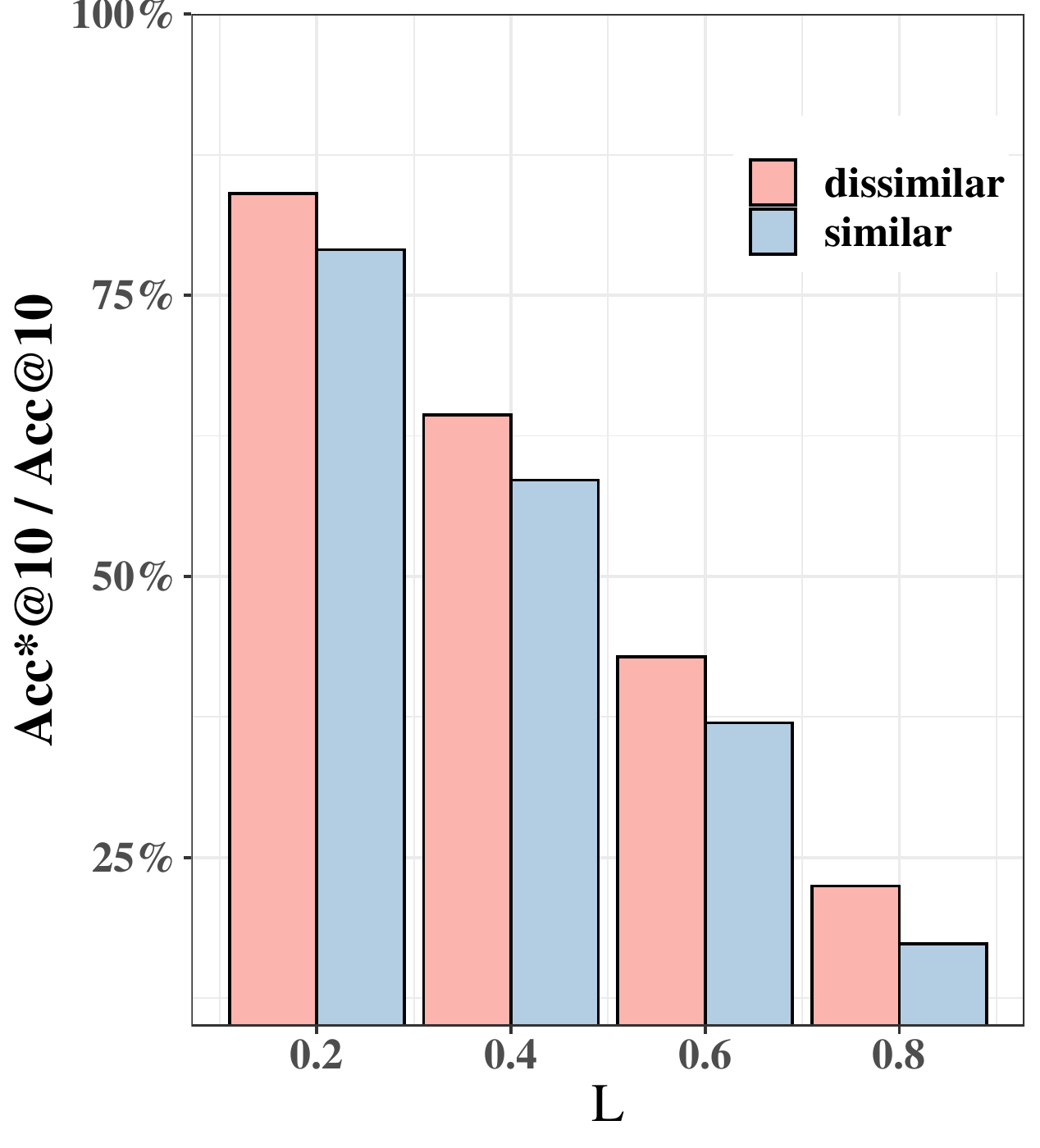}}
    \caption{Effect of the similarity-based position selection protection mechanism with uniform-based item replacement on MIND. L is the proportion of replacement. Rec*, NDCG* and Acc* denote the new attack recall, new attack NDCG and new recommendation accuracy, respectively.}
    \label{rq3-mind-similarityornot}
\end{figure*}
In this subsection, we conduct experiments to verify the effect of the protection mechanism. We show the attack performance regarding recall and NDCG\footnote{Results of MRR show the same trend with NDCG.}; we also report the new recommendation accuracy. Fig. \ref{rq3-zhihu-rand} and Fig. \ref{rq3-mind-rand} illustrate the results with different item replacement ratios using random-based position selection at the first stage. The results using similarity-based position selection with different item replacement ratios are shown in Fig. \ref{rq3-zhihu-sim}, and Fig. \ref{rq3-mind-sim}. 
We can see that with the increase of the replacement proportion $L$, the attack performance drops dramatically. Such observation demonstrates that random exposure can help to alleviate the user privacy leakage risk. However, we can also see from Fig. \ref{rq3-acc-zhihu-rand}, Fig. \ref{rq3-acc-mind-rand}, Fig. \ref{rq3-acc-zhihu-similarity} and Fig. \ref{rq3-acc-mind-similarity} that the recommendation accuracy decreases. It indicates a trade-off effect between the recommendation accuracy and the leakage risk. 
We can see from Fig. \ref{rq3-zhihu-rand} that random-based position selection method with uniform-based replacement mechanism alleviate about 40\% leakage risk while maintain about 80\% recommendation accuracy when the setting of the replacement ration $L$ to 0.2 in the Zhihu dataset.

Given the fact that users would only click very few items in the exposed list, the designed similarity-based position selection method, which can keep the items that the user would like to click and then replace the other items with random or more diverse items, achieve better recommendation accuracy.
We can see from the comparison between Fig. \ref{rq3-acc-zhihu-rand} and Fig. \ref{rq3-acc-zhihu-similarity} (also (Fig. \ref{rq3-acc-mind-rand} and Fig. \ref{rq3-acc-mind-similarity})) that compared with random-based position selection method, the similarity-based selection method maintain higher recommendation accuracy with the increase of the replacement ratio $L$, when using uniform-based item replacement.

Besides, Fig. \ref{rq3-zhihu-similarityornot} and Fig.\ref{rq3-mind-similarityornot} show the results of using similarity-based position selection at the first stage and uniform-based replacement at the second stage on two datasets. Different from the method of replacing the dissimilar position items (i.e., similarity-based position selection method with uniform-based replacement in Fig. \ref{rq3-zhihu-sim} and Fig. \ref{rq3-mind-sim}), the similar position selection method selects positions with items matching well with user preference. The results show that it can alleviate user privacy leakage risk at the cost of recommendation accuracy degradation compared with replacing the dissimilar positions items. The reason could be that more similar positions items reveal much more user behavior information.

Regarding the item replacement strategy, we can see that uniform-based replacement is the most effective one to downgrade the attack performance while in-bath popularity replacement and overall popularity replacement have the similar effect on attack performance.
For the recommendation accuracy, in-batch popularity replacement achieves the minimum accuracy sacrifice.

\section{Conclusions and Future Work}
\label{sec:conclusion}
In this paper, we have investigated the risk of user behavior privacy leakage in the field of recommender systems. We have focused on answering the question of whether the user past behavior privacy can be inferred from system behavior data.
We have conducted an attack model through an encoder-decoder architecture. We have proposed to utilize three different encoding methods to encode the system exposure data. 
Thereafter, we have presented the point-wise decoding and three sequence-wise decoders to infer user past behaviors from the encoded representation. 
Experimental results on two real-world datasets have verified a great danger of privacy leakage in recommender systems. 
To alleviate the risk, we have proposed a two-stage protection mechanism based on infusing random items into the exposed item sets. We first select exposure positions based on random or item similarity at the first stage, and then replace exposed items on the corresponding positions with uniform or popularity sampled items. 
Experimental results have demonstrated a trade-off effect between the recommendation accuracy and the privacy leakage risk. 

We hope that this work could raise more community concerns regarding the protection of recommender system behavior data other than just focusing on user perspectives.
Compared with the sparse user historical behavior, the large volume of system exposure data receives relatively less research attention. This work have proposed a new perspective regarding the attack and protection of the system behavior data.

Future work includes investigating more advanced encoding and decoding methods to conduct the attack. Besides, the exposure data is highly affected by various kinds of biases (e.g., the popularity bias) so how to conduct debiased attack model would also be a research direction.
More importantly, one of the promising future directions is the design of more advanced protection methods with little sacrifice of the recommendation accuracy. Finally, exploring the relationship across the recommendation accuracy, the privacy leakage risk, and the recommendation diversity or novelty would also be an interesting direction. 

\section*{Acknowledgements}
This work was supported by the Natural Science Foundation of China (62272274, 62202271, 61972234, 61902219, 62072279, 62102234), the Key Scientific and Technological Innovation Program of Shandong Province (2019JZZY010129), the Fundamental Research Funds of Shandong University, Meituan, the Tencent WeChat Rhino-Bird Focused Research Program (JR-WXG-2021411). All content represents the opinion of the authors, which is not necessarily shared or endorsed by their respective employers and/or sponsors. 

\clearpage
\bibliographystyle{ACM-Reference-Format}
\balance
\bibliography{main}


\begin{thebibliography}{67}


\ifx \showCODEN    \undefined \def \showCODEN     #1{\unskip}     \fi
\ifx \showDOI      \undefined \def \showDOI       #1{#1}\fi
\ifx \showISBNx    \undefined \def \showISBNx     #1{\unskip}     \fi
\ifx \showISBNxiii \undefined \def \showISBNxiii  #1{\unskip}     \fi
\ifx \showISSN     \undefined \def \showISSN      #1{\unskip}     \fi
\ifx \showLCCN     \undefined \def \showLCCN      #1{\unskip}     \fi
\ifx \shownote     \undefined \def \shownote      #1{#1}          \fi
\ifx \showarticletitle \undefined \def \showarticletitle #1{#1}   \fi
\ifx \showURL      \undefined \def \showURL       {\relax}        \fi
\providecommand\bibfield[2]{#2}
\providecommand\bibinfo[2]{#2}
\providecommand\natexlab[1]{#1}
\providecommand\showeprint[2][]{arXiv:#2}

\bibitem[Ammad-ud din et~al\mbox{.}(2019)]%
        {ammad-ud-din2019federated}
\bibfield{author}{\bibinfo{person}{Muhammad Ammad-ud din},
  \bibinfo{person}{Elena Ivannikova}, \bibinfo{person}{A.~Suleiman Khan},
  \bibinfo{person}{Were Oyomno}, \bibinfo{person}{Qiang Fu},
  \bibinfo{person}{Eeik~Kuan Tan}, {and} \bibinfo{person}{Adrian Flanagan}.}
  \bibinfo{year}{2019}\natexlab{}.
\newblock \showarticletitle{Federated Collaborative Filtering for
  Privacy-Preserving Personalized Recommendation System}.
\newblock \bibinfo{journal}{\emph{arXiv: Information Retrieval}}
  (\bibinfo{year}{2019}).
\newblock


\bibitem[Beigi et~al\mbox{.}(2020)]%
        {beigi2020privacy-aware}
\bibfield{author}{\bibinfo{person}{Ghazaleh Beigi}, \bibinfo{person}{Ahmadreza
  Mosallanezhad}, \bibinfo{person}{Ruocheng Guo}, \bibinfo{person}{Hamidreza
  Alvari}, \bibinfo{person}{Alexander Nou}, {and} \bibinfo{person}{Huan Liu}.}
  \bibinfo{year}{2020}\natexlab{}.
\newblock \showarticletitle{Privacy-Aware Recommendation with Private-Attribute
  Protection using Adversarial Learning}.
\newblock \bibinfo{journal}{\emph{WSDM '20: The Thirteenth ACM International
  Conference on Web Search and Data Mining Houston TX USA February, 2020}}
  (\bibinfo{year}{2020}), \bibinfo{pages}{34--42}.
\newblock


\bibitem[Chai et~al\mbox{.}(2021)]%
        {DBLP:journals/expert/SecureFMF}
\bibfield{author}{\bibinfo{person}{Di Chai}, \bibinfo{person}{Leye Wang},
  \bibinfo{person}{Kai Chen}, {and} \bibinfo{person}{Qiang Yang}.}
  \bibinfo{year}{2021}\natexlab{}.
\newblock \showarticletitle{Secure Federated Matrix Factorization}.
\newblock \bibinfo{journal}{\emph{{IEEE} Intell. Syst.}} \bibinfo{volume}{36},
  \bibinfo{number}{5} (\bibinfo{year}{2021}), \bibinfo{pages}{11--20}.
\newblock


\bibitem[Chen et~al\mbox{.}(2021)]%
        {DBLP:conf/sigir/ChenDQ0XCLY21}
\bibfield{author}{\bibinfo{person}{Jiawei Chen}, \bibinfo{person}{Hande Dong},
  \bibinfo{person}{Yang Qiu}, \bibinfo{person}{Xiangnan He},
  \bibinfo{person}{Xin Xin}, \bibinfo{person}{Liang Chen},
  \bibinfo{person}{Guli Lin}, {and} \bibinfo{person}{Keping Yang}.}
  \bibinfo{year}{2021}\natexlab{}.
\newblock \showarticletitle{AutoDebias: Learning to Debias for Recommendation}.
  In \bibinfo{booktitle}{\emph{{SIGIR}}}. \bibinfo{publisher}{{ACM}},
  \bibinfo{pages}{21--30}.
\newblock


\bibitem[Chen et~al\mbox{.}(2020)]%
        {DBLP:journals/corr/abs-2010-03240}
\bibfield{author}{\bibinfo{person}{Jiawei Chen}, \bibinfo{person}{Hande Dong},
  \bibinfo{person}{Xiang Wang}, \bibinfo{person}{Fuli Feng},
  \bibinfo{person}{Meng Wang}, {and} \bibinfo{person}{Xiangnan He}.}
  \bibinfo{year}{2020}\natexlab{}.
\newblock \showarticletitle{Bias and Debias in Recommender System: {A} Survey
  and Future Directions}.
\newblock \bibinfo{journal}{\emph{CoRR}}  \bibinfo{volume}{abs/2010.03240}
  (\bibinfo{year}{2020}).
\newblock


\bibitem[Chen et~al\mbox{.}(2019)]%
        {googlewsdmoffpolicycorrection}
\bibfield{author}{\bibinfo{person}{Minmin Chen}, \bibinfo{person}{Alex Beutel},
  \bibinfo{person}{Paul Covington}, \bibinfo{person}{Sagar Jain},
  \bibinfo{person}{Francois Belletti}, {and} \bibinfo{person}{Ed~H Chi}.}
  \bibinfo{year}{2019}\natexlab{}.
\newblock \showarticletitle{Top-k off-policy correction for a REINFORCE
  recommender system}. In \bibinfo{booktitle}{\emph{Proceedings of the Twelfth
  ACM International Conference on Web Search and Data Mining}}. ACM,
  \bibinfo{pages}{456--464}.
\newblock


\bibitem[Cheng et~al\mbox{.}(2016)]%
        {cheng2016wide}
\bibfield{author}{\bibinfo{person}{Heng-Tze Cheng}, \bibinfo{person}{Levent
  Koc}, \bibinfo{person}{Jeremiah Harmsen}, \bibinfo{person}{Tal Shaked},
  \bibinfo{person}{Tushar Chandra}, \bibinfo{person}{Hrishi Aradhye},
  \bibinfo{person}{Glen Anderson}, \bibinfo{person}{Greg Corrado},
  \bibinfo{person}{Wei Chai}, \bibinfo{person}{Mustafa Ispir}, {et~al\mbox{.}}}
  \bibinfo{year}{2016}\natexlab{}.
\newblock \showarticletitle{Wide \& deep learning for recommender systems}. In
  \bibinfo{booktitle}{\emph{Proceedings of the 1st workshop on deep learning
  for recommender systems}}. \bibinfo{pages}{7--10}.
\newblock


\bibitem[Cho et~al\mbox{.}(2014)]%
        {DBLP:conf/emnlp/ChoMGBBSB14}
\bibfield{author}{\bibinfo{person}{Kyunghyun Cho}, \bibinfo{person}{Bart van
  Merrienboer}, \bibinfo{person}{{\c{C}}aglar G{\"{u}}l{\c{c}}ehre},
  \bibinfo{person}{Dzmitry Bahdanau}, \bibinfo{person}{Fethi Bougares},
  \bibinfo{person}{Holger Schwenk}, {and} \bibinfo{person}{Yoshua Bengio}.}
  \bibinfo{year}{2014}\natexlab{}.
\newblock \showarticletitle{Learning Phrase Representations using {RNN}
  Encoder-Decoder for Statistical Machine Translation}. In
  \bibinfo{booktitle}{\emph{{EMNLP}}}. \bibinfo{publisher}{{ACL}},
  \bibinfo{pages}{1724--1734}.
\newblock


\bibitem[Deldjoo et~al\mbox{.}(2021)]%
        {DBLP:journals/corr/abs-2005-10322}
\bibfield{author}{\bibinfo{person}{Yashar Deldjoo}, \bibinfo{person}{Tommaso
  {Di Noia}}, {and} \bibinfo{person}{Felice~Antonio Merra}.}
  \bibinfo{year}{2021}\natexlab{}.
\newblock \showarticletitle{A survey on Adversarial Recommender Systems: from
  Attack/Defense strategies to Generative Adversarial Networks}.
\newblock \bibinfo{journal}{\emph{Comput. Surveys}} (\bibinfo{year}{2021}).
\newblock
\urldef\tempurl%
\url{https://doi.org/10.1145/3439729}
\showDOI{\tempurl}


\bibitem[Ding et~al\mbox{.}(2018a)]%
        {ding2018improved}
\bibfield{author}{\bibinfo{person}{Jingtao Ding}, \bibinfo{person}{Fuli Feng},
  \bibinfo{person}{Xiangnan He}, \bibinfo{person}{Guanghui Yu},
  \bibinfo{person}{Yong Li}, {and} \bibinfo{person}{Depeng Jin}.}
  \bibinfo{year}{2018}\natexlab{a}.
\newblock \showarticletitle{An improved sampler for bayesian personalized
  ranking by leveraging view data}. In \bibinfo{booktitle}{\emph{Companion
  Proceedings of the The Web Conference 2018}}. \bibinfo{pages}{13--14}.
\newblock


\bibitem[Ding et~al\mbox{.}(2018b)]%
        {10.1145/3184558.3186905}
\bibfield{author}{\bibinfo{person}{Jingtao Ding}, \bibinfo{person}{Fuli Feng},
  \bibinfo{person}{Xiangnan He}, \bibinfo{person}{Guanghui Yu},
  \bibinfo{person}{Yong Li}, {and} \bibinfo{person}{Depeng Jin}.}
  \bibinfo{year}{2018}\natexlab{b}.
\newblock \showarticletitle{An Improved Sampler for Bayesian Personalized
  Ranking by Leveraging View Data}. In \bibinfo{booktitle}{\emph{Companion
  Proceedings of the The Web Conference 2018}} (Lyon, France)
  \emph{(\bibinfo{series}{WWW '18})}. \bibinfo{publisher}{International World
  Wide Web Conferences Steering Committee}, \bibinfo{address}{Republic and
  Canton of Geneva, CHE}, \bibinfo{pages}{13–14}.
\newblock
\showISBNx{9781450356404}
\urldef\tempurl%
\url{https://doi.org/10.1145/3184558.3186905}
\showDOI{\tempurl}


\bibitem[Ding et~al\mbox{.}(2019)]%
        {DBLP:conf/ijcai/DingQ00J19}
\bibfield{author}{\bibinfo{person}{Jingtao Ding}, \bibinfo{person}{Yuhan Quan},
  \bibinfo{person}{Xiangnan He}, \bibinfo{person}{Yong Li}, {and}
  \bibinfo{person}{Depeng Jin}.} \bibinfo{year}{2019}\natexlab{}.
\newblock \showarticletitle{Reinforced Negative Sampling for Recommendation
  with Exposure Data}. In \bibinfo{booktitle}{\emph{{IJCAI}}}.
  \bibinfo{publisher}{ijcai.org}, \bibinfo{pages}{2230--2236}.
\newblock


\bibitem[Gao et~al\mbox{.}(2020)]%
        {DBLP:conf/sigir/GaoHLJ020}
\bibfield{author}{\bibinfo{person}{Chen Gao}, \bibinfo{person}{Chao Huang},
  \bibinfo{person}{Dongsheng Lin}, \bibinfo{person}{Depeng Jin}, {and}
  \bibinfo{person}{Yong Li}.} \bibinfo{year}{2020}\natexlab{}.
\newblock \showarticletitle{{DPLCF:} Differentially Private Local Collaborative
  Filtering}. In \bibinfo{booktitle}{\emph{{SIGIR}}}.
  \bibinfo{publisher}{{ACM}}, \bibinfo{pages}{961--970}.
\newblock


\bibitem[Ge et~al\mbox{.}(2021)]%
        {doi:10.1177/15501477211061250}
\bibfield{author}{\bibinfo{person}{Zhengqiang Ge}, \bibinfo{person}{Xinyu Liu},
  \bibinfo{person}{Qiang Li}, \bibinfo{person}{Yu Li}, {and}
  \bibinfo{person}{Dong Guo}.} \bibinfo{year}{2021}\natexlab{}.
\newblock \showarticletitle{PrivItem2Vec: A privacy-preserving algorithm for
  top-N recommendation}.
\newblock \bibinfo{journal}{\emph{International Journal of Distributed Sensor
  Networks}} \bibinfo{volume}{17}, \bibinfo{number}{12} (\bibinfo{year}{2021}).
\newblock
\urldef\tempurl%
\url{https://doi.org/10.1177/15501477211061250}
\showDOI{\tempurl}


\bibitem[Han et~al\mbox{.}(2021)]%
        {han2021deeprec}
\bibfield{author}{\bibinfo{person}{Jialiang Han}, \bibinfo{person}{Yun Ma},
  \bibinfo{person}{Qiaozhu Mei}, {and} \bibinfo{person}{Xuanzhe Liu}.}
  \bibinfo{year}{2021}\natexlab{}.
\newblock \showarticletitle{DeepRec: On-device Deep Learning for
  Privacy-Preserving Sequential Recommendation in Mobile Commerce}. In
  \bibinfo{booktitle}{\emph{Proceedings of the Web Conference 2021}}.
  \bibinfo{pages}{900--911}.
\newblock


\bibitem[Hao et~al\mbox{.}(2021)]%
        {hao2021largescale}
\bibfield{author}{\bibinfo{person}{Bin Hao}, \bibinfo{person}{Min Zhang},
  \bibinfo{person}{Weizhi Ma}, \bibinfo{person}{Shaoyun Shi},
  \bibinfo{person}{Xinxing Yu}, \bibinfo{person}{Houzhi Shan},
  \bibinfo{person}{Yiqun Liu}, {and} \bibinfo{person}{Shaoping Ma}.}
  \bibinfo{year}{2021}\natexlab{}.
\newblock \bibinfo{title}{A Large-Scale Rich Context Query and Recommendation
  Dataset in Online Knowledge-Sharing}.
\newblock
\newblock
\showeprint[arxiv]{2106.06467}~[cs.IR]


\bibitem[He et~al\mbox{.}(2019)]%
        {DBLP:conf/cvpr/HeZ0ZXL19}
\bibfield{author}{\bibinfo{person}{Tong He}, \bibinfo{person}{Zhi Zhang},
  \bibinfo{person}{Hang Zhang}, \bibinfo{person}{Zhongyue Zhang},
  \bibinfo{person}{Junyuan Xie}, {and} \bibinfo{person}{Mu Li}.}
  \bibinfo{year}{2019}\natexlab{}.
\newblock \showarticletitle{Bag of Tricks for Image Classification with
  Convolutional Neural Networks}. In \bibinfo{booktitle}{\emph{{CVPR}}}.
  \bibinfo{publisher}{Computer Vision Foundation / {IEEE}},
  \bibinfo{pages}{558--567}.
\newblock


\bibitem[He and Chua(2017)]%
        {he2017nfm}
\bibfield{author}{\bibinfo{person}{Xiangnan He} {and} \bibinfo{person}{Tat-Seng
  Chua}.} \bibinfo{year}{2017}\natexlab{}.
\newblock \showarticletitle{Neural factorization machines for sparse predictive
  analytics}. In \bibinfo{booktitle}{\emph{Proceedings of the 40th
  International ACM SIGIR conference on Research and Development in Information
  Retrieval}}. \bibinfo{pages}{355--364}.
\newblock


\bibitem[He et~al\mbox{.}(2020)]%
        {he2020lightgcn}
\bibfield{author}{\bibinfo{person}{Xiangnan He}, \bibinfo{person}{Kuan Deng},
  \bibinfo{person}{Xiang Wang}, \bibinfo{person}{Yan Li},
  \bibinfo{person}{Yongdong Zhang}, {and} \bibinfo{person}{Meng Wang}.}
  \bibinfo{year}{2020}\natexlab{}.
\newblock \showarticletitle{Lightgcn: Simplifying and powering graph
  convolution network for recommendation}. In
  \bibinfo{booktitle}{\emph{Proceedings of the 43rd International ACM SIGIR
  Conference on Research and Development in Information Retrieval}}.
  \bibinfo{pages}{639--648}.
\newblock


\bibitem[He et~al\mbox{.}(2017)]%
        {he2017ncf}
\bibfield{author}{\bibinfo{person}{Xiangnan He}, \bibinfo{person}{Lizi Liao},
  \bibinfo{person}{Hanwang Zhang}, \bibinfo{person}{Liqiang Nie},
  \bibinfo{person}{Xia Hu}, {and} \bibinfo{person}{Tat-Seng Chua}.}
  \bibinfo{year}{2017}\natexlab{}.
\newblock \showarticletitle{Neural collaborative filtering}. In
  \bibinfo{booktitle}{\emph{Proceedings of the 26th international conference on
  world wide web}}. \bibinfo{pages}{173--182}.
\newblock


\bibitem[Hidasi et~al\mbox{.}(2016)]%
        {DBLP:journals/corr/HidasiKBT15}
\bibfield{author}{\bibinfo{person}{Bal{\'{a}}zs Hidasi},
  \bibinfo{person}{Alexandros Karatzoglou}, \bibinfo{person}{Linas Baltrunas},
  {and} \bibinfo{person}{Domonkos Tikk}.} \bibinfo{year}{2016}\natexlab{}.
\newblock \showarticletitle{Session-based Recommendations with Recurrent Neural
  Networks}. In \bibinfo{booktitle}{\emph{{ICLR} (Poster)}}.
\newblock


\bibitem[Hidasi and Tikk(2016)]%
        {gmf}
\bibfield{author}{\bibinfo{person}{Bal{\'a}zs Hidasi} {and}
  \bibinfo{person}{Domonkos Tikk}.} \bibinfo{year}{2016}\natexlab{}.
\newblock \showarticletitle{General factorization framework for context-aware
  recommendations}.
\newblock \bibinfo{journal}{\emph{Data Mining and Knowledge Discovery}}
  \bibinfo{volume}{30}, \bibinfo{number}{2} (\bibinfo{year}{2016}),
  \bibinfo{pages}{342--371}.
\newblock


\bibitem[Hu et~al\mbox{.}(2018)]%
        {reinforce-e-commerce}
\bibfield{author}{\bibinfo{person}{Yujing Hu}, \bibinfo{person}{Qing Da},
  \bibinfo{person}{Anxiang Zeng}, \bibinfo{person}{Yang Yu}, {and}
  \bibinfo{person}{Yinghui Xu}.} \bibinfo{year}{2018}\natexlab{}.
\newblock \showarticletitle{Reinforcement learning to rank in e-commerce search
  engine: Formalization, analysis, and application}. In
  \bibinfo{booktitle}{\emph{Proceedings of the 24th ACM SIGKDD International
  Conference on Knowledge Discovery \& Data Mining}}. ACM,
  \bibinfo{pages}{368--377}.
\newblock


\bibitem[Inan et~al\mbox{.}(2017)]%
        {DBLP:conf/iclr/InanKS17}
\bibfield{author}{\bibinfo{person}{Hakan Inan}, \bibinfo{person}{Khashayar
  Khosravi}, {and} \bibinfo{person}{Richard Socher}.}
  \bibinfo{year}{2017}\natexlab{}.
\newblock \showarticletitle{Tying Word Vectors and Word Classifiers: {A} Loss
  Framework for Language Modeling}. In \bibinfo{booktitle}{\emph{{ICLR}
  (Poster)}}. \bibinfo{publisher}{OpenReview.net}.
\newblock


\bibitem[Jeckmans et~al\mbox{.}(2013)]%
        {jeckmans2013privacyinrec}
\bibfield{author}{\bibinfo{person}{Arjan~JP Jeckmans}, \bibinfo{person}{Michael
  Beye}, \bibinfo{person}{Zekeriya Erkin}, \bibinfo{person}{Pieter Hartel},
  \bibinfo{person}{Reginald~L Lagendijk}, {and} \bibinfo{person}{Qiang Tang}.}
  \bibinfo{year}{2013}\natexlab{}.
\newblock \showarticletitle{Privacy in recommender systems}.
\newblock In \bibinfo{booktitle}{\emph{Social media retrieval}}.
  \bibinfo{publisher}{Springer}, \bibinfo{pages}{263--281}.
\newblock


\bibitem[Jiang et~al\mbox{.}(2019)]%
        {10.1145/3306618.3314288}
\bibfield{author}{\bibinfo{person}{Ray Jiang}, \bibinfo{person}{Silvia
  Chiappa}, \bibinfo{person}{Tor Lattimore}, \bibinfo{person}{Andr\'{a}s
  Gy\"{o}rgy}, {and} \bibinfo{person}{Pushmeet Kohli}.}
  \bibinfo{year}{2019}\natexlab{}.
\newblock \showarticletitle{Degenerate Feedback Loops in Recommender Systems}.
  In \bibinfo{booktitle}{\emph{Proceedings of the 2019 AAAI/ACM Conference on
  AI, Ethics, and Society}} (Honolulu, HI, USA) \emph{(\bibinfo{series}{AIES
  '19})}. \bibinfo{publisher}{Association for Computing Machinery},
  \bibinfo{address}{New York, NY, USA}, \bibinfo{pages}{383–390}.
\newblock
\showISBNx{9781450363242}
\urldef\tempurl%
\url{https://doi.org/10.1145/3306618.3314288}
\showDOI{\tempurl}


\bibitem[Kabbur et~al\mbox{.}(2013)]%
        {kabbur2013fism}
\bibfield{author}{\bibinfo{person}{Santosh Kabbur}, \bibinfo{person}{Xia Ning},
  {and} \bibinfo{person}{George Karypis}.} \bibinfo{year}{2013}\natexlab{}.
\newblock \showarticletitle{Fism: factored item similarity models for top-n
  recommender systems}. In \bibinfo{booktitle}{\emph{Proceedings of the 19th
  ACM SIGKDD international conference on Knowledge discovery and data mining}}.
  \bibinfo{pages}{659--667}.
\newblock


\bibitem[Kang and McAuley(2018)]%
        {SASRec}
\bibfield{author}{\bibinfo{person}{Wang-Cheng Kang} {and}
  \bibinfo{person}{Julian McAuley}.} \bibinfo{year}{2018}\natexlab{}.
\newblock \showarticletitle{Self-attentive sequential recommendation}. In
  \bibinfo{booktitle}{\emph{2018 IEEE International Conference on Data Mining
  (ICDM)}}. IEEE, \bibinfo{pages}{197--206}.
\newblock


\bibitem[Khan et~al\mbox{.}(2021)]%
        {khan2021payload}
\bibfield{author}{\bibinfo{person}{Farwa~K Khan}, \bibinfo{person}{Adrian
  Flanagan}, \bibinfo{person}{Kuan~Eeik Tan}, \bibinfo{person}{Zareen Alamgir},
  {and} \bibinfo{person}{Muhammad Ammad-Ud-Din}.}
  \bibinfo{year}{2021}\natexlab{}.
\newblock \showarticletitle{A Payload Optimization Method for Federated
  Recommender Systems}. In \bibinfo{booktitle}{\emph{Recsys}}.
  \bibinfo{pages}{432--442}.
\newblock


\bibitem[Kim et~al\mbox{.}(2018)]%
        {DBLP:journals/tissec/KimKKYSK18}
\bibfield{author}{\bibinfo{person}{Jinsu Kim}, \bibinfo{person}{Dongyoung Koo},
  \bibinfo{person}{Yuna Kim}, \bibinfo{person}{Hyunsoo Yoon},
  \bibinfo{person}{Junbum Shin}, {and} \bibinfo{person}{Sungwook Kim}.}
  \bibinfo{year}{2018}\natexlab{}.
\newblock \showarticletitle{Efficient Privacy-Preserving Matrix Factorization
  for Recommendation via Fully Homomorphic Encryption}.
\newblock \bibinfo{journal}{\emph{{ACM} Trans. Priv. Secur.}}
  \bibinfo{volume}{21}, \bibinfo{number}{4} (\bibinfo{year}{2018}),
  \bibinfo{pages}{17:1--17:30}.
\newblock


\bibitem[Kingma and Ba(2015)]%
        {DBLP:journals/corr/KingmaB14}
\bibfield{author}{\bibinfo{person}{Diederik~P. Kingma} {and}
  \bibinfo{person}{Jimmy Ba}.} \bibinfo{year}{2015}\natexlab{}.
\newblock \showarticletitle{Adam: {A} Method for Stochastic Optimization}. In
  \bibinfo{booktitle}{\emph{{ICLR} (Poster)}}.
\newblock


\bibitem[Koren et~al\mbox{.}(2009)]%
        {koren2009mf}
\bibfield{author}{\bibinfo{person}{Yehuda Koren}, \bibinfo{person}{Robert
  Bell}, {and} \bibinfo{person}{Chris Volinsky}.}
  \bibinfo{year}{2009}\natexlab{}.
\newblock \showarticletitle{Matrix factorization techniques for recommender
  systems}.
\newblock \bibinfo{journal}{\emph{Computer}} \bibinfo{volume}{42},
  \bibinfo{number}{8} (\bibinfo{year}{2009}), \bibinfo{pages}{30--37}.
\newblock


\bibitem[Lian et~al\mbox{.}(2017)]%
        {10.1145/3124791.3124794}
\bibfield{author}{\bibinfo{person}{Jianxun Lian}, \bibinfo{person}{Fuzheng
  Zhang}, \bibinfo{person}{Min Hou}, \bibinfo{person}{Hongwei Wang},
  \bibinfo{person}{Xing Xie}, {and} \bibinfo{person}{Guangzhong Sun}.}
  \bibinfo{year}{2017}\natexlab{}.
\newblock \showarticletitle{Practical Lessons for Job Recommendations in the
  Cold-Start Scenario}. In \bibinfo{booktitle}{\emph{Proceedings of the
  Recommender Systems Challenge 2017}} (Como, Italy)
  \emph{(\bibinfo{series}{RecSys Challenge '17})}.
  \bibinfo{publisher}{Association for Computing Machinery},
  \bibinfo{address}{New York, NY, USA}, Article \bibinfo{articleno}{4},
  \bibinfo{numpages}{6}~pages.
\newblock
\showISBNx{9781450353915}
\urldef\tempurl%
\url{https://doi.org/10.1145/3124791.3124794}
\showDOI{\tempurl}


\bibitem[Lin et~al\mbox{.}(2020)]%
        {lin2020fedrecexplicit}
\bibfield{author}{\bibinfo{person}{Guanyu Lin}, \bibinfo{person}{Feng Liang},
  \bibinfo{person}{Weike Pan}, {and} \bibinfo{person}{Zhong Ming}.}
  \bibinfo{year}{2020}\natexlab{}.
\newblock \showarticletitle{Fedrec: Federated recommendation with explicit
  feedback}.
\newblock \bibinfo{journal}{\emph{IEEE Intelligent Systems}}
  (\bibinfo{year}{2020}).
\newblock


\bibitem[Liu et~al\mbox{.}(2020)]%
        {DBLP:conf/sigir/LiuCDHP020}
\bibfield{author}{\bibinfo{person}{Dugang Liu}, \bibinfo{person}{Pengxiang
  Cheng}, \bibinfo{person}{Zhenhua Dong}, \bibinfo{person}{Xiuqiang He},
  \bibinfo{person}{Weike Pan}, {and} \bibinfo{person}{Zhong Ming}.}
  \bibinfo{year}{2020}\natexlab{}.
\newblock \showarticletitle{A General Knowledge Distillation Framework for
  Counterfactual Recommendation via Uniform Data}. In
  \bibinfo{booktitle}{\emph{{SIGIR}}}. \bibinfo{publisher}{{ACM}},
  \bibinfo{pages}{831--840}.
\newblock


\bibitem[Minto et~al\mbox{.}(2021)]%
        {DBLP:conf/recsys/stronger}
\bibfield{author}{\bibinfo{person}{Lorenzo Minto}, \bibinfo{person}{Moritz
  Haller}, \bibinfo{person}{Benjamin Livshits}, {and} \bibinfo{person}{Hamed
  Haddadi}.} \bibinfo{year}{2021}\natexlab{}.
\newblock \showarticletitle{Stronger Privacy for Federated Collaborative
  Filtering With Implicit Feedback}. In \bibinfo{booktitle}{\emph{RecSys}}.
  \bibinfo{publisher}{{ACM}}, \bibinfo{pages}{342--350}.
\newblock


\bibitem[Muhammad et~al\mbox{.}(2020)]%
        {muhammad2020fedfast}
\bibfield{author}{\bibinfo{person}{Khalil Muhammad}, \bibinfo{person}{Qinqin
  Wang}, \bibinfo{person}{Diarmuid O'Reilly-Morgan}, \bibinfo{person}{Elias
  Tragos}, \bibinfo{person}{Barry Smyth}, \bibinfo{person}{Neil Hurley},
  \bibinfo{person}{James Geraci}, {and} \bibinfo{person}{Aonghus Lawlor}.}
  \bibinfo{year}{2020}\natexlab{}.
\newblock \showarticletitle{Fedfast: Going beyond average for faster training
  of federated recommender systems}. In \bibinfo{booktitle}{\emph{Proceedings
  of the 26th ACM SIGKDD International Conference on Knowledge Discovery \&
  Data Mining}}. \bibinfo{pages}{1234--1242}.
\newblock


\bibitem[Ning and Karypis(2011)]%
        {ning2011slim}
\bibfield{author}{\bibinfo{person}{Xia Ning} {and} \bibinfo{person}{George
  Karypis}.} \bibinfo{year}{2011}\natexlab{}.
\newblock \showarticletitle{Slim: Sparse linear methods for top-n recommender
  systems}. In \bibinfo{booktitle}{\emph{2011 IEEE 11th International
  Conference on Data Mining}}. IEEE, \bibinfo{pages}{497--506}.
\newblock


\bibitem[Otto(2018)]%
        {otto2018GDPR}
\bibfield{author}{\bibinfo{person}{Marta Otto}.}
  \bibinfo{year}{2018}\natexlab{}.
\newblock \showarticletitle{Regulation (EU) 2016/679 on the protection of
  natural persons with regard to the processing of personal data and on the
  free movement of such data (General Data Protection Regulation--GDPR)}. In
  \bibinfo{booktitle}{\emph{International and European Labour Law}}. Nomos
  Verlagsgesellschaft mbH \& Co. KG, \bibinfo{pages}{958--981}.
\newblock


\bibitem[Pazzani and Billsus(2007)]%
        {DBLP:conf/adaptive/PazzaniB07}
\bibfield{author}{\bibinfo{person}{Michael~J. Pazzani} {and}
  \bibinfo{person}{Daniel Billsus}.} \bibinfo{year}{2007}\natexlab{}.
\newblock \showarticletitle{Content-Based Recommendation Systems}. In
  \bibinfo{booktitle}{\emph{The Adaptive Web}} \emph{(\bibinfo{series}{Lecture
  Notes in Computer Science}, Vol.~\bibinfo{volume}{4321})}.
  \bibinfo{publisher}{Springer}, \bibinfo{pages}{325--341}.
\newblock


\bibitem[Press and Wolf(2017)]%
        {DBLP:conf/eacl/PressW17}
\bibfield{author}{\bibinfo{person}{Ofir Press} {and} \bibinfo{person}{Lior
  Wolf}.} \bibinfo{year}{2017}\natexlab{}.
\newblock \showarticletitle{Using the Output Embedding to Improve Language
  Models}. In \bibinfo{booktitle}{\emph{{EACL} {(2)}}}.
  \bibinfo{publisher}{Association for Computational Linguistics},
  \bibinfo{pages}{157--163}.
\newblock


\bibitem[Qi et~al\mbox{.}(2020)]%
        {qi2020fednews}
\bibfield{author}{\bibinfo{person}{Tao Qi}, \bibinfo{person}{Fangzhao Wu},
  \bibinfo{person}{Chuhan Wu}, \bibinfo{person}{Yongfeng Huang}, {and}
  \bibinfo{person}{Xing Xie}.} \bibinfo{year}{2020}\natexlab{}.
\newblock \showarticletitle{Privacy-Preserving News Recommendation Model
  Learning}. In \bibinfo{booktitle}{\emph{{EMNLP} (Findings)}}
  \emph{(\bibinfo{series}{Findings of {ACL}}, Vol.~\bibinfo{volume}{{EMNLP}
  2020})}. \bibinfo{publisher}{Association for Computational Linguistics},
  \bibinfo{pages}{1423--1432}.
\newblock


\bibitem[Rendle et~al\mbox{.}(2010)]%
        {DBLP:conf/www/RendleFS10}
\bibfield{author}{\bibinfo{person}{Steffen Rendle}, \bibinfo{person}{Christoph
  Freudenthaler}, {and} \bibinfo{person}{Lars Schmidt{-}Thieme}.}
  \bibinfo{year}{2010}\natexlab{}.
\newblock \showarticletitle{Factorizing personalized Markov chains for
  next-basket recommendation}. In \bibinfo{booktitle}{\emph{{WWW}}}.
  \bibinfo{publisher}{{ACM}}, \bibinfo{pages}{811--820}.
\newblock


\bibitem[Saito et~al\mbox{.}(2020)]%
        {DBLP:conf/wsdm/SaitoYNSN20}
\bibfield{author}{\bibinfo{person}{Yuta Saito}, \bibinfo{person}{Suguru
  Yaginuma}, \bibinfo{person}{Yuta Nishino}, \bibinfo{person}{Hayato Sakata},
  {and} \bibinfo{person}{Kazuhide Nakata}.} \bibinfo{year}{2020}\natexlab{}.
\newblock \showarticletitle{Unbiased Recommender Learning from
  Missing-Not-At-Random Implicit Feedback}. In
  \bibinfo{booktitle}{\emph{{WSDM}}}. \bibinfo{publisher}{{ACM}},
  \bibinfo{pages}{501--509}.
\newblock


\bibitem[Sak et~al\mbox{.}(2014)]%
        {DBLP:conf/interspeech/SakSB14}
\bibfield{author}{\bibinfo{person}{Hasim Sak}, \bibinfo{person}{Andrew~W.
  Senior}, {and} \bibinfo{person}{Fran{\c{c}}oise Beaufays}.}
  \bibinfo{year}{2014}\natexlab{}.
\newblock \showarticletitle{Long short-term memory recurrent neural network
  architectures for large scale acoustic modeling}. In
  \bibinfo{booktitle}{\emph{{INTERSPEECH}}}. \bibinfo{publisher}{{ISCA}},
  \bibinfo{pages}{338--342}.
\newblock


\bibitem[Sarwar et~al\mbox{.}(2001)]%
        {sarwar2001itemcf}
\bibfield{author}{\bibinfo{person}{Badrul Sarwar}, \bibinfo{person}{George
  Karypis}, \bibinfo{person}{Joseph Konstan}, {and} \bibinfo{person}{John
  Riedl}.} \bibinfo{year}{2001}\natexlab{}.
\newblock \showarticletitle{Item-based collaborative filtering recommendation
  algorithms}. In \bibinfo{booktitle}{\emph{Proceedings of the 10th
  international conference on World Wide Web}}. \bibinfo{pages}{285--295}.
\newblock


\bibitem[Shin et~al\mbox{.}(2018)]%
        {DBLP:journals/tkde/ShinKSX18}
\bibfield{author}{\bibinfo{person}{Hyejin Shin}, \bibinfo{person}{Sungwook
  Kim}, \bibinfo{person}{Junbum Shin}, {and} \bibinfo{person}{Xiaokui Xiao}.}
  \bibinfo{year}{2018}\natexlab{}.
\newblock \showarticletitle{Privacy Enhanced Matrix Factorization for
  Recommendation with Local Differential Privacy}.
\newblock \bibinfo{journal}{\emph{{IEEE} Trans. Knowl. Data Eng.}}
  \bibinfo{volume}{30}, \bibinfo{number}{9} (\bibinfo{year}{2018}),
  \bibinfo{pages}{1770--1782}.
\newblock


\bibitem[Shokri et~al\mbox{.}(2017)]%
        {shokri2017membershipblackbox}
\bibfield{author}{\bibinfo{person}{Reza Shokri}, \bibinfo{person}{Marco
  Stronati}, \bibinfo{person}{Congzheng Song}, {and} \bibinfo{person}{Vitaly
  Shmatikov}.} \bibinfo{year}{2017}\natexlab{}.
\newblock \showarticletitle{Membership inference attacks against machine
  learning models}. In \bibinfo{booktitle}{\emph{2017 IEEE Symposium on
  Security and Privacy (SP)}}. IEEE, \bibinfo{pages}{3--18}.
\newblock


\bibitem[Sotto et~al\mbox{.}(2010)]%
        {sotto2010privacy}
\bibfield{author}{\bibinfo{person}{Lisa~J Sotto}, \bibinfo{person}{Bridget~C
  Treacy}, {and} \bibinfo{person}{Melinda~L McLellan}.}
  \bibinfo{year}{2010}\natexlab{}.
\newblock \showarticletitle{Privacy and Data Security Risks in Cloud
  Computing.}
\newblock \bibinfo{journal}{\emph{World Communications Regulation Report}}
  \bibinfo{volume}{5}, \bibinfo{number}{2} (\bibinfo{year}{2010}),
  \bibinfo{pages}{38}.
\newblock


\bibitem[Sun et~al\mbox{.}(2019)]%
        {sun2019bert4rec}
\bibfield{author}{\bibinfo{person}{Fei Sun}, \bibinfo{person}{Jun Liu},
  \bibinfo{person}{Jian Wu}, \bibinfo{person}{Changhua Pei},
  \bibinfo{person}{Xiao Lin}, \bibinfo{person}{Wenwu Ou}, {and}
  \bibinfo{person}{Peng Jiang}.} \bibinfo{year}{2019}\natexlab{}.
\newblock \showarticletitle{BERT4Rec: Sequential recommendation with
  bidirectional encoder representations from transformer}. In
  \bibinfo{booktitle}{\emph{Proceedings of the 28th ACM international
  conference on information and knowledge management}}.
  \bibinfo{pages}{1441--1450}.
\newblock


\bibitem[Szegedy et~al\mbox{.}(2016)]%
        {DBLP:conf/cvpr/SzegedyVISW16}
\bibfield{author}{\bibinfo{person}{Christian Szegedy}, \bibinfo{person}{Vincent
  Vanhoucke}, \bibinfo{person}{Sergey Ioffe}, \bibinfo{person}{Jonathon
  Shlens}, {and} \bibinfo{person}{Zbigniew Wojna}.}
  \bibinfo{year}{2016}\natexlab{}.
\newblock \showarticletitle{Rethinking the Inception Architecture for Computer
  Vision}. In \bibinfo{booktitle}{\emph{{CVPR}}}. \bibinfo{publisher}{{IEEE}
  Computer Society}, \bibinfo{pages}{2818--2826}.
\newblock


\bibitem[Vaswani et~al\mbox{.}(2017)]%
        {Transformer}
\bibfield{author}{\bibinfo{person}{Ashish Vaswani}, \bibinfo{person}{Noam
  Shazeer}, \bibinfo{person}{Niki Parmar}, \bibinfo{person}{Jakob Uszkoreit},
  \bibinfo{person}{Llion Jones}, \bibinfo{person}{Aidan~N. Gomez},
  \bibinfo{person}{Lukasz Kaiser}, {and} \bibinfo{person}{Illia Polosukhin}.}
  \bibinfo{year}{2017}\natexlab{}.
\newblock \showarticletitle{Attention is All you Need}. In
  \bibinfo{booktitle}{\emph{{NIPS}}}. \bibinfo{pages}{5998--6008}.
\newblock


\bibitem[Wang et~al\mbox{.}(2021)]%
        {wang2021fast}
\bibfield{author}{\bibinfo{person}{Qinyong Wang}, \bibinfo{person}{Hongzhi
  Yin}, \bibinfo{person}{Tong Chen}, \bibinfo{person}{Junliang Yu},
  \bibinfo{person}{Alexander Zhou}, {and} \bibinfo{person}{Xiangliang Zhang}.}
  \bibinfo{year}{2021}\natexlab{}.
\newblock \showarticletitle{Fast-adapting and privacy-preserving federated
  recommender system}.
\newblock \bibinfo{journal}{\emph{The VLDB Journal}} (\bibinfo{year}{2021}),
  \bibinfo{pages}{1--20}.
\newblock


\bibitem[Wang et~al\mbox{.}(2019)]%
        {wang2019neural}
\bibfield{author}{\bibinfo{person}{Xiang Wang}, \bibinfo{person}{Xiangnan He},
  \bibinfo{person}{Meng Wang}, \bibinfo{person}{Fuli Feng}, {and}
  \bibinfo{person}{Tat-Seng Chua}.} \bibinfo{year}{2019}\natexlab{}.
\newblock \showarticletitle{Neural graph collaborative filtering}. In
  \bibinfo{booktitle}{\emph{Proceedings of the 42nd international ACM SIGIR
  conference on Research and development in Information Retrieval}}.
  \bibinfo{pages}{165--174}.
\newblock


\bibitem[Wang et~al\mbox{.}(2020)]%
        {wang2020reinforced}
\bibfield{author}{\bibinfo{person}{Xiang Wang}, \bibinfo{person}{Yaokun Xu},
  \bibinfo{person}{Xiangnan He}, \bibinfo{person}{Yixin Cao},
  \bibinfo{person}{Meng Wang}, {and} \bibinfo{person}{Tat-Seng Chua}.}
  \bibinfo{year}{2020}\natexlab{}.
\newblock \showarticletitle{Reinforced negative sampling over knowledge graph
  for recommendation}. In \bibinfo{booktitle}{\emph{Proceedings of The Web
  Conference 2020}}. \bibinfo{pages}{99--109}.
\newblock


\bibitem[Wu et~al\mbox{.}(2021)]%
        {wu2021fedgnn}
\bibfield{author}{\bibinfo{person}{Chuhan Wu}, \bibinfo{person}{Fangzhao Wu},
  \bibinfo{person}{Yang Cao}, \bibinfo{person}{Yongfeng Huang}, {and}
  \bibinfo{person}{Xing Xie}.} \bibinfo{year}{2021}\natexlab{}.
\newblock \showarticletitle{Fedgnn: Federated graph neural network for
  privacy-preserving recommendation}.
\newblock \bibinfo{journal}{\emph{arXiv preprint arXiv:2102.04925}}
  (\bibinfo{year}{2021}).
\newblock


\bibitem[Wu et~al\mbox{.}(2020)]%
        {DBLP:conf/acl/WuQCWQLLXGWZ20}
\bibfield{author}{\bibinfo{person}{Fangzhao Wu}, \bibinfo{person}{Ying Qiao},
  \bibinfo{person}{Jiun{-}Hung Chen}, \bibinfo{person}{Chuhan Wu},
  \bibinfo{person}{Tao Qi}, \bibinfo{person}{Jianxun Lian},
  \bibinfo{person}{Danyang Liu}, \bibinfo{person}{Xing Xie},
  \bibinfo{person}{Jianfeng Gao}, \bibinfo{person}{Winnie Wu}, {and}
  \bibinfo{person}{Ming Zhou}.} \bibinfo{year}{2020}\natexlab{}.
\newblock \showarticletitle{{MIND:} {A} Large-scale Dataset for News
  Recommendation}. In \bibinfo{booktitle}{\emph{{ACL}}}.
  \bibinfo{publisher}{Association for Computational Linguistics},
  \bibinfo{pages}{3597--3606}.
\newblock


\bibitem[Wu et~al\mbox{.}(2016)]%
        {wu2016collaborative}
\bibfield{author}{\bibinfo{person}{Yao Wu}, \bibinfo{person}{Christopher
  DuBois}, \bibinfo{person}{Alice~X Zheng}, {and} \bibinfo{person}{Martin
  Ester}.} \bibinfo{year}{2016}\natexlab{}.
\newblock \showarticletitle{Collaborative denoising auto-encoders for top-n
  recommender systems}. In \bibinfo{booktitle}{\emph{Proceedings of the Ninth
  ACM International Conference on Web Search and Data Mining}}.
  \bibinfo{pages}{153--162}.
\newblock


\bibitem[Xin et~al\mbox{.}(2020)]%
        {xin2020self}
\bibfield{author}{\bibinfo{person}{Xin Xin}, \bibinfo{person}{Alexandros
  Karatzoglou}, \bibinfo{person}{Ioannis Arapakis}, {and}
  \bibinfo{person}{Joemon~M Jose}.} \bibinfo{year}{2020}\natexlab{}.
\newblock \showarticletitle{Self-supervised reinforcement learning for
  recommender systems}. In \bibinfo{booktitle}{\emph{Proceedings of the 43rd
  International ACM SIGIR Conference on Research and Development in Information
  Retrieval}}. \bibinfo{pages}{931--940}.
\newblock


\bibitem[Yang et~al\mbox{.}(2018)]%
        {yang2018hop}
\bibfield{author}{\bibinfo{person}{Jheng-Hong Yang}, \bibinfo{person}{Chih-Ming
  Chen}, \bibinfo{person}{Chuan-Ju Wang}, {and} \bibinfo{person}{Ming-Feng
  Tsai}.} \bibinfo{year}{2018}\natexlab{}.
\newblock \showarticletitle{HOP-rec: high-order proximity for implicit
  recommendation}. In \bibinfo{booktitle}{\emph{Recsys}}.
  \bibinfo{pages}{140--144}.
\newblock


\bibitem[Yang et~al\mbox{.}(2019)]%
        {yang2019federated}
\bibfield{author}{\bibinfo{person}{Qiang Yang}, \bibinfo{person}{Yang Liu},
  \bibinfo{person}{Yong Cheng}, \bibinfo{person}{Yan Kang},
  \bibinfo{person}{Tianjian Chen}, {and} \bibinfo{person}{Han Yu}.}
  \bibinfo{year}{2019}\natexlab{}.
\newblock \showarticletitle{Federated learning}.
\newblock \bibinfo{journal}{\emph{Synthesis Lectures on Artificial Intelligence
  and Machine Learning}} \bibinfo{volume}{13}, \bibinfo{number}{3}
  (\bibinfo{year}{2019}), \bibinfo{pages}{1--207}.
\newblock


\bibitem[Yuan et~al\mbox{.}(2020)]%
        {yuan2020peterec}
\bibfield{author}{\bibinfo{person}{Fajie Yuan}, \bibinfo{person}{Xiangnan He},
  \bibinfo{person}{Alexandros Karatzoglou}, {and} \bibinfo{person}{Liguang
  Zhang}.} \bibinfo{year}{2020}\natexlab{}.
\newblock \showarticletitle{Parameter-efficient transfer from sequential
  behaviors for user modeling and recommendation}. In
  \bibinfo{booktitle}{\emph{Proceedings of the 43rd International ACM SIGIR
  Conference on Research and Development in Information Retrieval}}.
  \bibinfo{pages}{1469--1478}.
\newblock


\bibitem[Yuan et~al\mbox{.}(2019)]%
        {nextitnet}
\bibfield{author}{\bibinfo{person}{Fajie Yuan}, \bibinfo{person}{Alexandros
  Karatzoglou}, \bibinfo{person}{Ioannis Arapakis}, \bibinfo{person}{Joemon~M
  Jose}, {and} \bibinfo{person}{Xiangnan He}.} \bibinfo{year}{2019}\natexlab{}.
\newblock \showarticletitle{A Simple Convolutional Generative Network for Next
  Item Recommendation}. In \bibinfo{booktitle}{\emph{Proceedings of the Twelfth
  ACM International Conference on Web Search and Data Mining}}. ACM,
  \bibinfo{pages}{582--590}.
\newblock


\bibitem[Zhang et~al\mbox{.}(2021b)]%
        {DBLP:conf/ccs/ZhangRWRCHZ21}
\bibfield{author}{\bibinfo{person}{Minxing Zhang}, \bibinfo{person}{Zhaochun
  Ren}, \bibinfo{person}{Zihan Wang}, \bibinfo{person}{Pengjie Ren},
  \bibinfo{person}{Zhumin Chen}, \bibinfo{person}{Pengfei Hu}, {and}
  \bibinfo{person}{Yang Zhang}.} \bibinfo{year}{2021}\natexlab{b}.
\newblock \showarticletitle{Membership Inference Attacks Against Recommender
  Systems}. In \bibinfo{booktitle}{\emph{{CCS}}}. \bibinfo{publisher}{{ACM}},
  \bibinfo{pages}{864--879}.
\newblock


\bibitem[Zhang and Yin(2022)]%
        {DBLP:journals/corr/abs-2205-11857}
\bibfield{author}{\bibinfo{person}{Shijie Zhang} {and} \bibinfo{person}{Hongzhi
  Yin}.} \bibinfo{year}{2022}\natexlab{}.
\newblock \showarticletitle{Comprehensive Privacy Analysis on Federated
  Recommender System against Attribute Inference Attacks}.
\newblock \bibinfo{journal}{\emph{CoRR}}  \bibinfo{volume}{abs/2205.11857}
  (\bibinfo{year}{2022}).
\newblock


\bibitem[Zhang et~al\mbox{.}(2021c)]%
        {DBLP:conf/www/ZhangY0HC021}
\bibfield{author}{\bibinfo{person}{Shijie Zhang}, \bibinfo{person}{Hongzhi
  Yin}, \bibinfo{person}{Tong Chen}, \bibinfo{person}{Zi Huang},
  \bibinfo{person}{Lizhen Cui}, {and} \bibinfo{person}{Xiangliang Zhang}.}
  \bibinfo{year}{2021}\natexlab{c}.
\newblock \showarticletitle{Graph Embedding for Recommendation against
  Attribute Inference Attacks}. In \bibinfo{booktitle}{\emph{{WWW}}}.
  \bibinfo{publisher}{{ACM} / {IW3C2}}, \bibinfo{pages}{3002--3014}.
\newblock


\bibitem[Zhang et~al\mbox{.}(2021a)]%
        {DBLP:conf/sigir/ZhangF0WSL021}
\bibfield{author}{\bibinfo{person}{Yang Zhang}, \bibinfo{person}{Fuli Feng},
  \bibinfo{person}{Xiangnan He}, \bibinfo{person}{Tianxin Wei},
  \bibinfo{person}{Chonggang Song}, \bibinfo{person}{Guohui Ling}, {and}
  \bibinfo{person}{Yongdong Zhang}.} \bibinfo{year}{2021}\natexlab{a}.
\newblock \showarticletitle{Causal Intervention for Leveraging Popularity Bias
  in Recommendation}. In \bibinfo{booktitle}{\emph{{SIGIR}}}.
  \bibinfo{publisher}{{ACM}}, \bibinfo{pages}{11--20}.
\newblock


\end{thebibliography}

\end{document}